\documentclass[11pt]{article}
\pdfoutput=1

\usepackage[a4paper,margin=1in]{geometry}
\usepackage[T1]{fontenc}
\usepackage{lmodern}
\usepackage[protrusion=true,expansion=false]{microtype}
\usepackage{xspace}
\usepackage{amsmath,amssymb,amsfonts,amsthm}
\usepackage{mathtools}
\usepackage{algorithm}
\usepackage[noend]{algpseudocode}
\usepackage{booktabs}
\usepackage{multirow}
\usepackage{array}
\usepackage{caption}
\usepackage{enumitem}
\usepackage{graphicx}
\usepackage{xcolor}
\usepackage{authblk}
\usepackage[colorlinks=true,linkcolor=blue!50!black,citecolor=blue!50!black,urlcolor=blue!50!black]{hyperref}
\hypersetup{pdftitle={Ozaki 2.5: Engineering the Deconstruction Path of
FP64-Emulated Dense Matrix Multiplication on FP8 Tensor Cores},
pdfauthor={Satoshi Matsuoka},
pdfsubject={FP64 emulation on FP8/INT8 tensor cores ---
deconstruction-path engineering},
pdfkeywords={FP8, INT8, Ozaki scheme, CRT, tensor cores, GPU,
mixed precision}}

\newcommand{\eg}{e.g.,\xspace}
\newcommand{\ie}{i.e.,\xspace}
\newcommand{\fp}[1]{\textsc{fp#1}}
\newcommand{\inteight}{\textsc{int8}\xspace}
\newcommand{\intthirtytwo}{\textsc{int32}\xspace}

\newcommand{\tops}{TOPS}
\newcommand{\pflops}{PFLOPS}
\newcommand{\nbar}{\bar{n}}


\theoremstyle{definition}
\newtheorem{definition}{Definition}
\theoremstyle{plain}
\newtheorem{prediction}{Prediction}
\newtheorem{lemma}{Lemma}

\title{\LARGE\bfseries Ozaki 2.5: Engineering the Deconstruction Path of
\fp{64}-Emulated Dense Matrix Multiplication on FP8 Tensor
Cores\thanks{``Ozaki~2.5'' names an \emph{implementation-level} proposal:
the mathematics of Ozaki Scheme~II is unchanged on the published
modulus set (accuracy inheritance is proved for the round-to-nearest
variant; the implemented truncation rule is a stated obligation, and
codesigned variants carry per-set proof obligations,
Appendix~A), and the terminology will
be coordinated with the Ozaki-II/FP8-Ozaki-II authors.  All performance
numbers in this paper are conditional reduced-model projections; no new
kernel is measured here.  A companion paper, ``FP8 is All You Need''
(Part~1, arXiv:2606.06510), covers the shared tensor--memory equilibrium
model and hardware co-design proposal.}}

\author{%
  \large Satoshi Matsuoka\thanks{Correspondence:
  \href{mailto:matsu@acm.org}{\texttt{matsu@acm.org}}}\\
  \normalsize Director, RIKEN Center for Computational Science (R-CCS)\\
  \normalsize Kobe, Hyogo, Japan}
\date{Draft 27 of September 3, 2026}

\begin{document}
\maketitle

\begin{abstract}
\noindent
FP8 Ozaki~II emulates \fp{64} matrix multiplication with a fixed
schedule of low-precision tensor-core products over a CRT residue system.
Converting the \fp{64} operands into residue planes---the \emph{deconstruction}
charged by the fourth term of the Tensor--Memory Equilibrium (TME) model of
Part~1~\cite{matsuoka2026fp8part1}, identified in the NVIDIA technical
review~\cite{bayraktar2026tme}---consumes integer-pipe and memory resources
before any tensor instruction issues.  This paper engineers that path.  It is
a companion to Part~1 but an independent contribution: Part~1 argues a
platform-coverage thesis and prices deconstruction as one cost term among four,
never opening the conversion path itself, whereas every technical result
below---the split closures, the two-limb kernels, the modulus/encoding
codesign, the route menu and its crossover algebra, the reconstruction-aware
engine, and the closed-form reach floor with its hardware asks---appears only
here.  We make
four contributions, all conditional reduced-model projections pending the
measurement campaign we specify; no new kernel is measured here.

\emph{(1)~A deconstruction-aware performance model, with the honest ceiling
stated up front.}  Deconstruction competes with the low-precision stream through
the harmonic mean $\nbar = 2mn/(m{+}n)$ of the output dimensions: below a size
crossover the emulated rate rises with $\nbar$ (the branch through NVIDIA's
preliminary $\sim\!200$-TFLOPS ``Emulated DGEMM'' at model-inferred
$\nbar\approx500$--$900$) and approaches the raw arithmetic roof
$P_{\fp{8}}/(3r{+}1)$ ($\approx\!473$~TFLOPS at Rubin's $17.5$-\pflops{} rate,
$r{=}12$) only within a single thread-block cluster (a ${\le}256^2$ output).
A larger output is re-split on the fly, once per cluster, holding the achieved
rate at the \emph{deconstruction-$\lambda$ floor}, $\approx\!235$~TFLOPS on
Rubin for cluster-aligned large squares---half that $473$-TFLOPS roof, the
one-half being a derived ratio of three design integers (cluster reach,
pipe-provisioning ratio, plane-formation cost), not a fitted
efficiency---and the
regime a DGEMM
benchmark such as HPL runs in (${\approx}1.9$~EFLOPS \fp{64} on a
$10{,}000$-GPU cluster).  That figure is route- and shape-conditional and we
state the condition up front: it is the codesigned hybrid set~E at
$\textrm{NB}\approx1024$, the upper end of the cited
$892$--$1024$ GPU-HPL panel-width range; the published set~S, which
carries the round-to-nearest Ozaki~II theorem, floors at
$\approx\!182$~TFLOPS ($0.38$ of its $473$-TFLOPS roof,
${\approx}1.4$--$1.5$~EFLOPS).  Ragged,
non-aligned sizes dip further, to a disclosed rigid-schedule
worst case of ${\approx}0.36$ of that same roof---the only minimum we
construct.  A
ragged-edge-aware (predicated) schedule is \emph{modeled} to recover
${\approx}0.54$ there; we log that as a Part-3 validation target, not a
second guaranteed floor.
Crossover and floor are one curve, drawn together (Fig.~\ref{fig:knee}), not a
late reveal.  The positive corollary is that the floor is a \emph{large-square}
phenomenon: the tall/skinny and small-batch shapes that dominate real solvers
(block-Krylov, batched GEMMs, panel factorisations) re-split their big operand
only once ($\lambda_A{=}1$, bounded $\lambda_{\mathrm{eff}}{\approx}1.25$--$1.5$)
and stay near the crossover---only lightly clipped ($0.89$--$0.94\times$), never
at the floor---so Ozaki~2.5 is already worth ${\approx}1.6$--$1.9\times$
(${\approx}2\times$ on block-Krylov shapes) over simple deconstruction on Rubin
today, with no hardware change (Table~\ref{tab:apps}).

\emph{(2)~The Ozaki~2.5 method.}  Leaving the Ozaki~II reconstruction framework
unchanged, we engineer the deconstruction path around the \emph{actual}
9--11-bit FP8 modulus set: convert-once residue-plane workspaces, an exact
\emph{two-limb} constant-reduction GEMM on integer tensor pipes (or its
pure-SIMT \texttt{dp4a} realisation), a ${\approx}6.3$-instruction SIMT residual,
and conversion pipelined behind the MMA stream---moving the Rubin crossover from
${\approx}1{,}211$ to ${\approx}480$--$730$ and modeled throughput at
$\nbar{=}512$ from ${\approx}200$ to $364$--$438$~TFLOPS (a conditional
$1.8$--$2.2\times$ convert-once envelope).

\emph{(3)~Modulus/encoding codesign.}  A script-checked study shows the modulus
set is itself a performance parameter: an all-byte system (15 coprime moduli
$\le256$) legalises a one-pass reduction and is projected \emph{faster} below
$\nbar\approx540$ despite a $20\%$ lower roof; a hybrid (squares $\le33^2$ plus a
byte tail) concedes only $7.5\%$ asymptotically while dominating below
$\nbar\approx620$; and two exhaustive supply bounds (squares cap at $94.1$ bits,
sub-64 systems at $89.9$, both short of the $111.8$ required) close the
space---motivating a regime-switched modulus dispatch.

\emph{(4)~Potential hardware co-design.}  Because the floor is the closed-form
$R\,P_{\text{int}}/(c_q r)$---a per-element deconstruction load
$c_q r/\mathrm{reach}$---it names its own escape: we specify cluster reach, TMEM
capacity, and L2 service bandwidth for dense DGEMM (with a target table), and the
deconstruction-cost ($c_q$) datapath of Part~1---minimally an in-flight-convert
copy engine, which also un-binds the conversion-bound sparse kernels---each a
concrete, measurable ask that lifts the emulated rate toward the arithmetic roof,
none yet present in silicon.

None of this is Rubin-specific: instantiated at Blackwell~Ultra (GB300) rates the
same machinery yields a $135$-TFLOPS roof already at its own floor (roof-bound),
a ${\sim}166$-TOPS residual-\inteight{} cap that reshuffles the route ranking
(the all-byte set leading via pure-SIMT \texttt{dp4a}), and, against a
${\approx}1.4$-TFLOPS native \fp{64} pipe, every modeled route above native for
$\nbar\ge32$---so the method is present-tense on shipping hardware.  Measured
GEMM/SYRK call-shape traces across four application classes (LOBPCG, multifrontal
LU, CCSD, blocked LAPACK) ground the shape analysis.  Every rate is flagged
\emph{achieved today}---meaning modeled on shipping silicon with no hardware
change, a reduced-model projection, \emph{not} a measured kernel---or
\emph{convert-once envelope} (clipping-limited) where it
appears; four falsifiable predictions with a decisive size sweep, baselines
(GEMMul8~\cite{gemmul8_github}, fused-kernel work~\cite{emugemm2026}), and
negative controls define the test, and all scripts, traces, and the parameter
file generating every figure and table are supplied with this submission as
a reproduction artifact (arXiv ancillary files); an archival Zenodo DOI
and commit hash will be added in the first arXiv revision.
\end{abstract}

\vspace{0.4em}
\noindent\textbf{Keywords:} FP64 emulation; Ozaki Scheme II; Ozaki 2.5;
FP8 tensor cores; deconstruction cost; dense matrix multiplication;
hardware co-design; Tensor--Memory Equilibrium model; NVIDIA Blackwell (GB300); NVIDIA Rubin.

\bigskip

\section{Introduction}
\label{sec:intro}

Double-precision dense matrix multiplication is entering its emulated
era.  On NVIDIA's Rubin generation, native \fp{64} matrix throughput is
no longer the headline; instead the official specifications list
\emph{``Emulated DGEMM''} as a first-class column---$\sim\!200$~TFLOPS
per GPU~\cite{nvidia_rubin_blog,lockwood_rubin}---a preliminary,
``up to'' specification whose algorithm, mode, and benchmark
dimensions are not published; the natural candidate is the Ozaki~II
scheme~\cite{ozaki2025_scheme2} on \fp{8} tensor
cores~\cite{uchino2026_fp8,mukunoki2025_fp8}, which NVIDIA's public
material describes as the direction of its emulation
work~\cite{nvidia_cublas_emulation}.  NVIDIA thus positions emulated DGEMM as the primary high-throughput
\fp{64} matrix path, and its projected rate is a product
specification.

That specification invites a comparison the companion Part~1
paper~\cite{matsuoka2026fp8part1} makes precise.  At the official
17.5-\pflops{} dense \fp{8} rate of the DGX Rubin NVL8
specification~\cite{nvidia_dgx_rubin_nvl8}, the analytic ceiling of
\fp{8} Ozaki~II at $r{=}12$ moduli---$3r{+}1 = 37$ \fp{8} MMA passes
per \fp{64} product---is
\begin{equation*}
\frac{P_{\fp{8}}}{3r+1} \;=\; \frac{17{,}500}{37}
\;\approx\; 473~\text{TFLOPS},
\end{equation*}
so the announced specification is $\approx\!42\%$ of the raw
arithmetic quotient.  That $473$-TFLOPS quotient is a peak-operation \emph{roof},
not an attainable ceiling---tile efficiency, launches, reconstruction,
and resource contention all take their share, and B200 calibration
suggests a sustained-\fp{8} efficiency well below unity at large
sizes~\cite{uchino2026_fp8}---so a delivered efficiency of ${\sim}40\%$
(a ${\sim}60\%$ shortfall) could be unremarkable.  But part of the margin could also have a specific,
identifiable, and largely removable cause, with a size-dependent
signature that the alternative lacks.

\textbf{The answer, up front.}  The removable part is the deconstruction
cost, and honesty requires stating its ceiling \emph{before} the constructions
that chip at it.  For a \emph{small} output the emulated rate rises with $\nbar$
and can approach the roof; but a \emph{large} dense output---the regime a DGEMM
benchmark such as HPL runs in---spans many thread-block clusters, on-the-fly
conversion repeats once per cluster, and the achieved rate settles at the
\emph{deconstruction-$\lambda$ floor}---${\approx}235$~TFLOPS on Rubin for
cluster-aligned sizes, half the $473$-TFLOP roof, with ragged sizes dipping to
a disclosed rigid-schedule worst case (Fig.~\ref{fig:knee}, drawn \emph{with} that
floor so it is visible early, with the crossover, not only at the end).
Ozaki~2.5 thus does \emph{not} deliver the raw roof for large DGEMM on today's
silicon; it delivers about half---still ${\approx}8\times$ native \fp{64},
consistent with NVIDIA's ${\sim}200$-TFLOP figure---and, because the floor is
set by the load $c_q r/\mathrm{reach}$, a co-design coordinate we lift back to the roof
(\S\ref{sec:hw}).  The equally important positive is that the floor is a
\emph{large-square} phenomenon: the tall/skinny and small-batch matrices that
dominate real solvers (block-Krylov, batched GEMMs, panel factorisations)
re-split their big operand only once ($\lambda_A{=}1$; bounded
$\lambda_{\mathrm{eff}}{\approx}1.25$--$1.5$), stay near the crossover (a mild
$0.89$--$0.94\times$, not the floor), and are already worth
${\approx}1.6$--$1.9\times$ (${\approx}2\times$ on block-Krylov shapes) over
simple deconstruction with no hardware change (Table~\ref{tab:apps}).  Every
performance number below is flagged \emph{achieved today} or \emph{convert-once
envelope} (clipping-limited) where it appears---where \emph{achieved today}
means \emph{modeled to be attainable on today's silicon with no hardware
change}, a reduced-model projection like every rate here, not a measured
kernel.  The rival explanation for the
margin---a size-independent efficiency artefact---carries an orthogonal
signature that a single size sweep separates.

\subsection{Prior work}
\label{sec:prior}
The Ozaki scheme writes a high-precision product as a sum of low-precision inner
products over a splitting/residue system; the error-free \inteight{} slicing of
the Ozaki~I lineage~\cite{ozaki2012} and the CRT-based Scheme~II of
Ozaki~et~al.~\cite{ozaki2025_scheme2} are the two realisations relevant here.
Its use for \fp{64} emulation on \fp{8} tensor cores has been developed and
measured by Uchino~et~al.~\cite{uchino2026_fp8} and
Mukunoki~et~al.~\cite{mukunoki2025_fp8}, with accuracy guarantees proved under
round-to-nearest hypotheses~\cite{ozaki_error_analysis_2026,schwarz2025};
GEMMul8~\cite{gemmul8_github} and fused-kernel emulation~\cite{emugemm2026} are
the public baselines.  NVIDIA's material positions Ozaki-style emulation as the
direction of its ``Emulated DGEMM'' path~\cite{nvidia_cublas_emulation}, and its
technical review~\cite{bayraktar2026tme} first isolated the per-input
deconstruction cost that the companion
Part~1~\cite{matsuoka2026fp8part1} folds into the TME performance model as a
fourth term.  This body of work establishes the scheme, proves its accuracy, and
reports achieved throughput.  What it does not do---and what this paper
adds---is engineer the deconstruction path around the \emph{actual} FP8 modulus
set (the accuracy contract inherits only for the round-to-nearest variant; the
shipping truncation rule and the codesigned sets carry stated obligations,
Appendix~\ref{app:obligations}), treat the modulus set as a runtime performance
variable, and localise the large-DGEMM ceiling as the closed-form
$R\,P_{\text{int}}/(c_q r)$ floor that names its own hardware escape.  The technical
recap the rest of the paper builds on---the Ozaki schemes and the
\inteight{}$\to$\fp{8} substrate transition that created the deconstruction
cost---is \S\ref{sec:ozaki2}.

\subsection{Contributions and roadmap}
\label{sec:contrib}
The name is chosen deliberately: the method adds no new mathematics to
Ozaki~II---the moduli and the reconstruction framework are untouched---and stops
short of the hardware datapaths whose evaluation belongs to follow-up
co-design work.  It is the missing half-step, \emph{Ozaki~II with the conversion
path engineered as deliberately as the multiplication path always has been}, an
engineering need that did not arise before the substrate transition.
Concretely:

\begin{itemize}
\item \textbf{A deconstruction-aware model with an honest ceiling}
(\S\ref{sec:tme}--\S\ref{sec:puzzle}).  We recall the four-term TME model of
Part~1 and show that its fourth term---the per-input deconstruction cost $c_q$,
counted at $c_q{=}16$ instructions per modulus per element on the shipping
path~\cite{bayraktar2026tme} (Appendix~\ref{app:memo})---gates dense GEMM behind
a size crossover $n^{*}=c_q r P_{\fp{8}}/(\alpha P_{\text{int}})\approx1{,}211$
on Rubin, and that a large multi-cluster output is pinned at the
deconstruction-$\lambda$ floor ($0.50$ of the $473$-TFLOPS roof---a derived
ratio of three design integers, Eq.~\eqref{eq:whyhalf}, not a fitted
efficiency).  The reduced
model passes
through the published $200$-TFLOPS figure at model-inferred
$\nbar\approx500$--$900$; we state this as a falsifiable hypothesis, not a
finding.
\item \textbf{The Ozaki~2.5 method} (\S\ref{sec:ozaki25}).  Ozaki~II unchanged
in its mathematics, with the deconstruction path engineered around the actual
modulus set (convert-once workspaces, an exact two-limb integer-tensor
reduction, a ${\approx}6.3$-instruction SIMT residual, pipelined conversion;
Algorithm~\ref{alg:ozaki25}), moving the Rubin crossover from ${\approx}1{,}211$
to ${\approx}480$--$730$.
\item \textbf{Modulus/encoding codesign} (\S\ref{sec:codesign}).  A
script-checked study treating the modulus set as a design variable: an all-byte
system that legalises the one-pass reduction, a hybrid that concedes only $7.5\%$
asymptotically, and two exhaustive supply bounds---motivating a regime-switched
modulus dispatch.
\item \textbf{Potential hardware co-design} (\S\ref{sec:hw}).  Because the floor
is the closed-form $R\,P_{\text{int}}/(c_q r)$, it names its escape: cluster reach, TMEM
capacity, and L2 bandwidth for dense DGEMM, and a $c_q$-removing copy-engine
datapath that also un-binds the sparse kernels---each a concrete, measurable ask.
\end{itemize}

\paragraph{Relation to Part 1, and why this is a separate paper.}
Part~1~\cite{matsuoka2026fp8part1} argues a platform thesis: that an
FP8-dominated datapath, priced by the four-term TME model, can serve the
scientific dwarfs at large.  Ozaki emulation enters there as one workload
among many, and the deconstruction term enters as a cost to be priced, not a
path to be engineered.  This paper takes the opposite cut: it holds the
workload fixed---\fp{64} GEMM emulation---and engineers the single term
Part~1 could only price.  Every technical result below is new to this paper
and appears in no FP8-platform or Ozaki-scheme publication we know of: the
carry-corrected redundant split that closes the six nonsquare S-tail moduli
$487$--$511$ in \textsc{e4m3} (Lemma~\ref{lem:carry}); the exact two-limb
byte-identity reduction kernels and their pure-SIMT \texttt{dp4a}
realisation; the modulus/encoding codesign study---sets E/A/D with per-set
proof obligations (Appendix~\ref{app:obligations})---and its two exhaustive
supply bounds; the route menu with per-route numerical contracts and the
$k{\approx}6777$ route-crossover algebra; the reconstruction-aware
per-record engine with $N_{\text{acc}}$ accounting, grounded by measured
\texttt{LD\_PRELOAD} call-shape traces; and the closed-form reach floor of
Eq.~\eqref{eq:decfloor} with its quantified hardware asks.  From Part~1 we
import exactly two things, both by citation: the TME cost model
(\S\ref{sec:tme} recalls it) and the memo-derived constants of
Appendix~\ref{app:memo}.  Nothing here duplicates Part~1's coverage
analysis, and nothing there anticipates the modulus, kernel, or floor
results; the two papers share a cost model the way two instruction-set
studies share an ISA manual.

Every projection is instantiated for the Blackwell generation (in particular
GB300) as well as Rubin; because GB300 pairs a ${\approx}1.4$-TFLOPS native
\fp{64} pipe with a $135$-TFLOPS emulated roof, the method is a present-tense
proposition on shipping hardware, not one that waits for Rubin.  We close
(\S\ref{sec:proj}--\S\ref{sec:validation}) with measured call-shape traces that
ground the shape analysis and four falsifiable predictions whose decisive
experiment is a DGEMM size sweep, with a slicing-based \inteight{} kernel on B200
as the control arm---identifying the responsible term whether the branch
hypothesis survives measurement or falls to it, in the same two-way spirit as
the TME model itself~\cite{matsuoka2026fp8part1}.

Because every result below carries one of a small set of
epistemic labels, we fix them here in Table~\ref{tab:claimstatus};
each results-table caption repeats its label, and the detailed
ledger is in \S\ref{sec:validation}.

\begin{table}[!ht]
\caption{Claim-status labels used throughout the paper.}
\label{tab:claimstatus}
\centering
\footnotesize
\begin{tabular}{lp{0.66\linewidth}}
\toprule
label & what carries it \\
\midrule
proved algebraically & the harmonic-mean crossover (reduced model); the two-limb byte identity; the square-modulus and Karatsuba plane identities and the accumulator floor they imply; the \textsc{e4m3} layout envelope ($m\le321$ nonsquare under a canonical split, $m\le577$ under a redundant one, $m\le1089$ square); the carry-corrected three-plane layout of Lemma~\ref{lem:carry} and its plane bounds; the exact-accumulation bound $K_{\text{slab}}\max|\text{term}|\le2^{24}$; the operand-bandwidth ratio $p_{\text{store}}/\alpha$ and its \textsc{dsm} counterpart $(1{-}1/c)\,p_{\text{store}}/\alpha$ \\
script-checked & supply bounds; coprimality and exact CRT products; centred digit maps and \textsc{e4m3} representability of all A/E/D planes and the six square S moduli.  The six nonsquare S-tail moduli ($487$--$511$) are closed: no \emph{canonical} balanced split places them in \textsc{e4m3} at any base, but the carry-corrected redundant split of Lemma~\ref{lem:carry} does, verified exhaustively over all $m^{2}$ ordered centred pairs per modulus (\texttt{verify/stail\_layout.py}) \\
vendor-announced & the \fp{8}/\inteight{}/\fp{64} platform rates and the $\sim$200-TFLOPS emulated-DGEMM figure \\
memo-derived & $c_q{=}16$ and the $P_{\text{int}}{\approx}75$-\tops{} normalisation (Appendix~\ref{app:memo}) \\
modeled projection & every route knee, envelope, bracket, and composite speedup (Tables~\ref{tab:codesign}--\ref{tab:apps}) \\
measured & the \texttt{LD\_PRELOAD} call-shape traces (geometry only)---and nothing else \\
\bottomrule
\end{tabular}
\end{table}

\section{Background: the Ozaki Schemes and the Substrate Transition}
\label{sec:ozaki2}

\paragraph{From error-free splitting to modular arithmetic.}  The
original Ozaki scheme~\cite{ozaki2012}---Ozaki~I in the present
numbering---computes a high-precision matrix product by
\emph{error-free splitting}: each operand is split into a short sum of
lower-precision slices such that every pairwise slice product is exact
in the target arithmetic, and the exact partial products are summed.
Its cost grows quadratically in the slice count, which is what makes
it attractive on \inteight{} (few wide slices) and prohibitive on
\fp{8} (many narrow ones).  Ozaki Scheme~II~\cite{ozaki2025_scheme2}
replaces the splitting by \emph{residue arithmetic}: one exact integer
product is computed per modulus of a Chinese Remainder Theorem
(CRT) system, so the cost grows \emph{linearly} in the number of
moduli.  Because the integer core is exact, the scheme delivers
componentwise \fp{64}-grade accuracy---the ``Grade~A'' contract in
Part~1's terminology, \ie \fp{64}-equivalent componentwise error
bounds---with the error analysis and the automatic
precision-escalation machinery (ESC/ADP) supplied
by~\cite{ozaki_error_analysis_2026,uchino2026_fp8,schwarz2025}.

\paragraph{The scheme in equations.}  Let $C = AB$ with $A \in
\mathbb{R}^{m\times k}$, $B \in \mathbb{R}^{k \times n}$ in \fp{64}.
\emph{(i)~Scale to integers.}  Choose per-row exponents
$\sigma_1..\sigma_m$ for $A$ and per-column exponents
$\tau_1..\tau_n$ for $B$, keeping $t$ significant bits, and set
\begin{equation}
A' = \big\lfloor \operatorname{diag}(2^{\sigma}) \, A \big\rceil,
\qquad
B' = \big\lfloor B \, \operatorname{diag}(2^{\tau}) \big\rceil,
\qquad
C \;\approx\; \operatorname{diag}(2^{-\sigma})\, (A'B')\,
\operatorname{diag}(2^{-\tau}),
\label{eq:scale}
\end{equation}
where $\lfloor\cdot\rceil$ denotes the integer quantiser
\emph{selected by the named variant} (Table~\ref{tab:variants}):
round-to-nearest for Ozaki-II-RN, under which the cited error
theorem~\cite{ozaki_error_analysis_2026} holds within its hypotheses;
\emph{truncation toward zero} for Ozaki-II-TZ, the rule the shipping
FP8 implementation uses~\cite{uchino2026_fp8} and the one our
projections assume.  The round-to-nearest theorem is \emph{not}
claimed for the truncation rule: TZ does not automatically inherit
the RN bound, and TZ-grade accuracy remains an explicit validation
obligation rather than an established result.  The integer product
$C' = A'B'$ is computed exactly under either quantiser, so
floating-point rounding enters only in the scaling of
Eq.~\eqref{eq:scale} and in the single final conversion of each
output to \fp{64}.
\emph{(ii)~Choose the residue system.}  The entries of $C'$ are
bounded, $|c'_{ij}| \le k\, 2^{2t} \eqqcolon \mu$; choose pairwise
coprime moduli $m_1, \dots, m_r$ (coprimality, not primality, is what
CRT requires; the shipping FP8 implementation uses 9--11-bit square and
near-prime moduli, $1089, 1024, 961, 841, 625, 529, 511, 509, 503,
499, 491, 487$ at $r{=}12$~\cite{uchino2026_fp8,gemmul8_github}) with
\begin{equation}
M \;=\; \prod_{i=1}^{r} m_i \;>\; 2\mu ,
\label{eq:moduli}
\end{equation}
so that $C'$ is uniquely determined by its residues, taken in the
half-open symmetric range $[-M/2,\, M/2)$.  The \emph{same}
convention is used for every per-modulus residue in this paper:
intervals $[-x/2, x/2)$, with the tie map stated once and for all as
``for even $x$ the endpoint $-x/2$ is included and $+x/2$
excluded.''  \emph{(iii)~One exact low-precision product
per modulus.}  For each modulus form the residue operands and their
exact products,
\begin{equation}
A^{(i)} = A' \bmod m_i, \quad
B^{(i)} = B' \bmod m_i, \qquad
C^{(i)} = A^{(i)} B^{(i)} \bmod m_i ,
\label{eq:residueprod}
\end{equation}
where every entry of $A^{(i)}, B^{(i)}$ lies in $[0, m_i)$ (or,
in the implementations, in the centred range $[-m_i/2, m_i/2)$, which
is what the digit decompositions below require); exactness
of the residue products is obtained \emph{through the low-precision
decomposition of the next paragraph}, not by representing full residue
products directly, and rounding-free FP32 accumulation holds for
$k \le 2^{16}$, with $k$-blocking (and its recombination cost) beyond
that bound~\cite{uchino2026_fp8}.  This is where the tensor cores do
all the heavy work.  \emph{(iv)~Reconstruct by CRT in Garner form.}  With
precomputed constants $w_j = (m_1 m_2 \cdots m_{j-1})^{-1} \bmod
m_j$~\cite{garner1959,knuth_taocp2}, the mixed-radix digits of each
output element and the element itself are
\begin{equation}
v_1 = c^{(1)}, \qquad
v_j = \Big( c^{(j)} - \big(v_1 + v_2 m_1 + \cdots +
v_{j-1} m_1 \cdots m_{j-2}\big) \Big)\, w_j \bmod m_j ,
\label{eq:garner}
\end{equation}
\begin{equation}
c' \;=\; v_1 + m_1\big(v_2 + m_2( v_3 + \cdots + m_{r-1} v_r )\big),
\label{eq:horner}
\end{equation}
evaluated in bounded, fixed-width multi-limb \intthirtytwo{}
arithmetic (Horner form), lifted to the symmetric range, unscaled per
Eq.~\eqref{eq:scale}, and rounded \emph{once} to \fp{64}.
Reconstruction touches only the $mn$ \emph{outputs} and is amortised
over the inner dimension $k$; deconstruction---the residue formation
on the left of Eq.~\eqref{eq:residueprod}---touches every
\emph{input}.  That asymmetry is what this paper is about.

\paragraph{The \fp{8} variant.}  On \fp{8} tensor cores a residue
modulo a 9--11-bit modulus does not fit a single \textsc{e4m3}
significand.  The source method therefore represents each residue by
\emph{two} limbs at a piece width $b$,
\begin{equation}
d \;=\; d_1\, 2^{b} + d_0 ,
\qquad 0 \le d_0 < 2^{b},
\label{eq:pieces}
\end{equation}
with $b$ chosen so that every pairwise limb product is exact in the
\fp{8} MMA datapath.  Two limbs give three products, and the two
modulus kinds reach them differently.  For a square modulus $m=s^2$ the
$d_1e_1$ term is annihilated modulo $s^2$, so the two \emph{stored}
planes suffice directly, with epilogue coefficients $\{1,s,s\}$; for a
nonsquare modulus a third, \emph{sum} plane $d_0{+}d_1$ is stored and
the Karatsuba identity is
evaluated~\cite{uchino2026_fp8}.  Either way the count is three
plane-product MMAs per modulus, so the stored-plane load
$p_{\text{store}}$ (two or three bytes per modulus per scalar) and the
issued-MMA count $p_{\text{mma}}=3r$ are \emph{different} quantities and
are kept apart throughout this paper; the companion Part~1 freezes the
resulting kernel, including the accumulator ledger these identities
imply.  In the accurate mode one
additional \fp{8} GEMM estimates an input magnitude bound (it is
\emph{not} a Karatsuba pass), and fast/accurate modes trade the
modulus count $r$ at comparable accuracy~\cite{uchino2026_fp8}.  The
accurate-mode schedule thus costs
\begin{equation}
\alpha \;=\; 3r + 1 \ \ \text{\fp{8} MMA passes per \fp{64} product}
\qquad (= 37 \ \text{at } r{=}12),
\label{eq:alpha}
\end{equation}
the working point that suffices for \fp{64}-equivalent accuracy on
well-scaled data; the source analysis reports, under its definition,
${\approx}55$ effective significand bits at $r{=}12$, ${\approx}59$
at $13$, and ${\approx}64$ at $14$, with the precise mode- and
set-specific condition given
there~\cite{uchino2026_fp8,mukunoki2025_fp8}.  Dividing the
official dense \fp{8} rates by $37$ gives the emulation ceilings used
throughout: $\approx\!135$~TFLOPS on GB300 ($5$~\pflops{} \fp{8})
and $\approx\!473$~TFLOPS on Rubin ($17.5$~\pflops{}; the primary
HGX specification also lists $250$-\tops{} dense \inteight{},
$33$-TFLOPS \fp{64}, and the $200$-TFLOPS emulated-DGEMM figure
itself~\cite{nvidia_hgx,nvidia_dgx_rubin_nvl8}).  The GB300 rates
used throughout are \emph{NVL72-derived per-GPU} values---NVIDIA's
NVL72 rack specification with the sparsity
convention unpacked: $720$~\pflops{} sparse \fp{8} and
$24$~POPS sparse \inteight{} over $72$ GPUs give $5$~\pflops{} and
$166.7$~\tops{} dense per GPU, and $100$~TFLOPS rack \fp{64} gives
${\approx}1.4$ per GPU~\cite{nvidia_gb300_nvl72}; the air-cooled
HGX~B300 SKU carries different per-GPU ratings, so ``GB300'' in this
paper always means the NVL72-derived per-GPU operating point.

\paragraph{The substrate transition, and why it matters here.}
Blackwell-class emulation ran on the \emph{\inteight{}} tensor
substrate: B200 supplies $4{,}500$~\tops{} of dense \inteight{}, and
the shipping cuBLAS emulated DGEMM delivered
$\approx\!150$~TFLOPS on it~\cite{nvidia_cublas_emulation}.  The
\inteight{} substrate is deconstruction-friendly in two distinct
ways.  In the error-free-slicing realisation of the Ozaki~I
lineage~\cite{ozaki2012}, deconstruction is nearly free: slicing a
scaled integer into 8-bit slices \emph{is byte extraction}---a scale,
a round, and a type-pun, a few SIMT instructions per \emph{element}
in total, with no per-modulus arithmetic at all.  Even the CRT
realisation is cheaper \emph{per modulus} on \inteight{}: the centred
residue of a byte modulus ($m \le 256$) fits an \inteight{} operand
directly, so the Karatsuba piece split (and its converts) disappears,
leaving $c_q \approx 13$ against the \fp{8} route's~$16$.  Per modulus
only, however, and that qualification is not cosmetic: an
\inteight{}-native set must be all-byte, fourteen byte moduli cannot
reach the published set's $111.8$ bits (\S\ref{sec:codesign}), and so
$r_{\textsc{i}} = 15$---candidate~A, $117.8$ bits.  The deconstruction
load is then $c_q r_{\textsc{i}} \approx 195$ against the \fp{8}
route's $16 \cdot 12 = 192$, a wash.  What the \inteight{} substrate
actually buys is a smaller \emph{divisor} in the arithmetic roof
$P/\alpha$: $\alpha_{\inteight} = r_{\textsc{i}}{+}1 = 16$ against
$3r{+}1 = 37$, a $2.31\times$ advantage per unit of substrate rate;
the deconstruction advantage cancels itself.
From the Blackwell-Ultra generation onward, however, silicon area has
been redirected to low-precision floating point: the \inteight{}
tensor rate falls to a residual $\sim\!166$--$250$~\tops{} while
\fp{8} scales to 5--17.5~\pflops{}, so the only high-throughput
emulation route on Rubin-class parts runs through the CRT/\fp{8}
variant~\cite{matsuoka2026fp8part1}---whose deconstruction is
\emph{not} byte extraction: every streamed element must be reduced
modulo each of $r$ moduli and split into three \fp{8} pieces before a
single MMA can issue.  The cost of that step is the subject of this
paper.  (Algorithm attribution of the shipping path requires care:
the released cuBLAS GEMM emulation is publicly described as
Scheme-I slicing~\cite{emugemm2026}, while NVIDIA's cuEST emulation
stack now exposes \emph{both} schemes---slice-count and
modulus-count controls, selected by compute
capability~\cite{nvidia_cuest}.  The memo-counted conversion
SASS~\cite{bayraktar2026tme} exhibits per-modulus reduction
structure, i.e.\ a CRT path; whether that path is the one behind the
published B200 figure is a provenance question the memo's authors
have been asked to confirm.  Where this paper reads the B200 figure
through CRT constants, the reading is a sensitivity illustration
conditional on that
attribution; the slicing realisation matters below as the clean
experimental \emph{control}, \S\ref{sec:validation}.)

\section{The Updated TME Model and the Fourth Term}
\label{sec:tme}

Part~1's Tensor--Memory Equilibrium model~\cite{matsuoka2026fp8part1},
as corrected by the NVIDIA technical review~\cite{bayraktar2026tme},
bounds the execution time of an emulated kernel by four terms:
\begin{equation}
T_{\text{serial}} \;=\;
\max\!\left(
\frac{\alpha\,W_{\text{mma}}}{P_{\text{low}}},\;
\frac{\beta\,Q_0}{B_{\text{mem}}},\;
\frac{c_q\, r\, \varphi\, n_{\text{in}}}{P_{\text{int}}}
\right) \;+\; \gamma\, n_{\text{out}},
\label{eq:tme}
\end{equation}
the tensor time ($\alpha = 3r{+}1$), the memory time ($\beta\ge1$ the
bandwidth multiplier on the native reference bytes $Q_0$), the \emph{deconstruction} time---$c_q$
integer-pipe instructions per modulus for each of the $\varphi
n_{\text{in}}$ converted input elements---and the reconstruction
latency $\gamma$ per output.  This displayed quantity is
$T_{\text{serial}}$, a \emph{service-tail endpoint}: the
tensor, memory, and deconstruction \emph{maxima} (already mutually overlapped)
followed by an unoverlapped reconstruction tail $\gamma\,n_{\text{out}}$.  It
is \emph{not} fully serialised---the genuine no-overlap endpoint is the sum
$T_{\fp{8}}{+}T_{\text{mem}}{+}T_{\text{dec}}{+}\gamma n_{\text{out}}$---and,
being a $\max$, it does not by itself upper-bound measured time.  We keep it
distinct from the \emph{ideal service-overlap} time that the tables actually
use,
\begin{equation}
T_{\text{svc}} \;=\; \max\!\Big(
T_{\fp{8}},\; T_{\text{mem}},\;
\tfrac{N^{I}_{\text{dec}}{+}N^{I}_{\text{rec}}}{P_{I}},\;
\tfrac{N^{S}_{\text{dec}}{+}N^{S}_{\text{rec}}}{P_{S}}\Big),
\label{eq:tsvc}
\end{equation}
in which deconstruction and reconstruction are charged to their host
pipes ($I$ the integer/tensor deconstruction pipe of rate $P_{I}$, $S$
the SIMT reconstruction pipe of rate $P_{S}$) and overlapped with the
\fp{8} and memory times.  These two are model \emph{endpoints}, not rigorous bounds on measured time:
with perfect service overlap the time is $T_{\text{svc}}$, and imperfect
overlap moves $T_{\text{meas}}$ toward---and, absent the assumed
service-independence, potentially past---the tail estimate.
Table~\ref{tab:apps} and the companion engine
(\texttt{\_compute\_rate}) evaluate $T_{\text{svc}}$, while
Eq.~\eqref{eq:tme} is retained as the service-tail sensitivity
$T_{\text{serial}}$; a measured schedule is what settles where in the band
$T_{\text{meas}}$ falls.  The fourth term is the review's
first-order correction: reconstruction is charged per output, but
deconstruction is charged per \emph{input}, and for memory-bound
kernels inputs outnumber outputs by orders of magnitude.  On the
shipping cuBLAS conversion path the review counts $c_q = 16$
instructions per modulus per element ($13$ counted from the shipping
conversion SASS plus $3$ for the \fp{8} Karatsuba piece
converts), against an integer-pipe budget of $P_{\text{int}} \approx
75$~\tops{} counted on GB300 (instruction throughput for the specific conversion
sequence; vendor ``TOPS'' conventions differ, and the normalisation
must accompany any fitted value) and presumed unchanged on Rubin
pending the pipe census of the follow-up measurement work~\cite{bayraktar2026tme,
matsuoka2026fp8part1}.

\begin{figure}[!t]
\centering
\includegraphics[width=\linewidth]{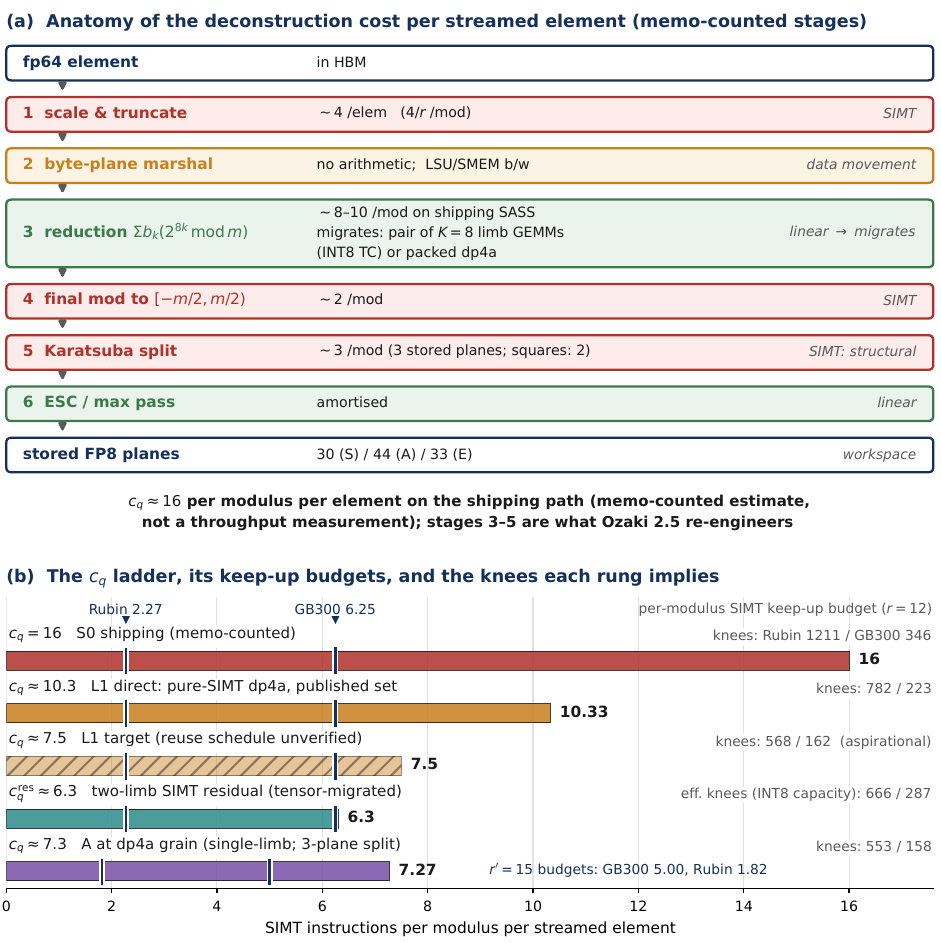}
\caption{Anatomy of the deconstruction cost $c_q$ per streamed element
(reproduced from Part~1~\cite{matsuoka2026fp8part1}).  Panel (a): the
six-stage conversion pipeline, colour-coded by disposition---red
stages are elementwise-nonlinear and stay on SIMT, green stages are
linear and migrate to the tensor pipes, amber is data movement.  Panel
(b): the five audited $c_q$ rungs, each drawn against the per-modulus
SIMT keep-up budget $8P_{\text{int}}/(r B_{\text{mem}})$ that applies
to \emph{that} rung (the paired ticks; the strip above names the two
columns).  The budget is not one number, because $r$ is not: the four
$r{=}12$ rungs are charged against $6.25$ instructions per modulus on
GB300 and $2.27$ on Rubin, while the all-byte $A$ rung runs at
$r'{=}15$ and is charged against $5.00$ and $1.82$, which is why its
ticks stand to the left of the others.  Every bar ends to the right of
both of its own ticks; the narrowest miss is the two-limb residual,
whose $6.3$ all but coincides with GB300's $6.25$.  Ozaki~2.5 is the
engineering of this pipeline toward its SIMT-residual count,
Eq.~\eqref{eq:floor}.}
\label{fig:anatomy}
\end{figure}

Two consequences of Part~1 frame what follows.  First, a streamed
kernel that saturates HBM consumes elements at $B_{\text{mem}}/8$ per
second, so the SIMT pipe can spend at most $8P_{\text{int}}/(r
B_{\text{mem}})$ instructions per modulus per element without falling
behind: $6.25$ on GB300 but only $2.3$ on Rubin---the budget
\emph{shrinks} with the generation, because bandwidth grew
$2.75\times$ while the integer pipe did not.  Second, dissecting the
$c_q = 16$ into its six stages (Figure~\ref{fig:anatomy}) shows the
bulk of the cost---the per-modulus reduction, $\sim$8--10 of the 13
counted instructions---is \emph{linear over the byte planes} of the
scaled integer, $x \bmod m = \sum_k b_k (2^{8k} \bmod m) \bmod m$, and
therefore executable as matrix arithmetic on the tensor side; what
must remain on SIMT (scale/truncate, the final reduction to the
centred residue interval, and
the structural Karatsuba split) sums to a proposed SIMT-residual count of
\begin{equation}
c_q^{\text{res}} \;\approx\; \underbrace{\tfrac{4}{r}}_{\text{scale/truncate}}
+ \underbrace{1}_{\text{limb recomb.}}
+ \underbrace{2}_{\text{final mod}}
+ \underbrace{3}_{\text{piece split}}
\;\approx\; 6.3
\quad\text{instr.\ per modulus at } r{=}12,
\label{eq:floor}
\end{equation}
where the limb-recombination term is required by the exact two-limb
encoding of \S\ref{sec:ozaki25} (the actual moduli exceed one byte),
and each per-instruction count is an instruction-ledger projection
to be verified at SASS level, not a measured value.
(The lazy-reduction variant of Part~1 lowers this to ${\approx}4.3$
for the published set---${\approx}3.3$ on the one-pass all-byte
set---but doubles the tensor cost to $\alpha' \approx 6r{+}1$; benign
for memory-bound kernels, it would halve the dense ceiling and is
therefore \emph{excluded} throughout this paper.)

\paragraph{The reconstruction term, quantified.}
\label{sec:recon}
The fourth term $\gamma\,n_{\text{out}}$ cannot be waved away with an
``$O(r^2/k)$, negligible'' aside, and we give it a rate bound.
Garner reconstruction is $N_{\gamma} = r(r{-}1)/2 + 2r \approx 90$
narrow integer operations per output at $r = 12$.  Hosted on the SIMT
pipe, its service time as a fraction of the emulated GEMM's is
\begin{equation}
\frac{T_{\gamma}}{T_{\text{mma}}}
\;=\; \frac{N_{\gamma}/2 \cdot P_{\text{emu}}}{k\, P_{\text{int}}}
\;=\; \frac{45\,P_{\text{emu}}}{k\, P_{\text{int}}},
\label{eq:recongamma}
\end{equation}
which at $k = 1000$ is $8.1\%$ on GB300 and $28.4\%$ on Rubin---
\emph{not} $<\!1\%$---and falls below $1\%$ only at
$k \gtrsim 8.1{\times}10^3$ (GB300) / $2.84{\times}10^4$ (Rubin).  The
review is right that SIMT-hosted Garner is a small-$k$ wall.  We
therefore \emph{adopt} the SIMT-Garner cost as the paper's
\emph{conservative reconstruction cost ceiling}---the most expensive
implementable reconstruction, and hence a \emph{performance floor}
\emph{within the model}---a lower bound on the modeled rate, not a measured
one, since every other term in the composite remains a projection---and report every trace
composite (Table~\ref{tab:apps}) at it.  The ${\approx}r$-op
\inteight{} hosting is now presented as a \emph{codesign target}, not
a delivered count: an exact reconstruction of the ${\approx}112$-bit
residue state (the supply lemma needs $111.8$ bits) requires either a
multi-limb integer recombination (${\approx}rL$ byte-MACs, $L \approx 14$;
by op-count $rL{\approx}168$ exceeds Garner's ${\sim}90$ narrow ops, so the
comparison is of \emph{host-pipe times}, $rL/P_{\text{byteMAC}}$ vs
$N_\gamma/P_{\text{narrow}}$, which the engine finds comparable to or below
Garner---no decisive gain), or a floating-point fractional-CRT recombination
(${\approx}r$ \fp{32} MACs) whose \fp{64}-exactness is a Part-3
validation obligation.  Until that validation the ${\approx}r$
thresholds ($k \gtrsim 977$ (GB300) $/\,2270$ (Rubin)) are reported as
a target/sensitivity, and the conservative thresholds are the
SIMT-Garner values above.  Both hostings are carried in the per-record
engine (\texttt{params.py}: \texttt{\_compute\_rate}, dispatching the
reconstruction host like the storage mode); the trace composites of
Table~\ref{tab:apps} are computed at the conservative SIMT-Garner cost floor,
applied to \emph{both} the emulated envelope and the shipping
baseline.  Relative to the ${\approx}r$ codesign target they are
\emph{unchanged for the compute-advantaged records} ($k \gg$
threshold) but \emph{lower for the small-$k$ records}: some
shipping-relative ratios move between the target and
guaranteed modes---the guaranteed values being those tabulated in
Table~\ref{tab:apps} (e.g.\ UMFPACK GB300 $1.00\to1.19$; dense-QR Rubin
$2.06\to1.60$)---while the vs-native multipliers for the small-$k$
multifrontal and dense-LU fronts on Rubin fall more (there
${\approx}4\to1$ and ${\approx}9\to3$); the codesign target would
restore these, pending Part-3 validation (\S\ref{sec:apps}).  The formal
fourth term and the companion's per-record engine now use the
\emph{same} overlap semantics ($T_{\text{svc}}$): reconstruction demand
is added to its host pipe's service time and overlapped with the \fp{8}
and memory times (\texttt{\_compute\_rate}), with the serialised
$T_{\text{serial}}$ (Eq.~\eqref{eq:tme}) retained only as the
worst-case upper bound.

\section{The Dense Puzzle: a Size Crossover, and Where 200 TFLOPS Falls}
\label{sec:puzzle}

\paragraph{Amortisation, quantified.}  For dense GEMM the folk
intuition is that deconstruction cannot matter: $O(n^2)$ operand
elements against $O(n^3)$ multiply work.  The intuition is
asymptotically right and quantitatively misleading.  For a
$m{\times}k$ by $k{\times}n$ product with each operand element
deconstructed once and its residue planes reused thereafter, the
conversion time is $c_q r (mk{+}kn)/P_{\text{int}}$ against a tensor
time $\alpha\,2mkn/P_{\fp{8}}$; conversion stops binding
when
\begin{equation}
\nbar \;\coloneqq\; \frac{2mn}{m+n} \;\;\ge\;\;
n^{*} \;=\; \frac{c_q\, r\, P_{\fp{8}}}{\alpha\, P_{\text{int}}},
\label{eq:crossover}
\end{equation}
the harmonic mean of the two \emph{output} dimensions clearing a
critical size---one independent of the inner dimension $k$.  At the
memo-counted $c_q{=}16$ ($r{=}12$, $\alpha{=}37$,
$P_{\text{int}}{\approx}75$~\tops{}):
\begin{equation*}
n^{*} \approx 346 \ \ \text{on GB300} \qquad\qquad
n^{*} \approx 1{,}211 \ \ \text{on Rubin,}
\end{equation*}
the knee moving $3.5\times$ outward in one generation because
$P_{\fp{8}}$ grew while $P_{\text{int}}$ did not.  Below the
crossover, delivered throughput under conversion--MMA overlap is
\begin{equation}
P_{\text{DGEMM}}(\nbar) \;=\;
\min\!\left(\frac{P_{\fp{8}}}{3r+1},\;
\frac{\nbar\, P_{\text{int}}}{c_q\, r}\right),
\label{eq:pdense}
\end{equation}
and with conversion fully serialised the ceiling is divided by
$(1+n^{*}/\nbar)$; real kernels land between the brackets.

\paragraph{The convert-once premise, and its reach.}  Eq.~\eqref{eq:pdense}
assumes each operand element is deconstructed \emph{once} and its planes
reused thereafter ($\lambda{=}1$). That is exact only while the output fits
one thread-block cluster---an output edge we call the \emph{reach} $R$,
equal to $256$ for the shipping $4{\times}4$ cluster of $64^2$ tiles and
given in closed form as a function of cluster shape by
Eq.~\eqref{eq:reach} in \S\ref{sec:hw}.  A larger output spans
several clusters, and on-chip planes cannot be shared across them, so each
operand is re-split $\lambda_A = \lceil n/R \rceil$, $\lambda_B = \lceil
m/R \rceil$ times, and the per-element deconstruction load
$c_q r\,\lambda/\bar n$ stops falling---it pins at $c_q r/R$.  Substituting
$\nbar \to R$ in Eq.~\eqref{eq:pdense} therefore gives the plateau in
closed form, and the substitution is worth doing explicitly because of what
cancels:
\begin{equation}
P_{\text{dec}}^{\text{floor}}
 \;=\; R \cdot \min\!\left(
   \frac{P_{\text{I8TC}}/2}{a_{\inteight}},\;
   \frac{I_{\text{SIMT}}}{c_q^{\text{res}}\, r}\right),
\qquad
\text{capped at }\; \frac{P_{\fp{8}}}{3r+1},
\label{eq:decfloor}
\end{equation}
in the symbols of Eq.~\eqref{eq:tensorknee}: $a_{\inteight}$ the useful
\inteight{} MACs per streamed element, $P_{\text{I8TC}}/2$ the \inteight{}
tensor MAC rate, $I_{\text{SIMT}}$ the SIMT instruction rate, and
$c_q^{\text{res}}$ the SIMT residual of
Table~\ref{tab:codesign}.  \textbf{The \fp{8} peak has
cancelled.}  Below the cap, the floor is a product of exactly two things:
the reach $R$, which is a geometry the hardware fixes, and a per-modulus
cost the modulus system fixes.  Raising $P_{\fp{8}}$ moves it not at all.
That is the whole co-design argument in one line, and it is the reason
\S\ref{sec:hw} spends its effort on reach rather than on peak.  The two
terms bind on different routes---S and E are \inteight{}-capacity bound
($0.71$ and $0.92$~TFLOPS per unit of reach), A is SIMT-residual bound
($0.95$)---so the ranking of routes is itself a function of which host pipe
is scarce.  Eq.~\eqref{eq:decfloor} reproduces the engine exactly at every
rung of Table~\ref{tab:rungs} (\texttt{verify/cluster\_map.py}).

The convert-once crossover is therefore an \emph{upper envelope}, realised up
to $\bar n{\approx}R$; beyond it the achieved dense-square rate plateaus
at this \emph{deconstruction-$\lambda$ floor} for cluster-aligned sizes (a
sawtooth in $\lceil E/R\rceil$, not a hard constant; the ragged worst case
is reported below). So the
arithmetic roof ($473$~TFLOPS on Rubin) at $\bar n \gtrsim n^{*}$ is attained
only in the
convert-once (single-cluster, or materialised) regime, and the honest
large-DGEMM ceiling is that floor---which \S\ref{sec:hw} quantifies
(${\approx}0.50$ of that same $473$-TFLOPS roof for the codesigned
routes), shows to
be a co-design coordinate, and lifts back to the roof
(Fig.~\ref{fig:codesign}). We flag it here so the crossover of
Eq.~\eqref{eq:pdense} and the floor of \S\ref{sec:hw} read as one
deconstruction curve, not two claims.  Note that
Eq.~\eqref{eq:decfloor} is a $P_{\text{dec-only}}$ statement
(\S\ref{sec:twofunc}): it charges no reconstruction, and at $R{=}256$ it
gives $243$~TFLOPS for~A against the $235$ that the
reconstruction-aware $P_{\text{svc}}$ reports for~E as the winning route at
$k{=}4096$, the HPL-relevant depth of Table~\ref{tab:hplnb} (the ordering
inverts beyond $k{\approx}6777$; Table~\ref{tab:algs}).

\paragraph{Why the floor lands at one half, exactly---and why minor
hardware recovers all of it.}
The $0.50$ reads like a fitted efficiency, the kind of number a benchmark
produces; it is not.  Write the floor of Eq.~\eqref{eq:decfloor} for the
winning hybrid route~E, whose \inteight{}-tensor term binds:
$P^{\text{floor}} = R\cdot P_{\text{I8TC-MAC}}/a_{\inteight} = 256 \times
125\,\text{T}/136 = 235$~TFLOPS.  Divide by the $473$-TFLOPS roof
$P_{\fp{8}}/(3r{+}1)$ and every \emph{rate} cancels into a ratio of three
design integers:
\begin{equation}
\frac{P^{\text{floor}}}{P^{\text{roof}}}
 \;=\;
 \underbrace{R\vphantom{\frac{P}{P}}}_{256}
 \,\times\,
 \underbrace{\frac{P_{\text{I8TC-MAC}}}{P_{\fp{8}\text{-MAC}}}}_{1/140}
 \,\times\,
 \underbrace{\frac{3r+1}{a_{\inteight}}}_{37/136}
 \;=\; \frac{256\times 37}{140\times 136} \;=\; 0.497.
\label{eq:whyhalf}
\end{equation}
The reach $R{=}256$ is cluster geometry (Eq.~\eqref{eq:reach}); $140$ is
Rubin's provisioning ratio between the \fp{8} and \inteight{}-tensor MAC
pipes; $136$ is route~E's two-limb plane-formation cost in \inteight{}
MACs per streamed element.  Nothing is fitted, and that the product lands
within $0.3$ percentage points of one half is arithmetic coincidence:
against route~E's
\emph{own} $437.5$-TFLOPS roof the same floor is $0.54$, and the published
set~S floors at $0.38$ of its $473$-TFLOPS roof.  The number moves exactly
as the
integers move---doubling the \inteight{}-tensor provisioning lifts the
formula to $471$~TFLOPS, where route~E's own roof caps it at $0.92$ of the
common $473$-TFLOPS roof; halving it drops the floor to $0.25$; and between
cluster-aligned sizes the achieved value follows the $\lceil E/R\rceil$
sawtooth.  The same three integers explain why Blackwell~Ultra already
\emph{has} full recovery: at GB300's provisioning ratio (${\approx}60$,
not $140$) the formula gives $156$~TFLOPS---above route~E's $125$-TFLOPS
roof there---so the $\min$ of Eq.~\eqref{eq:decfloor} is taken by the roof
and the floor never binds (the engine reports~E attaining its GB300 roof
exactly; the $135$~TFLOPS GB300 floor the tables quote is set~S's
$\alpha{=}37$ roof, which the \texttt{dp4a} route~L1d attains at the memo's
$P_{\text{int}}$---see the sensitivity in Appendix~\ref{app:memo}).  Rubin's one half is thus a \emph{provisioning} statement, not a
method limit: one generation widened the \fp{8}:\inteight{}-tensor ratio
from ${\approx}60$ to $140$ while plane formation stayed charged to the
narrow pipe.

\begin{figure}[t]
\centering
\includegraphics[width=\linewidth]{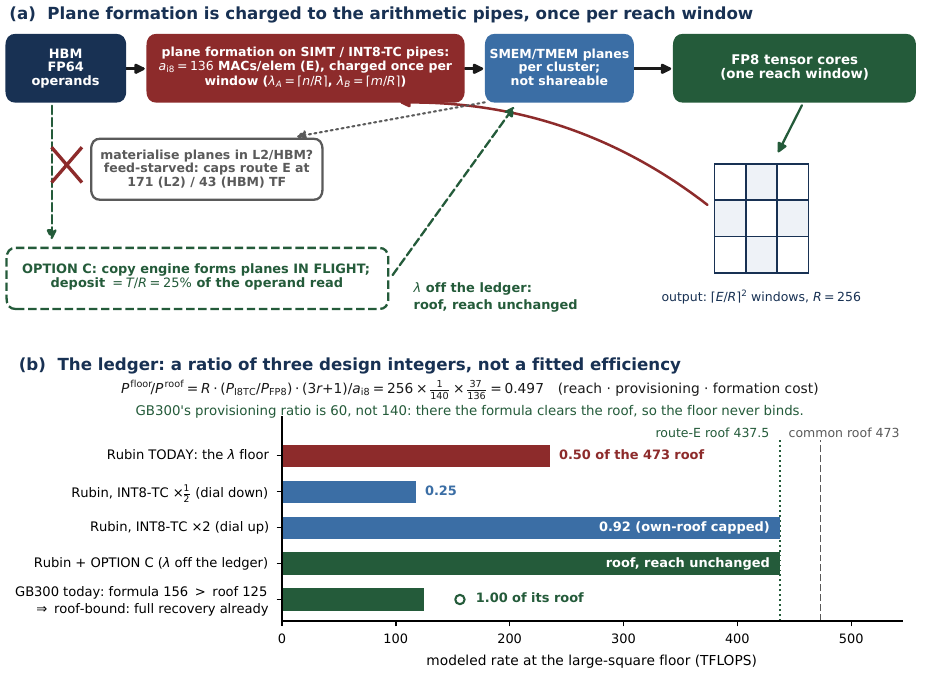}
\caption{Why the large-square floor is one half on Rubin---and why a minor
datapath addition recovers all of it.  \emph{(a)}~Data movement at the
floor: each reach-sized output window re-splits its operand panels into
residue planes on the arithmetic pipes (the $\lambda$ charges), because
planes cannot be shared across clusters and streaming materialised planes
from L2/HBM is feed-starved, capping the rate below the floor itself; the
dashed Option-C path forms planes on the copy engine instead, removing the
charge.  \emph{(b)}~The ledger: Rubin's floor as the derived product of
Eq.~\eqref{eq:whyhalf} against the two roofs; the provisioning dial
(halved/doubled \inteight{}-tensor rate) moving the plateau; GB300, where
the formula exceeds the roof and the floor never binds; and Option~C
returning Rubin to the roof at unchanged reach.  All quantities are
computed by the artifact engine (\texttt{render\_lambda\_floor.py}); no
value is fitted.}
\label{fig:lambdafloor}
\end{figure}

Figure~\ref{fig:lambdafloor} draws the data movement behind that ledger.
The floor exists only because plane \emph{formation} is charged to the
arithmetic pipes: an output larger than one cluster's reach re-splits its
operands ($\lambda_A{=}\lceil n/R\rceil$, $\lambda_B{=}\lceil m/R\rceil$)
because on-chip planes cannot be shared across clusters, and materialising
them through memory instead is \emph{feed-starved}: at the $64\times64$
output tile a materialised plane stream costs $p/\mathrm{TILE}$ bytes per
useful flop, capping route~E at $171$~TFLOPS from L2 and $43$ from
HBM---both below the $235$ the on-the-fly schedule holds
(\S\ref{sec:storagemodes}).  The minor datapath addition of \S\ref{sec:hw}
(Option~C, an in-flight-convert copy engine) moves formation onto the copy
path at stream rate: deconstruction leaves the arithmetic ledger entirely,
the $\lambda$ term vanishes rather than shrinks, and the rate returns to
$\min(\text{roof},\text{memory bound})$---the roof, at unchanged
reach---while the added deposit traffic is a fraction of an operand read
the kernel already performs (\S\ref{sec:hw}).  Full recovery from a minor
addition, because the obstruction was an accounting assignment, not a
bandwidth shortage.

\begin{figure}[tp]
\centering
\includegraphics[width=\linewidth]{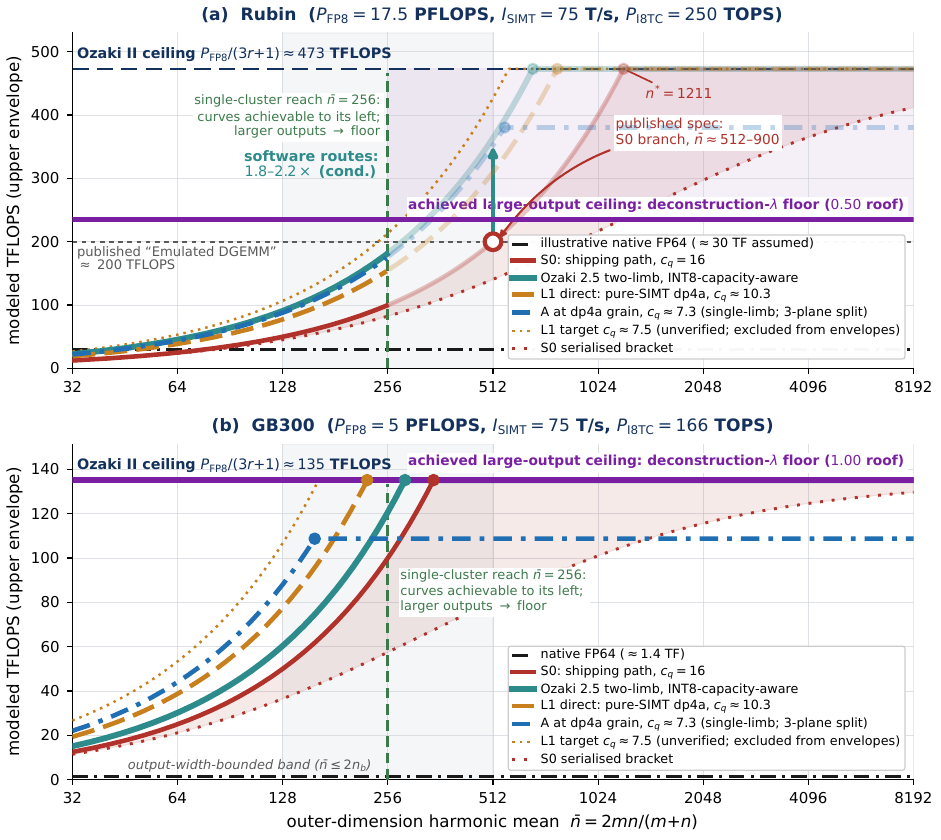}
\caption{Modeled emulated-DGEMM upper envelopes versus problem size
on Rubin \emph{(a)} and GB300 \emph{(b)}, Eq.~\eqref{eq:pdense}, at
the memo-counted reference constants and under the Ozaki~2.5 ladder
(conservative \texttt{dp4a} counts; generated from
\texttt{params.py}).  \textbf{The rising curves are the
\emph{convert-once} upper envelope, achievable only up to the single-cluster
reach} (green dashed, $\nbar{=}256$); a larger multi-cluster output is
re-split on the fly, so its achieved rate is capped at the \textbf{solid
brick deconstruction-$\lambda$ floor} (${\approx}235$~TFLOPS, $0.50$ of
set~S's $473$-TFLOPS roof on
Rubin; equal to GB300's $135$-TFLOPS set-S roof, via \texttt{dp4a}: roof-bound),
and the shaded
gap between floor and roof is unreachable without the co-design of
\S\ref{sec:hw}. We draw the floor \emph{here}, with the crossover, so the
honest large-output ceiling is visible at the outset rather than disclosed
late.  The shaded band is the overlap bracket of the
shipping path (S0, $c_q{=}16$); the published $\sim$200-TFLOPS figure
lies on the S0 branch at $\nbar{=}512$ (overlapped) to
$\nbar{\approx}900$ (serialised)---a model-inferred size, not a
published one.  The two-limb route is drawn at its
\inteight{}-capacity-aware effective knee (${\approx}670$ on Rubin,
${\approx}290$ on GB300, both at $\eta_{\text{red}}{=}1$); L1 direct
($c_q{\approx}10.3$) and A at \texttt{dp4a} grain
($c_q{\approx}7.3$) use no tensor capacity; the aspirational L1
target ($c_q{\approx}7.5$, unverified reuse schedule) is the thin
dotted curve, excluded from all envelopes.  The flat dash-dotted line
is an \emph{illustrative} native-\fp{64} reference (Rubin:
$\approx\!30$~TFLOPS, the 33-TFLOPS vector specification at
${\sim}0.9$ blocked efficiency; GB300: ${\approx}1.4$~TFLOPS)---the
fallback path emulation does not accelerate.  The shaded vertical
band marks the $\nbar$ range of \emph{output-width-bounded} kernels
(block-vector GEMMs and width-bounded panels, $\nbar \lesssim 2 n_b$
for widths $n_b = 64$--$256$); the measured traces of
\S\ref{sec:apps} place LOBPCG and QR-panel work in this band and
multifrontal fronts at $\nbar \approx 400$--$900$.  In the band the
switched dispatch of Figure~\ref{fig:switch} is worth
${\approx}1.9$--$2.4\times$ over S0 (platform-dependent; band
dispatch---distinct from the schedule-filtered trace composites of
Table~\ref{tab:apps})
(independent tensor service assumed; the serialised bracket and the
\inteight{}-tensor-contention-free \texttt{dp4a} floor of
\S\ref{sec:proj} and the
storage-mode accounting of \S\ref{sec:storagemodes} apply).  On GB300
every plotted route exceeds the native reference throughout the
evaluated range $\nbar \ge 32$.}
\label{fig:knee}
\end{figure}

\paragraph{Where the published figure falls.}
Figure~\ref{fig:knee} plots Eq.~\eqref{eq:pdense}.  Reading it at
$c_q = 16$: the modeled rate is $200$~TFLOPS at $\nbar = 512$
overlapped, and $200$~TFLOPS at $\nbar \approx 900$
serialised.  NVIDIA's published Emulated-DGEMM figure is reproduced by
the deconstruction term \emph{alone}, with the \fp{8} tensor pipes
idle $58\%$ of the time, for benchmark sizes anywhere in the
$\nbar \approx 500$--$900$ window.  We state the reading precisely:

\begin{quote}\itshape
Hypothesis: the published $\sim$200-TFLOPS Rubin Emulated-DGEMM figure
lies on the deconstruction-limited branch of Eq.~\eqref{eq:pdense}---a
deconstruction-bound, not tensor-bound, operating point at a
model-inferred size.
\end{quote}

\noindent
This is a hypothesis, not a finding, and the alternative is real: a
flat ${\sim}40\%$ delivered efficiency (tile efficiency,
sustained-versus-boost clocks, workspace traffic) explains the same
single number if the benchmark was large.  The B200 figure cuts both
ways, and we read it as a \emph{sensitivity illustration, not an
anchor}.  Its \inteight{}-substrate emulated DGEMM
delivered $\approx\!150$ of a $281$-TFLOPS ceiling
($\approx\!53\%$)---a ratio so similar to Rubin's that a common,
generic library-efficiency margin is the parsimonious first reading.
A CRT accounting can nonetheless be laid over it: at
$c_q \approx 13$ (the memo's Karatsuba-free count) and
$\alpha_{\inteight} = r_{\textsc{i}}{+}1 = 16$ on the all-byte set
($r_{\textsc{i}}{=}15$; \S\ref{sec:ozaki2})---with the
$P_{\text{int}} \approx 75$-\tops{} normalisation \emph{transferred}
from the GB300 memo, not measured on B200---the B200 path would have
a knee of its own at
$n^{*} \approx 13 \cdot 15 \cdot 4{,}500/(16 \cdot 75) \approx 730$,
and its published $150$~TFLOPS would fall on \emph{its} serialised
branch at $\nbar \approx 840$: the two generations' margins are then
jointly consistent with the fourth-term reading at a single, common
benchmark size in the upper half of that window,
$\nbar \approx 840$--$900$.  This illustrative joint
reading rests on transferred constants and is conditional on
the B200 figure's algorithm being the CRT path (\S\ref{sec:ozaki2}
provenance note); if the shipping path is Scheme-I
slicing~\cite{emugemm2026}, its deconstruction is byte extraction,
the 53\% margin is generic library efficiency, and only the Rubin
side of the joint reading survives---it validates nothing either
way.  The two explanations thus cannot be
separated by the published numbers alone---but their \emph{signatures}
are orthogonal.  The fourth term predicts a delivered rate
\emph{linear in $\nbar$} up to a knee whose position moves with each
path's $c_q$ (${\approx}730$ on B200, ${\approx}1{,}211$ on Rubin),
and predicts \emph{no} knee at practical sizes for an
error-free-slicing \inteight{} kernel, whose per-element byte
extraction puts its crossover below $\nbar \approx 25$; efficiency
artifacts are flat in $\nbar$ on every path.  A DGEMM size sweep on
Rubin and B200, with a slicing-based \inteight{} kernel as the
control arm, decides the question and fits each path's $c_q$ from its
knee position in the same run (\S\ref{sec:validation}).

With the ceiling stated, the rest of the paper is comprehensive about
\emph{how far each deconstruction algorithm gets}---and honest about where
each stops. Table~\ref{tab:algs} is the menu the switched dispatch chooses
from: \S\ref{sec:ozaki25}--\S\ref{sec:codesign} design its entries (the
modulus sets and reduction routes that bring $c_q$ from the shipping $16$
toward ${\approx}5$--$6$ and set both the small-size rate and the floor's
height), and \S\ref{sec:hw} collects the limitations that bound them
(Table~\ref{tab:limits}) and the preferred co-design target that removes the
dominant one. Read together, the two tables are the whole design space on
one page.

\begin{table}[t]
\centering\footnotesize
\setlength{\tabcolsep}{4pt}
\caption{The Ozaki~2.5 route menu (Rubin), dispatched by size and platform.
``pipe'' = where deconstruction runs; ``floor'' = the $16$-CTA large-output
deconstruction-$\lambda$ ceiling (Fig.~\ref{fig:knee}) as a fraction of that
route's \emph{own} roof; ``roof@$C$'' = cluster size at which \emph{this
route} reaches its own roof. Tensor routes carry a high per-element deconstruction on
the fast INT8 pipe (higher floors); \texttt{dp4a} routes carry a tiny count
on the slower SIMT pipe (lower floors, but \emph{zero} integer-tensor
demand---decisive on GB300 and deployable on any CUDA core today). No route
reaches the raw roof for large DGEMM without the co-design of \S\ref{sec:hw}.
The floor entries below are evaluated at $(m,n,k){=}(4096,4096,4096)$,
the HPL-relevant panel depth of Table~\ref{tab:hplnb} (reported GPU HPL
practice is $\mathrm{NB}{\approx}892$--$1024$).  The floor is a
\emph{large-$k$} quantity, and this reorders the routes
relative to the single-cluster crossover of Table~\ref{tab:proj}: at
HPL-scale reduction depth the Garner reconstruction epilogue is not yet
amortised, so the all-byte A route---which peaks at $243$~TF at $\nbar{=}256$
(its \emph{crossover}) and rises to that only as $k{\to}\infty$---delivers a
\emph{floor} of $231$ (its $0.61$ column, $=231/380$), \emph{below} the
hybrid E floor of $235$; E is therefore the best floor \emph{in this regime},
while A keeps the best roof-\emph{fraction}.  This ordering is itself
finite-$k$: route A's finite-$k$ rate keeps rising past route E's plateau
and overtakes it at $k{\approx}6777$ (Rubin; the crossover is invariant to
$m{=}n$ at this scale, including the $4096$ this table's floors use),
roughly
$6$--$7\times$ beyond any panel width this paper reports as practiced, so E
is the better choice at every HPL-relevant depth but not as $k\to\infty$. The
$\nbar{=}256$ crossover value and
the large-square floor are thus distinct numbers for a route and must not be
conflated, and neither should be read as holding for arbitrarily large $k$.}
\label{tab:algs}
\begin{tabular}{@{}llcccc>{\raggedright\arraybackslash}p{4.5cm}@{}}
\toprule
route & set ($r$) & pipe & roof & floor & roof@ & where it wins / what limits it \\
      &           &      & (TF) & /roof & $C$   & \\
\midrule
S0  & S ($12$) & SIMT & 473 & 0.21 & -- & shipping baseline; high $c_q{=}16$ \\
S2L & S ($12$) & INT8-tens.\ & 473 & 0.38 & 128 & highest roof, full FP64; deconstruction-heavy \\
Et  & E ($13$) & INT8-tens.\ & 438 & 0.54 & 64 & \textbf{best floor at HPL-relevant $k$} ($235$ TF); hybrid moduli \\
At  & A ($15$) & INT8-tens.\ & 380 & 0.61 & 64 & \textbf{best roof-fraction}; one-pass reduction; lower roof \\
Dt  & D ($17$) & INT8-tens.\ & 337 & -- & -- & \emph{not carried}: lowest roof; needs $n_{\text{blk}}{=}3$ ($288{>}256$ KiB at $2$) \\
L1d & S ($12$) & SIMT \texttt{dp4a} & 473 & 0.32 & 256 & zero int-tensor, any CUDA core today; SIMT-bound \\
Adp & A ($15$) & SIMT \texttt{dp4a} & 380 & 0.45 & 128 & zero int-tensor; \textbf{GB300 leader}; SIMT-bound \\
\bottomrule
\end{tabular}
\end{table}

\section{The Ozaki 2.5 Method}
\label{sec:ozaki25}

\begin{definition}[Ozaki 2.5]
\label{def:ozaki25}
\emph{Ozaki~2.5-S} is the Ozaki~II algorithm on the \emph{published}
modulus set---MMA schedule and reconstruction mathematics unchanged;
the accuracy contract inherits as proved for the round-to-nearest
variant, the implemented truncation rule being a stated obligation
(Appendix~\ref{app:obligations})---with the deconstruction path engineered to its proposed
SIMT residual and overlapped behind the tensor stream; the
\emph{codesigned variants} (A/E/D, \S\ref{sec:codesign}) retain the
reconstruction framework but carry the per-set obligations of
Appendix~\ref{app:obligations}.  We write ``Ozaki~2.5'' for the
family.  The engineering discipline is: (O1)~each operand element is converted \emph{once} where the
storage mode permits ($\lambda{=}1$; fused modes admit
$\lambda \ge 1$, \S\ref{sec:storagemodes}) and its
residue planes reused thereafter; (O2)~the linear per-modulus
byte-plane reduction leaves scalar code for dot-product
grain---a small constant GEMM on an integer tensor pipe, or packed
\texttt{dp4a} on SIMT (route L1; Eq.~\eqref{eq:dp4a})---a route
\emph{choice} dispatched per platform; (O3)~the SIMT residual
(scale/truncate, final reduction, Karatsuba split) is implemented at
\texttt{DP4A}/narrow-integer grain, at the floor of
Eq.~\eqref{eq:floor}; (O4)~conversion of tile $t{+}1$ is pipelined
behind the MMA of tile $t$ through the asynchronous copy path; and
(O5)~the lazy-reduction trade is \emph{not} taken, preserving the full
$P_{\fp{8}}/(3r{+}1)$ ceiling.
\end{definition}

\begin{figure}[!t]
\centering
\includegraphics[width=\linewidth]{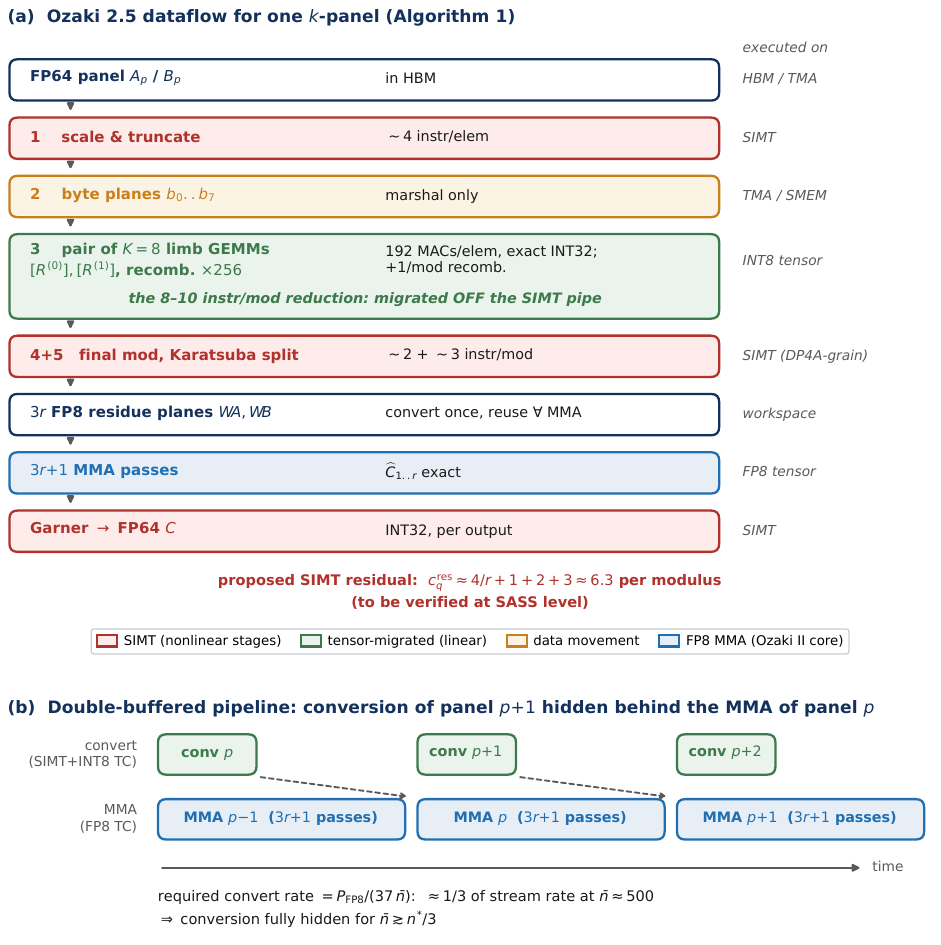}
\caption{The Ozaki~2.5 method.  \emph{(a)}~Dataflow for one $k$-panel
(Algorithm~\ref{alg:ozaki25}): the \fp{64} panel is scaled and
truncated to integers on SIMT, its byte planes are marshaled, the per-modulus
reduction runs as a \emph{pair} of exact $8{\times}r$ constant limb
GEMMs (recombined with weight $256$) on the \inteight{} tensor pipe (the 8--10 instructions per modulus that
dominate the memo-counted $c_q$, migrated off the SIMT pipe---or, as
route L1, kept on SIMT as packed \texttt{dp4a} dot products,
Eq.~\eqref{eq:dp4a}, when the integer-tensor rate is the binding
resource, as on GB300), the final
mod and the Karatsuba split---the irreducible nonlinearities, the
$c_q^{\text{res}} \approx 6.3$ proposed SIMT residual (incl.\ limb recombination)---remain on SIMT, and
the resulting stored \fp{8} residue planes ($30$ for the published
set) are written once per operand lifetime (call-scoped in mode M;
persistent in mode P; never written in modes F/split-$k$,
\S\ref{sec:storagemodes}) and reused by every MMA pass.  \emph{(b)}~Double-buffered pipelining hides the
conversion of panel $p{+}1$ behind the $(3r{+}1)$-pass MMA of panel
$p$; the required conversion rate is $P_{\fp{8}}/(37\nbar)$, about a
third of the full stream rate at $\nbar \approx 500$.}
\label{fig:pipeline}
\end{figure}

\noindent
The name marks its place: mathematically it \emph{is} Ozaki~II (hence
not ``III''), but as an implementation discipline it is the missing
half-step between the shipping realisation and the
hardware-assisted deconstruction datapaths whose evaluation is
follow-up codesign work.  Its net effect
on the model constants is
\begin{align*}
c_q&:\ 16 \;\longrightarrow\; {\approx}\,6.3\ \text{(two-limb SIMT residual), hence}\\
n^{*}\ \text{(Rubin)}&:\ {\approx}\,1{,}211 \longrightarrow {\approx}\,480\text{--}730,\\
n^{*}\ \text{(GB300)}&:\ {\approx}\,346 \longrightarrow {\approx}\,130\text{--}290,
\end{align*}
the ranges spanning the SIMT-residual and \inteight{}-capacity
limits (GB300's residual \inteight{} tensor rate,
${\sim}166$~\tops{}, caps harder than Rubin's ${\sim}250$).
Figure~\ref{fig:pipeline} illustrates the dataflow and the
pipelining; Algorithm~\ref{alg:ozaki25} specifies the method.

\begin{algorithm}[tp]
\caption{\textsc{Ozaki25-Dgemm}$(A, B, C, r)$ --- Ozaki~II with the
deconstruction path at its proposed SIMT residual.  Per-stage costs in
comments; stages tagged \textbf{[SIMT]}, \textbf{[TENSOR]},
\textbf{[TMA]}.}
\label{alg:ozaki25}
\begin{algorithmic}[1]
\State \textbf{input:} $A, B, C, r$; storage mode
   $\in \{$F, M-L2, M-HBM, P$\}$; conversion multiplicities
   $\lambda_A, \lambda_B$; workspace location and lifetime
   \Comment{per \S\ref{sec:storagemodes}}
\State \textbf{setup:} coprime moduli $m_1..m_r$ (the published
   $9$--$11$-bit set---six squares $\le 33^2$ plus a near-$2^9$
   tail---or a codesigned set of \S\ref{sec:codesign});
   constant $\mathcal{R} \in \mathbb{Z}^{8\times r}$, $\mathcal{R}_{ki} = 2^{8k} \bmod m_i$;
   Garner constants; ESC exponent scan of $A, B$
   \Comment{once per call; amortised}
\State allocate plane workspaces \emph{by storage mode} (stored
   planes: 30 S / 44 A / 33 E):
   \textbf{F}/split-$k$---ephemeral CTA/cluster SMEM only, no HBM
   object, conversion multiplicities $\lambda_A,\lambda_B$ charged;
   \textbf{M-L2}/\textbf{M-HBM}---call-scoped workspace of intended
   residency (L2 or HBM), written once and read back;
   \textbf{P}---reuse the existing operand-stationary object, its
   one-time setup charged separately; \textbf{hybrid}---materialise
   the smaller operand's planes, stream the larger
   \Comment{O1: $N_{\text{conv}} = \lambda_A mk + \lambda_B kn$;
   ledger and byte ledger kept separate, \S\ref{sec:storagemodes}}
\For{each $k$-panel $p$ \textbf{ async, double-buffered}}
   \Comment{O4: overlaps MMA of panel $p{-}1$}
   \State \textbf{[SIMT]}\ $S \gets \mathrm{trunc}(D_A\, A_p)$
      \Comment{scale + truncate (source convention): $\sim\!4$ instr/element}
   \State \textbf{[TMA]}\ marshal byte planes
      $\mathcal{B}[0..7] \gets \mathrm{bytes}(S)$
      \Comment{data movement, no arithmetic}
   \State \textbf{[TENSOR/\inteight{} or SIMT-\texttt{dp4a}]}\ $U \gets \mathcal{B} \times \mathcal{R}$
      \Comment{O2 route choice: two-limb reduction GEMMs
      ($16r \approx 192$ MACs/elem, exact \intthirtytwo{} accum.\ $+$
      recomb.)---or route L1, packed \texttt{dp4a}
      (Eq.~\eqref{eq:dp4a}); dispatched per \S\ref{sec:codesign}}
   \State \textbf{[SIMT]}\ $V_i \gets U_i - m_i \lfloor U_i / m_i \rfloor$
      \Comment{final mod to centred $[-m_i/2, m_i/2)$: $\sim$2 instr/mod}
   \State \textbf{[SIMT]}\ $W\!A \gets \mathrm{karatsuba3}(V_i)$
      \Comment{$3$ \fp{8} pieces per nonsquare modulus ($2$/square): $\sim\!3$ instr/modulus}
   \State same for $B_p \to W\!B$
      \Comment{SIMT residual incl.\ limb recomb.: $c_q^{\mathrm{res}} \approx 4/r + 1 + 2 + 3
      \approx 6.3$/modulus}
\EndFor
\For{$i = 1..r$; piece-pairs of the $(3r{+}1)$-pass schedule of~\cite{uchino2026_fp8}}
   \State \textbf{[TENSOR/\fp{8}]}\ $\widehat{C}_i \mathrel{+}=
      \mathrm{MMA}_{\fp{8}}(W\!A_i, W\!B_i)$
      \Comment{exact: \textsc{e4m3}-exact Karatsuba pieces, bounded
      accumulation; the compute roof, $P_{\fp{8}}/(3r{+}1)=473$~TFLOPS on Rubin}
\EndFor
\State \textbf{[\inteight-TENSOR $+$ SIMT]}\ mixed-radix
   reconstruction of $\widehat{C}_{1..r}$: ${\approx}\,r$ MACs/output
   as a length-$r$ recombination on the \inteight{} tensor pipe (the
   codesigned hosting) plus a few SIMT carry/normalise ops; single
   rounding to \fp{64}
   \Comment{$\gamma$ per output, tax $\propto 1/k$ (Eq.~\eqref{eq:recongamma});
   \inteight-hosted $<\!1\%$ at $k \gtrsim 977/2270$ (GB300/Rubin);
   SIMT-only Garner ($90$ ops) $<\!1\%$ only at $k \gtrsim 8.1{\times}10^3/2.84{\times}10^4$}
\State \textbf{[SIMT]}\ ESC check; native-\fp{64} fallback for
   out-of-range rows \Comment{accuracy contract
   of~\cite{uchino2026_fp8,schwarz2025} per variant status,
   Appendix~\ref{app:obligations}}
\end{algorithmic}
\end{algorithm}

\paragraph{O1: convert-once residue-plane workspace.}  Deconstruction
is charged per \emph{converted} element, and the conversion count is
schedule-dependent:
$N_{\text{conv}} = \lambda_A\, mk + \lambda_B\, kn$, with per-operand
multiplicities $\lambda_A, \lambda_B \ge 1$ set by the storage mode
of \S\ref{sec:storagemodes}---the materialised and persistent modes
(M/P) achieve $\lambda = 1$ but pay plane traffic, while the fused
modes (F) generate no plane traffic but admit $\lambda > 1$.  The
conversion-instruction ledger and the HBM byte ledger are therefore
\emph{separate} ledgers, related only under a specific schedule;
neither ``convert once'' nor ``no plane traffic'' holds
unconditionally.  Each $k$-panel
of $A$ and $B$ is converted into its stored \fp{8}
operand planes---$30$ for the published set (two per square modulus,
three per nonsquare~\cite{uchino2026_fp8}; $44$ for candidate A, $33$
for E), a ${\approx}3.8\times$ transient footprint over the \fp{64}
panel for S, bounded by the panel size, not the matrix---and the
planes are reused by every MMA that touches the panel.  For
operand-stationary workloads---Krylov and block-Krylov solvers with a
fixed $A$, repeated application in iterative refinement---the planes
persist \emph{across calls}, amortising the conversion to zero over
the iteration count.  Retained preprocessing of this kind is already
exposed by the public GEMMul8 interface~\cite{gemmul8_github}; O1
adopts it as a baseline requirement (panel-granular and cross-call
persistent) rather than claiming it as new.  The discipline's storage
cost and conversion multiplicity depend on \emph{where} the planes
live---fused in the owning CTA, transiently materialised, or
persistent across calls---a coupling with the traffic model made
explicit, with its mode boundaries, in \S\ref{sec:storagemodes}.

\paragraph{O2: the byte-plane reduction as tensor work---the exact
two-limb encoding.}  The dominant
stage of the memo-derived $c_q$ is the per-modulus reduction, and it
is linear over the byte planes: $x \bmod m_i = \sum_{k} b_k\,(2^{8k}
\bmod m_i) \bmod m_i$.  A one-pass \inteight{} GEMM against
$\mathcal{R}_{ki} = 2^{8k} \bmod m_i$ is \emph{not} exact for the actual
moduli: with $m_i$ up to $1089$, the constants reach $826$, far
outside both signed and unsigned 8-bit range.  The exact realisation
splits each constant into two unsigned byte limbs,
$\mathcal{R} = \mathcal{R}^{(0)} + 256\,\mathcal{R}^{(1)}$ with $\mathcal{R}^{(1)}_{ki} \le 4$, and computes
two \inteight{}$\to$\intthirtytwo{} GEMMs (unsigned bytes times
unsigned limbs; all dot products bounded by $8 \cdot 255 \cdot 255 <
2^{19}$, hence exact), recombined as $U = U^{(0)} + 256\,U^{(1)}$ at
one SIMT instruction per modulus per element (the $+1$ of
Eq.~\eqref{eq:floor}).  The cost is $2 \cdot 8r \approx 192$ MACs
per element; a signed-input correction (the scaled operands are
two's-complement) \emph{subtracts} the precomputed
$2^{64} \bmod m_i$---equivalently, adds the stored constant
$(-2^{64}) \bmod m_i$---in the final reduction, with residues carried
in the centred interval $[-m_i/2, m_i/2)$ (for even $m_i$ the endpoint
$-m_i/2$ is included and $+m_i/2$ excluded; for $m{=}256$ this
represents $-128$); the sign and the boundary cases join the
exhaustive-per-modulus residue unit test of \S\ref{sec:validation}.  A
one-pass single-limb variant becomes exact only under a sub-256
modulus system---a modulus/encoding \emph{codesign} (more moduli,
cheaper formation); \S\ref{sec:codesign} takes this up and produces
concrete candidate systems.  Against the $37\nbar$ \fp{8} operations per element of the
main MMA stream this is nominally small, but the reduction GEMM must
run on an integer datapath (bytes are not exactly representable in
\textsc{e4m3}), and its service demand is first-order in the
conversion-bound regime: at the SIMT-residual knee the two-limb route
demands ${\sim}380$~\tops{} (two-operation convention),
\emph{exceeding} the preliminary $250$-\tops{} Rubin \inteight{}
tensor rate.  Confined to that pipe, the reduction itself limits the
knee to $\nbar \approx 730$ rather than ${\approx}480$; splitting
the reduction between the \inteight{} tensor and the SIMT
\texttt{dp4a} route below (the GEMMul8 baseline
machinery~\cite{gemmul8_github}) is a scheduling question the
measurements must settle.  Logical-shape efficiency is a further open
cost: the exact logical shapes are a \emph{pair} of $K{=}8, N{=}r$
GEMMs (or one $K{=}8, N{=}2r$ product with the two limb outputs kept
separate, then recombined with weight $256$---concatenating the limbs
along $K$ would sum them unweighted and is \emph{not} the identity);
such narrow shapes do not map to dense MMA tiles at peak, and the
vendor \inteight{} figure is a dense-tile rate, not a proven
concurrent capacity beside near-peak \fp{8} work (the two-endpoint
concurrency bracket of \S\ref{sec:proj} quantifies both extremes of
that uncertainty).  Writing
$\eta_{\text{red}}(M,K,N) \le 1$ for the issued-versus-useful
efficiency of this kernel, the capacity knees below assume
$\eta_{\text{red}} = 1$; we therefore quote knee and throughput
\emph{ranges}, not single values.  Formally, the tensor-migrated
routes obey
\begin{equation}
n^{*}_{\text{tensor}} \;=\;
\max\!\left(\frac{c_q^{\text{res}}\, r\, P_{\text{roof}}}{I_{\text{SIMT}}},\
\frac{a_{\inteight}\, P_{\text{roof}}}{\eta_{\text{red}}\, P_{\text{I8TC}}/2}\right),
\qquad P_{\text{roof}} = \frac{P_{\fp{8}}}{\alpha},
\label{eq:tensorknee}
\end{equation}
with $a_{\inteight}$ the useful reduction MACs per element and
$P_{\text{I8TC}}/2$ the MAC rate (the vendor \tops{} figure counts
two operations per MAC); Eq.~\eqref{eq:tensorknee} instantiates the
fourth term of the TME model for these routes and is the form
evaluated throughout (\texttt{params.py}).

\paragraph{The same dot product on SIMT: the \texttt{dp4a}
instruction and route L1.}  The tensor pipe is not the only
dot-product engine on the die.  Since Pascal, the SIMT integer pipe
exposes (PTX \texttt{dp4a})
\begin{equation}
\texttt{dp4a}(a, b, c) \;=\; c \,+\, \sum_{j=0}^{3} a_j\, b_j ,
\label{eq:dp4a}
\end{equation}
one instruction that reads $a$ and $b$ as four packed bytes each and
accumulates their exact dot product into \intthirtytwo{}: four MACs
per issue slot---precisely the grain of the stage-3 sum
$\sum_k b_k\,(2^{8k} \bmod m_i)$.  A worked instance at $m{=}487$,
four planes: the constants $(1, 256, 278, 66)$ split into byte limbs
$(1, 0, 22, 66) + 256 \cdot (0, 1, 1, 0)$; for
$x = \texttt{0xDEADBEEF}$, planes $(239, 190, 173, 222)$, two
\texttt{dp4a} return $18{,}697$ and $363$, and
$18{,}697 + 256 \cdot 363 = 111{,}625 \equiv 102 = x \bmod 487$.
Eight planes cost two \texttt{dp4a} per limb, and a \texttt{dp4a}
returns one scalar dot product per instruction, so narrow high-limb
coefficients do not make the second limb free.  The honest
\emph{direct} count for the published set is therefore
\begin{equation*}
c_q^{\text{L1}} \;\approx\; \underbrace{4/r}_{\text{scale}}
+ \underbrace{2}_{R^{(0)}} + \underbrace{2}_{R^{(1)}}
+ \underbrace{1}_{\text{recombine}} + \underbrace{2}_{\text{mod}}
+ \underbrace{3}_{\text{split}} \;\approx\; 10.3 ,
\end{equation*}
and this is the reference L1 count used in every table and figure.
The high-limb vectors are narrow (entries $\le 4$) and partially
shared---$509$ and $487$ have identical ones---so an explicit
common-subexpression schedule could plausibly reach
$c_q \approx 7.5$; we carry that figure only as an \emph{unverified
optimisation target}, excluded from the dispatch envelope until a
reuse schedule and SASS count exist.  The distinction dissolves for
byte-range constants: for candidate A of \S\ref{sec:codesign} every
constant fits one byte, the second limb and its recombination vanish,
and $c_q \approx 4/15 + 2 + 2 + 3 \approx 7.3$ with \emph{no}
unverified term---the split is charged at three, because three operand
planes ($d_0$, $d_1$, $d_0{+}d_1$) are stored per modulus, matching
$\alpha' = 3r'{+}1$ (an accounting consistency caught in external
review: two \emph{digits} do not imply two \emph{stored
operands})---the fastest rigorously-counted pure-SIMT route in this
paper.  L1 and its A-grain
sibling are thus O2's architectural alternatives, not rivals in
mathematics: the same limb identities, executed on the SIMT pipe
instead of the tensor pipe, trading \inteight{}-tensor capacity for
SIMT issue slots; which trade wins is a platform question
\S\ref{sec:codesign} resolves by rate ratio.

\paragraph{O3: the proposed SIMT residual.}  What remains
on SIMT is the elementwise-nonlinear residue of
Figure~\ref{fig:anatomy}: diagonal scale and truncate-to-integer
($\sim\!4$ per element, amortised over the $r$ moduli), the final
reduction of the GEMM-produced partial residues to the centred
interval $[-m_i/2, m_i/2)$
($\sim\!2$ per modulus), and the Karatsuba split into three \fp{8}
pieces ($\sim\!3$ per modulus, structural for the \fp{8} substrate).
Together with the limb recombination of O2: Eq.~\eqref{eq:floor},
$c_q^{\text{res}} \approx 6.3$---an instruction-ledger projection to
be verified at SASS level (constant division compiles to
multiply-high sequences, not single instructions), implemented at
\texttt{dp4a}/narrow-integer grain (Eq.~\eqref{eq:dp4a}) where the
ISA allows.

\paragraph{O3$'$: closing the nonsquare S-tail layout.}  One line of
that ledger---``the Karatsuba split into three \fp{8} pieces''---was
until now an \emph{assumption} for six of the twelve published moduli,
and it is the line the whole roof rests on: three stored planes per
nonsquare modulus is what makes $\alpha = 3r{+}1 = 37$ and hence
$473$~TF, rather than $\alpha = 43$ and $407$~TF.  Under the canonical
balanced split of Eq.~\eqref{eq:pieces} the six nonsquare S-tail moduli
$487$--$511$ have no \textsc{e4m3}-exact layout: their balanced digit
sums $d_0{+}d_1$ reach $\pm22$ and pass through $\pm17, \pm19, \pm21,
\pm23$, none of which is representable.  Earlier drafts of this paper
recorded that as a proved \emph{impossibility} and carried the six
layouts as an open \texttt{SKIP}.  That was too strong a reading of
its own negative result---it proved only that the \emph{canonical}
split fails---and Lemma~\ref{lem:carry} closes the gap constructively.

\begin{lemma}[Carry-corrected three-plane layout]
\label{lem:carry}
Let $\mathbb{E}$ denote the integers exactly representable in
\textsc{e4m3}, \ie those of the form $M\!\cdot\!2^{e}$ with
$|M| \le 15$ and magnitude at most $448$.  Let $m$ be a nonsquare
modulus with $m \le 513$ and let $d$ range over the centred residues
$[-m/2, m/2)$.  Take the canonical balanced base-$16$ split
$d_0 \equiv d \pmod{16}$, $d_0 \in [-8,7]$, $d_1 = (d-d_0)/16$; if
$d_0{+}d_1 \notin \mathbb{E}$, replace $(d_0, d_1)$ by
$(d_0 - 16\varsigma,\ d_1 + \varsigma)$ where
$\varsigma = \operatorname{sign}(d_0)$.  Then $d = d_0 + 16\,d_1$ still
holds identically, and all three stored planes
$d_0,\, d_1,\, d_0{+}d_1$ lie in $\mathbb{E}$, with $|d_0| \le 14$,
$|d_1| \le 16$, $|d_0{+}d_1| \le 22$.  Every pairwise plane product is
therefore bounded by $22^2 = 484$, so the Karatsuba epilogue
\begin{equation}
d\,e \;=\; (1{-}16)\,d_0 e_0 \;+\; 16\,(d_0{+}d_1)(e_0{+}e_1)
\;+\; (16^2{-}16)\,d_1 e_1
\label{eq:karcarry}
\end{equation}
accumulates exactly in \fp{32} to depth
$K \le \lfloor 2^{24}/484 \rfloor = 34\,663$---\ie at
$K_{\text{slab}} = 4096$ with $N_{\text{acc}} = 3$, with an order of
magnitude to spare.  The bound $m \le 513$ is sharp for this rule:
$m = 514$ fails.
\end{lemma}

The correction is a single predicated add-pair, taken by
$28$ of $511$ residues at $m{=}511$; it costs ${\approx}1$ SIMT
instruction per modulus and is already inside the
$c_q^{\text{res}} \approx 6.3$ ledger, because the split was charged at
three from the outset.  The construction is \emph{redundant} rather
than canonical---$|d_0|$ may exceed $8$---which is precisely why it
escapes the canonical envelope; the price is that $d_0$ is no longer
the unique base-$16$ digit of $d$, which nothing downstream requires.
Verification is exhaustive, not sampled: for each of the six moduli we
check Eq.~\eqref{eq:karcarry} over \emph{all} $m^2$ ordered centred
pairs ($237\,169$ to $261\,121$ per modulus), obtaining zero
mismatches and confirming plane representability at every residue
(\texttt{verify/stail\_layout.py}).  Two consequences are worth
stating plainly.  The published set's roof of $473$~TF no longer rests
on an unexamined property of a third party's stored-plane
construction.  And the ``$m \le 289$ nonsquare'' \textsc{e4m3}
envelope that earlier drafts quoted in Table~\ref{tab:claimstatus}
was doubly wrong: it is not tight even for canonical splits (the
true canonical bound is $m \le 321$, at base $17$), and it does not
bound redundant ones at all (the reachable optimum is $m \le 577$, at base $18$).
The table now carries both corrected numbers.

Algorithms~\ref{alg:deconstruct} and~\ref{alg:reconstruct} write out
the two stages that Algorithm~\ref{alg:ozaki25} compresses into
comments.  They are given separately because they are the two stages
the cost model actually charges---the first sets the deconstruction
floor, the second the fourth term $\gamma\,n_{\text{out}}$ of
Eq.~\eqref{eq:tme}---and because the per-line instruction ledger is
the object a SASS-level audit has to reproduce.

\begin{algorithm}[tp]
\caption{\textsc{Deconstruct}$(X, \{m_i\}, D_X)$ --- one operand panel
to its stored \fp{8} planes; lines~5--11 of
Algorithm~\ref{alg:ozaki25} written out.  This is the stage the
deconstruction floor charges for: the right-hand comments are the
per-element instruction ledger that sums to $c_q$
(Eq.~\eqref{eq:floor}).  Charges are per element \emph{per modulus}
unless marked ``/el''.}
\label{alg:deconstruct}
\begin{algorithmic}[1]
\State \textbf{input:} \fp{64} panel $X$; moduli $m_1..m_r$; ESC
   diagonal scale $D_X$; byte-plane constants pre-split into unsigned
   limbs, $\mathcal{R}^{(0)}_{ki} + 256\,\mathcal{R}^{(1)}_{ki} =
   2^{8k} \bmod m_i$ with $\mathcal{R}^{(1)}_{ki} \le 4$; the
   signed-input correction $\sigma_i = (-2^{64}) \bmod m_i$
\State \textbf{output:} \fp{8} operand planes---$2$ per square modulus,
   $3$ per nonsquare; $p = 30$ stored bytes per scalar for set~S ($44$
   for A, $33$ for E)
\ForAll{elements $x$ of $X$ \textbf{in parallel}}
   \State \textbf{[SIMT]}\ $z \gets \operatorname{trunc}(D_X\, x)$;\quad
      $\mathcal{B}[0..7] \gets \operatorname{bytes}(z)$
      \Comment{scale $+$ truncate: ${\sim}4$/el, \ie $4/r$ per modulus}
   \ForAll{moduli $m_i$, $i = 1..r$}
      \State \textbf{[TENSOR/\inteight{} \emph{or} SIMT-\texttt{dp4a}]}\
         $U^{(0)}_i \gets \textstyle\sum_k \mathcal{B}[k]\,\mathcal{R}^{(0)}_{ki}$,\ \
         $U^{(1)}_i \gets \textstyle\sum_k \mathcal{B}[k]\,\mathcal{R}^{(1)}_{ki}$
         \Comment{O2. Route S2L: $2{\cdot}8$ \inteight{} MACs, both
         partials $< 2^{19}$, hence exact. Route L1: $2{+}2$
         \texttt{dp4a} (Eq.~\eqref{eq:dp4a})}
      \State \textbf{[SIMT]}\ $U_i \gets U^{(0)}_i + 256\,U^{(1)}_i
         - [\,z < 0\,]\,\sigma_i$
         \Comment{limb recombination $+$ two's-complement correction: $1$}
      \State \textbf{[SIMT]}\ $d \gets U_i - m_i \lfloor U_i/m_i \rfloor$,
         re-centred to $[-m_i/2,\, m_i/2)$
         \Comment{O3, final reduction: ${\sim}2$ (multiply-high, not a divide)}
      \If{$m_i = s^2$ is square}
         \State \textbf{[SIMT]}\ $(d_1, d_0) \gets$ balanced base-$s$
            split of $d$;\ \ \textbf{store} $(d_0, d_1)$
            \Comment{$2$ planes; $d_1e_1$ dies mod $s^2$; epilogue
            $\{1, s, s\}$: ${\sim}2$}
      \Else
         \State \textbf{[SIMT]}\ $d_0 \gets$ balanced $d \bmod 16 \in [-8,7]$;\quad
            $d_1 \gets (d - d_0)/16$
         \If{$d_0 + d_1 \notin \mathbb{E}$}
            \Comment{fails only for
            $d_0{+}d_1 \in \{\pm17, \pm19, \pm21, \pm23\}$: $28$ of
            $511$ residues at $m{=}511$}
            \State \textbf{[SIMT]}\ $\varsigma \gets \operatorname{sign}(d_0)$;\quad
               $d_0 \mathrel{-}{=} 16\varsigma$;\quad
               $d_1 \mathrel{+}{=} \varsigma$
               \Comment{one predicated carry, Lemma~\ref{lem:carry}}
         \EndIf
         \State \textbf{[SIMT]}\ \textbf{store} $(d_0,\, d_1,\, d_0{+}d_1)$
            \Comment{$3$ planes; epilogue
            $\{-15,\, 16,\, 240\}$ (Eq.~\eqref{eq:karcarry}): ${\sim}3$
            incl.\ the carry}
      \EndIf
   \EndFor
\EndFor
\State \textbf{invariant:} every stored plane lies in $\mathbb{E}$ and
   every pairwise product is ${\le}484$, so \fp{32} accumulation is
   exact to $K = 34\,663 \gg K_{\text{slab}}$
   \Comment{$N_{\text{acc}} = 3$ for the nonsquare tail}
\State \textbf{ledger:} ${\sim}4/r + 1 + 2 + 3 \approx 6.3$ SIMT
   instructions per element per modulus $= c_q^{\text{res}}$;
   the reduction of line~6 is the term that migrates off SIMT
   \Comment{S row refines this to $5.8$, \S\ref{sec:codesign}}
\end{algorithmic}
\end{algorithm}

\begin{algorithm}[tp]
\caption{\textsc{Reconstruct}$(\{\widehat{C}_i\}, \{m_i\}, D, E)$ ---
mixed-radix CRT recovery of one output tile; line~16 of
Algorithm~\ref{alg:ozaki25} written out.  This is the fourth term
$\gamma\,n_{\text{out}}$ of Eq.~\eqref{eq:tme}: unlike deconstruction
it is charged per \emph{output} element, so its tax falls as $1/k$
(Eq.~\eqref{eq:recongamma}) and it is a small-$k$ wall, not a
large-$k$ one.}
\label{alg:reconstruct}
\begin{algorithmic}[1]
\State \textbf{input:} exact per-modulus accumulators $\widehat{C}_i$,
   $i = 1..r$; moduli $m_1..m_r$; ESC scales $D, E$; precomputed
   inverses $\mu_{ij} = (m_i)^{-1} \bmod m_j$ for $i<j$
\State \textbf{choose host} $\in \{\inteight\text{-tensor},\
   \textsc{simt}\}$
   \Comment{dispatched per \S\ref{sec:proj}; both hostings are carried
   per record}
\If{host $=$ \inteight-tensor}
   \State \textbf{[TENSOR/\inteight{}]}\ $y \gets \textstyle\sum_{i}
      w_i\,\widehat{C}_i$ as a length-$r$ recombination against the
      precomputed mixed-radix weight vector $w$
      \Comment{${\approx}r$ MACs per output; the codesigned hosting}
   \State \textbf{[SIMT]}\ carry-normalise $y$ against the $m_i$ ladder
      \Comment{a few ops; $\gamma$ small}
\Else
   \State \textbf{[SIMT]}\ $v_1 \gets \widehat{C}_1$
   \For{$j = 2..r$}
      \State $v_j \gets \bigl(\widehat{C}_j - \textstyle\sum_{i<j}
         v_i \prod_{l<i} m_l\bigr)\,\mu_{j-1,j} \bmod m_j$
         \Comment{Garner: $r(r{-}1)/2$ multiply--reduce}
   \EndFor
   \State $y \gets \textstyle\sum_j v_j \prod_{l<j} m_l$
      \Comment{$2r$ further ops; $N_\gamma = r(r{-}1)/2 + 2r \approx 90$
      at $r{=}12$}
\EndIf
\State \textbf{[SIMT]}\ rescale by $D^{-1}E^{-1}$ and round \emph{once}
   to \fp{64}
   \Comment{single rounding: the accuracy contract of
   \cite{uchino2026_fp8}}
\State \textbf{[SIMT]}\ ESC range check; re-issue out-of-range rows on
   the native-\fp{64} path
   \Comment{Appendix~\ref{app:obligations}}
\State \textbf{cost:} \inteight-hosted, $<\!1\%$ of service time at
   $k \gtrsim 977$ (GB300) / $2270$ (Rubin); SIMT-only Garner,
   $<\!1\%$ only at $k \gtrsim 8.1{\times}10^3$ /
   $2.84{\times}10^4$
   \Comment{Table~\ref{tab:apps} is priced at the \emph{conservative}
   SIMT-Garner floor}
\end{algorithmic}
\end{algorithm}

\paragraph{O4: pipelining.}  The conversion of panel $t{+}1$ runs
concurrently with the MMA passes over panel $t$, double-buffered
through the TMA/asynchronous-copy path.  The required conversion
element rate is $P_{\fp{8}}/(37\nbar)$---at $\nbar \approx 500$ about
one third of the full HBM stream rate---so the SMEM marshaling budget
that binds fully streamed kernels~\cite{matsuoka2026fp8part1} is not
binding here; overlap is what moves a real kernel from the serialised
toward the overlapped bracket of Eq.~\eqref{eq:pdense}.

\paragraph{O5: what Ozaki 2.5 deliberately does not do.}  The
lazy-reduction trade (deferring the final mod into the accumulation at
$\alpha' \approx 6r{+}1 \approx 73$) buys $c_q \approx 4.3$ on the
published set (${\approx}3.3$ on the one-pass all-byte set) but halves
the dense ceiling to ${\approx}240$~TFLOPS on Rubin; it is the right
rung for memory-bound streamed kernels and the wrong one for dense
GEMM, and Ozaki~2.5 excludes it.

\subsection{Modulus/Encoding Codesign: Sub-256 Systems}
\label{sec:codesign}

The two-limb encoding of O2 accepts the published modulus set and pays
for it.  The converse design question---posed by the representability
problem itself---is whether a \emph{different} modulus system can make
the one-pass reduction exact.  We report a script-checked design
study (constraints: pairwise coprimality; CRT range $\ge$ the current
set's $111.8$ bits; every residue digit \emph{and} every Karatsuba sum
exactly representable in \textsc{e4m3} \emph{under the centred-residue
convention} (residues carried in the symmetric range
$[-m/2, m/2)$, as the symmetric CRT lift already requires); reduction constants
$2^{8k} \bmod m_i \le 255$ for a one-pass GEMM).  Two supply bounds,
computed by \emph{exhaustive} dynamic programming over prime-factor
masks with no cardinality cap (script and result manifest to be
released with the paper; an earlier draft reported greedy constructions as
bounds, an error caught in external review), close the space.
\emph{(i)}~Pairwise-coprime squares under the digit constraint
($s \le 33$) supply at most $94.1$ bits (optimum, attained at
$r{=}11$): squares alone cannot span the \fp{64} range's $111.8$
bits, so some tail is mandatory and the codesign question is only
\emph{which} tail.  \emph{(ii)}~Moduli $\le 64$ supply at most
$89.9$ bits (optimum at $r{=}18$; single-digit moduli $\le 16$ at
most $19.5$): one-pass-per-modulus schedules of that granularity are
unreachable at \fp{64} accuracy---CRT supply, not representability,
is the hard wall.  The word \emph{exhaustive} is scoped in two
parts: the two supply bounds are exhaustive---feasibility upper
bounds over their precisely defined admissible universes
(pairwise-coprime squares with $s \le 33$; moduli $\le 64$)---whereas
the chosen S/A/E/D performance designs are \emph{not} exhaustive:
they are selected points in the far larger, unswept space of mixed
moduli, schedules, and encodings.  Table~\ref{tab:codesign} collects
those selected points and their \texttt{dp4a} reference routes on both
platforms; it is the rate table the rest of the paper reads from.

\begin{table}[!ht]
\caption{Candidate modulus systems and \texttt{dp4a} reference routes on
both platforms (reduced model, $\eta_{\fp{8}}{=}1$; script-generated
conditional projections, produced---with every figure---from the
\texttt{params.py}, supplied in the artifact).  ``mod.@$\nbar$'' = modeled
upper-envelope TFLOPS of useful \fp{64} work, evaluated with
$P_{\text{dec-only}}(\nbar)$---the deconstruction-limited envelope at
$k\!\to\!\infty$, carrying \emph{no} reconstruction charge
(\S\ref{sec:twofunc}).  Every rate column of this table uses that one
function; none of its numbers may be read against a $P_{\text{svc}}$
figure quoted elsewhere, and the two rank routes differently at small
$\nbar$.  S = published set with
the two-limb reduction; A = all-byte; E = hybrid (square moduli $s^2$
with $s \le 33$, plus byte tail); D = 7-bit.  Italicised rows are the
pure-SIMT \texttt{dp4a} realisations of \S\ref{sec:ozaki25}.  Values
assume independent \fp{8}/\inteight{} tensor service; the serialised
bracket and the \inteight{}-tensor-contention-free \texttt{dp4a}
floor are given in
the text ($\theta$-bracket, \S\ref{sec:proj}).  All route values are
ideal-overlap envelopes; schedule eligibility (storage mode, shape,
capacity) per \S\ref{sec:storagemodes}; serialisation endpoints in
the text.  Claim status: modeled projection.}
\label{tab:codesign}
\centering
\footnotesize
\setlength{\tabcolsep}{3.5pt}
\begin{tabular}{lccccccccc}
\toprule
system & $r'$ & $\alpha'$ & roof (TF) & reduction & MACs/el & $c_q^{\text{res}}$ & knee $n^{*}$ & mod.@256 & mod.@512 \\
\midrule
\multicolumn{10}{@{}l}{\emph{Rubin ($17.5$-\pflops{} \fp{8}; \inteight{} cap $250$~\tops{}; native ${\approx}30$~TF illustrative)}} \\
\addlinespace[1pt]
S: published $+$ two-limb & 12 & 37 & 473 & two-limb & 176 & 5.8 & 666 & 182 & 364 \\
A: all-byte & 15 & 46 & 380 & \textbf{one-pass} & 112 & 5.3 & 401 & 243 & 380 \\
E: hybrid squares$+$byte & 13 & 40 & 438 & mixed & 136 & 5.3 & 476 & 235 & 438 \\
D: 7-bit & 17 & 52 & 337 & one-pass & 128 & 5.2 & 400 & 216 & 337 \\
\emph{S $+$ \texttt{dp4a} (L1 direct)} & \emph{12} & \emph{37} & \emph{473} & \emph{pure SIMT} & --- & 10.3$^{\dagger}$ & \emph{782} & \emph{155} & \emph{310} \\
\emph{A $+$ \texttt{dp4a} (single-limb)} & \emph{15} & \emph{46} & \emph{380} & \emph{pure SIMT} & --- & 7.3$^{\dagger}$ & \emph{553} & \emph{176} & \emph{352} \\
\midrule
\multicolumn{10}{@{}l}{\emph{GB300 ($5$-\pflops{} \fp{8}; \inteight{} cap $166$~\tops{}; native ${\approx}1.4$~TF)}} \\
\addlinespace[1pt]
S: published $+$ two-limb & 12 & 37 & 135 & two-limb & 176 & 5.8 & 287 & 121 & 135 \\
A: all-byte & 15 & 46 & 109 & \textbf{one-pass} & 112 & 5.3 & 147 & 109 & 109 \\
E: hybrid squares$+$byte & 13 & 40 & 125 & mixed & 136 & 5.3 & 205 & 125 & 125 \\
D: 7-bit & 17 & 52 & 96 & one-pass & 128 & 5.2 & 148 & 96 & 96 \\
\emph{S $+$ \texttt{dp4a} (L1 direct)} & \emph{12} & \emph{37} & \emph{135} & \emph{pure SIMT} & --- & 10.3$^{\dagger}$ & \emph{223} & \emph{135} & \emph{135} \\
\emph{A $+$ \texttt{dp4a} (single-limb)} & \emph{15} & \emph{46} & \emph{109} & \emph{pure SIMT} & --- & 7.3$^{\dagger}$ & \emph{158} & \emph{109} & \emph{109} \\
 
\bottomrule
\end{tabular}

\smallskip
\noindent{\scriptsize $^{\dagger}$Total $c_q$, entire conversion on
SIMT.  L1 direct $=10.3$: $4/r + 2{+}2$~\texttt{dp4a} (two limbs)
${}+1$ recombine ${}+2$ final mod ${}+3$ split; an explicit high-limb
reuse schedule could lower this toward ${\approx}7.5$, an unverified
target (\S\ref{sec:ozaki25}).  A's byte constants need no second limb;
its $7.3$ (three-plane split charged, matching $\alpha' = 3r'{+}1$)
carries no unverified term.  Knees are model outputs at
stated rates; read them as ${\approx}$two-significant-figure bands.}
\end{table}

\emph{Accounting note:} the S row of Table~\ref{tab:codesign} refines
the uniform $6.3/192$
accounting used elsewhere by counting per modulus (squares split at
$\sim$2 instructions, $1024$ reduces by masking), giving $5.8/176$
and knees ${\approx}440/666$; the uniform and refined figures bracket the same
conclusion.

\paragraph{Three plane counts, not one.}
\label{sec:planecounts}
The single symbol ``number of planes'' is used in the literature for
three quantities that this design separates, because on Rubin they take
three different values and each prices a different resource.  Let $r$ be
the number of moduli and let a modulus be \emph{square} when
$m = s^{2}$.

\begin{description}
\item[$p_{\text{store}}$] \emph{stored} planes per scalar: what the
  operand path must carry from L2 or across the DSM crossbar.  A square
  modulus stores two limbs $d = d_1 s + d_0$; a nonsquare modulus stores
  three ($d_0$, $d_1$, and the Karatsuba sum plane $d_0{+}d_1$).  It
  prices \emph{bandwidth}.
\item[$p_{\text{mma}} = 3r$] operand \emph{feeds} into the tensor
  pipe.  A square modulus has weights $\{1, s, s\}$, so its high limb is
  fed twice; the count is therefore three per modulus regardless of how
  many distinct planes were stored, and $p_{\text{mma}} > p_{\text{store}}$
  exactly on the squares.  With the single reduction pass this gives the
  familiar $\alpha = 3r + 1$.  It prices \emph{issue}.
\item[$N_{\text{acc}}$] fp32 accumulator tiles that must stay resident
  across a whole $k$-slab, hence in \textsc{tmem}.  It is $2$ per modulus
  wherever the merging $-2^{b}$ pre-scale is \emph{taken} and $3$ where it
  is not, and it is \emph{not} $p_{\text{store}}$, $p_{\text{mma}}$, or
  $2r$ in general.  Whether the merge is taken is a per-modulus decision
  with two gates---the pre-scaled plane must be substrate-exact, and the
  merged term must leave accumulator margin at $K_{\text{slab}}$---and on
  S's 9-bit tail it is the second gate, not the first, that we decline
  (below).  It prices \emph{capacity}.
\end{description}

\noindent Table~\ref{tab:tuples} gives the three counts for the three
tensor routes.  The S row is the one that resists a mnemonic: its six
9-bit nonsquare tail moduli ($487$--$511$) stay at $N_{\text{acc}}{=}3$,
so S totals $6(2)+6(3) = 30$, not $2r = 24$, while E and A do reach
$2r$, at $26$ and $30$.  Earlier revisions justified that exception by
\inteight{} range---the pre-scaled plane $-2^{b}d_1$ leaves
$[-128,127]$---which is true of \inteight{} and beside the point here,
because every Ozaki \emph{product} pass in the menu runs on the \fp{8}
substrate.  On \textsc{e4m3} the merge is in fact available: under
Lemma~\ref{lem:carry} the pre-scaled plane reaches exactly
$-256 = -2^{8} \in \mathbb{E}$ on all six, and regrouping
\eqref{eq:karcarry} as
$d\,e = (1{-}16)\,[\,d_0e_0 - 16\,d_1e_1\,] + 16\,(d_0{+}d_1)(e_0{+}e_1)$
reproduces $d\,e$ over all $m^{2}$ ordered centred pairs of all six with
zero mismatches.  What the merge costs is accumulator headroom: the
merged term reaches $4095$, so exact \fp{32} accumulation holds only to
$K \le \lfloor 2^{24}/4095 \rfloor = 4097$, and the schedule is exact at
$K_{\text{slab}} = 4096$ \emph{by a single term}---against $34\,663$ for
the unmerged three planes of Lemma~\ref{lem:carry} and $15\,420$ for the
merged byte-modulus accumulator.  We decline a margin of one and keep
the tail at three (\texttt{verify/stail\_merge.py}).  The exception is
therefore a design choice with a stated margin, not an impossibility;
either way the consequence is that S and A make the \emph{same}
\textsc{tmem} ask from opposite directions, and that $N_{\text{acc}}$
cannot be inferred from $r$ alone---it has to be counted per modulus
(\texttt{verify/nacc\_table.py}).

\begin{table}[!ht]
\caption{The three plane counts and what each prices, for the three
tensor routes at $T{=}64$ on the \fp{8} product substrate.  ``squares''
counts moduli of the form $s^{2}$; $p_{\text{store}}$ is $2$ per square
and $3$ per nonsquare; $p_{\text{mma}} = 3r$ always; $N_{\text{acc}}$ is
counted per modulus---$2$ wherever the merging pre-scale is taken, $3$ on
S's six 9-bit tail moduli, where it is available but declined for
accumulator margin (see text).  ``op.\ ratio'' is the operand
demand at the route's own roof, $p_{\text{store}}/\alpha$
B/FLOP-per-B/FLOP, which is tile- and platform-independent; the
corresponding remote (\textsc{dsm}) share is $(1{-}1/c)$ times it, $0.75$
at $c{=}4$, and is not tabulated separately.  ``\textsc{tmem}
$n_{\text{blk}}{=}2$'' is the live tile count under modulus blocking
(eq.~\ref{eq:nblk}), at $16$~KiB per $64^{2}$ fp32 tile, against a
$256$~KiB budget.  Claim status: counted, script-checked.}
\label{tab:tuples}
\centering
\footnotesize
\setlength{\tabcolsep}{4pt}
\begin{tabular}{lcccccccc}
\toprule
route & $r$ & squares & $p_{\text{store}}$ & $p_{\text{mma}}$ & $\alpha$
 & $N_{\text{acc}}$ & op.\ ratio & \textsc{tmem} $n_{\text{blk}}{=}2$ \\
\midrule
S (published) & 12 & 6 & 30 & 36 & 37 & $6(2){+}6(3) = 30$ & $0.81$ & $15$ tiles, $240$~KiB \\
E (hybrid)    & 13 & 6 & 33 & 39 & 40 & $13(2) = 26$       & $0.83$ & $14$ tiles, $224$~KiB \\
A (all-byte)  & 15 & 1 & 44 & 45 & 46 & $15(2) = 30$       & $0.96$ & $16$ tiles, $256$~KiB \\
\bottomrule
\end{tabular}
\end{table}

\noindent Read across the rows, the table is also the case against~A as a
design point, and it is worth being blunt about it because A is the
route this paper's own deconstruction analysis makes look best.  A holds
exactly one square in fifteen moduli ($256 = 16^{2}$), against six in
twelve for~S and six in thirteen for~E, so it pays $p_{\text{store}} =
44$ against $p_{\text{mma}} = 45$: one modulus in fifteen is amortised on
the operand path and the other fourteen are not.  Its operand ratio
$0.96$ leaves four percent of margin at its own roof, where S and E leave
nineteen and seventeen.  And its blocked \textsc{tmem}
footprint is $256$~KiB against a $256$~KiB budget---exactly zero
headroom, so any co-resident use of \textsc{tmem}, any tile larger than
$64^{2}$, or any less-than-perfect allocator makes A infeasible rather
than merely slow.  And its roof is the lowest of the three to begin
with, $380$~TF against S's $473$.  Three independent margins---operand
bandwidth, \textsc{tmem} capacity, and the roof itself---close on the
same route, and not one of them is the deconstruction cost that makes A
look attractive in the first place.

\paragraph{Two performance functions, kept apart.}
\label{sec:twofunc}
Route comparisons in this section and the next are reported through two
distinct functions, and a good deal of confusion---including in earlier
drafts of this paper and its companion---comes from reading a number
computed by one as though it were the other.  They are:
\begin{description}
\item[$P_{\text{dec-only}}(\nbar)$] the \emph{deconstruction-limited
  envelope}: the rate at which residue formation and the MMA schedule can
  supply a problem of edge $\nbar$ in the limit $k\!\to\!\infty$, with
  \textbf{no} reconstruction charge.  This is the right function for asking
  which modulus set deconstructs most cheaply, and it is what the knees,
  crossovers and envelope figures of this section report.
\item[$P_{\text{svc}}(m,n,k)$] the \emph{reconstruction-aware service
  rate} at finite $k$: the same schedule with Garner reconstruction
  charged against the $mn$ outputs it actually touches.  This is the right
  function for asking what a library call would deliver, and it is what
  the floor, the rung table, and every ``delivered'' figure report.
\end{description}
The two agree only as $k/\nbar\!\to\!\infty$, and they can rank routes
\emph{differently}: at $\nbar{=}256$ the all-byte set~A leads on
$P_{\text{dec-only}}$ and trails on cubic $P_{\text{svc}}$, for the reason
given below.  No table row in this paper mixes them, and each is labelled
with the function it uses.

\paragraph{Candidate A (all-byte).}  Fifteen coprime moduli
$\{256$, $255$, $253$, $251$, $247$, $241$, $239$, $233$, $229$,
$227$, $223$, $217$, $211$, $199$, $197\}$ give $117.8$ bits; the largest reduction constant is
$243$, so the one-pass \inteight{} GEMM is exact as written
($112$~MACs per element; $8 \cdot 255 \cdot 243 < 2^{19}$).  Byte
moduli also make Karatsuba representability clean, \emph{provided
residues are carried centred}: two balanced base-16 digits span
$|d_1 \cdot 16 + d_0| \le 136$, which covers the centred range
$|x| \le 128$ of every byte modulus (it would \emph{not} cover
uncentred residues up to $255$---the digits would overflow).  Under
the canonical tie rule $d_0 \in [-8, 7]$, $d_1 = (x - d_0)/16$ the
decomposition is unique, the digits lie in $[-8,8]$, the stored sum
plane $d_0{+}d_1$ lies in $[-16,15]$, and \emph{every} integer in
these ranges is \textsc{e4m3}-exact
(\texttt{verify\_candidates.py})---unlike
the published nonsquare tail, whose balanced digit sums reach
$\pm 22$, where $17, 19, 21$ are not representable.  We should be
precise about how much this buys, because Lemma~\ref{lem:carry} has
narrowed it: the tail is no longer \emph{unlayoutable}, only no longer
\emph{canonically} layoutable.  A's advantage over S here is one
predicated add-pair on ${\approx}5\%$ of residues, not a feasibility
gap; the honest case for byte moduli rests on the capacity and
deconstruction terms below, not on representability.  The price is the
pass count:
$\alpha' = 46$ ($+24\%$), roof $473 \to 380$~TFLOPS ($-20\%$),
and three stored planes per modulus (the split charged at three
accordingly).  The return is the deconstruction side improving
substantially, the effective knee falling $666 \to 401$ (now
SIMT-residual-bound; the \inteight{}-capacity knee is $341$); below
$\nbar \approx 540$, A is projected \emph{faster} than the
published set despite the lower roof ($243$ vs $182$~TFLOPS at
$\nbar{=}256$, $+34\%$).

That last comparison is a statement about $P_{\text{dec-only}}$, the
deconstruction-limited envelope at $k\!\to\!\infty$, and it does not
survive transfer to the reconstruction-aware service rate---so we make the
distinction explicit here rather than let the reader carry the wrong
number forward.  A's advantage is real where $k$ is deep: at $\nbar{=}256$
and $k{=}4096$, $P_{\text{svc}}$ gives $231$ for~A against $182$ for~S,
$+27\%$.  In the \emph{cubic} regime $k{=}\nbar$ it inverts, and sharply:
$131$ for~A against $168$ for~S, $-22\%$.  The mechanism is the modulus
count A buys its exactness with---Garner reconstruction grows like $r^2$,
so $r{=}15$ carries a residual that $k{=}\nbar{=}256$ cannot amortise.  A
is therefore a deep-$k$ candidate, not a small-square one, and the two
functions are tabulated separately throughout
(\S\ref{sec:twofunc}); no row of this paper mixes them.

\paragraph{Candidate E (hybrid)---the projected sweet spot.}  Keep
the six square moduli ($1089$ down to $529$: squares of $s \le 33$,
where the square shortcut needs no third plane and, again centred,
two balanced base-$s$ digits with $|d_i| \le 16$ span
$16s{+}16 \ge (s^2{-}1)/2$---exactly tight at $s{=}33$, which is
presumably why $33^2$ tops the published set), and replace only the nonsquare tail with seven
byte moduli $\{251, 247, 241, 239, 233, 229, 227\}$: $r' = 13$,
$\alpha' = 40$.  The roof concedes $7.5\%$ ($473 \to 438$~TF), the knee
falls to $476$, modeled throughput at $\nbar{=}512$ rises
$364 \to 438$, and on the deconstruction envelope E dominates S
everywhere below $\nbar \approx 620$ (above which the two-limb S passes
E's roof).  The reconstruction-aware picture is again different, and in
E's case it is the \emph{more} favourable of the two at scale: under
$P_{\text{svc}}$ at $k{=}4096$, E leads S at every $\nbar$ we evaluate
($235$ vs $182$ at $\nbar\ge256$), and in the cubic regime E leads S from
$\nbar \approx 768$ upward while S leads below.  E is thus the route we
recommend for large problems on either accounting, and the one place S
retains an edge---small cubic shapes---is precisely where the
regime-switched dispatch below sends the work to S anyway.  Two-limb work survives for five moduli only
($1024$ reduces by bit masking, free).

\begin{figure}[!t]
\centering
\includegraphics[width=\linewidth]{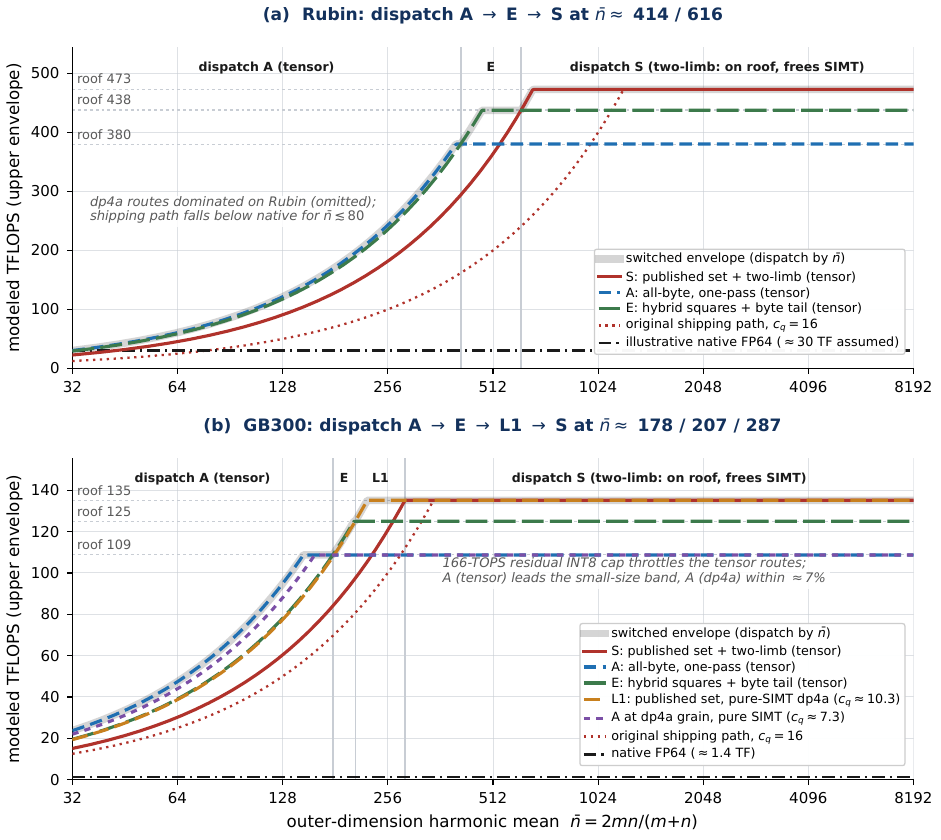}
\caption{The regime-switched dispatch on both platforms (reduced
model, \inteight{}-capacity-aware effective rates, conservative
\texttt{dp4a} counts; modeled \emph{convert-once upper envelopes}, not
delivered performance---realised only up to the single-cluster reach
($\nbar{\lesssim}256$); a large multi-cluster DGEMM is capped at the
deconstruction-$\lambda$ floor ($0.50$ of the $473$-TFLOPS roof on Rubin),
Fig.~\ref{fig:knee}
and \S\ref{sec:hw}).  \emph{(a)}~Rubin: below $\nbar \approx 410$ the
all-byte set A is fastest (one-pass exact reduction, steepest
effective slope); the hybrid E carries the middle band to
${\approx}620$; the two-limb S takes the dispatch above, on the roof
from ${\approx}670$.  The pure-SIMT \texttt{dp4a} routes are
dominated on Rubin and omitted for clarity.  \emph{(b)}~GB300: the
${\sim}166$-\tops{} residual \inteight{} cap throttles the
tensor-migrated reductions and compresses the small-size band:
tensor A leads to its $109$-TFLOPS roof (its \texttt{dp4a}
realisation, shown dotted, within ${\approx}7\%$ with no
\inteight{}-tensor use), E carries ${\approx}180$--$210$, L1 direct
reaches the full $135$-TFLOPS roof at $\nbar \approx 220$, and the
two-limb S ties from ${\approx}290$ (freeing SIMT).  The shaded envelope is the dispatched performance:
no single-route curve exceeds it anywhere.  For comparison, the
original shipping path ($c_q{=}16$, dotted) and the native-\fp{64}
reference lines (illustrative, flat) are shown: on Rubin the shipping
path falls \emph{below} native for $\nbar \lesssim 80$ while the
switched envelope clears it from $\nbar \approx 32$ (${\approx}35$
against the official $33$-TFLOPS figure~\cite{nvidia_hgx}); on GB300
every plotted route exceeds the native reference throughout the
evaluated range $\nbar \ge 32$.}
\label{fig:switch}
\end{figure}

\paragraph{Regime-switched dispatch.}  The modulus set and the
reduction engine are runtime parameters, so a library need not
choose: dispatch by $\nbar$, recovering the sub-crossover region at
$\le\!7.5\%$ asymptotic cost with the Ozaki-II reconstruction framework
unchanged (CRT range matched or exceeded in every candidate; per-set
proof obligations in Appendix~\ref{app:obligations}).  The dispatch
menu comprises the tensor-migrated routes (two-limb S; one-pass A;
mixed E) and the pure-SIMT \texttt{dp4a} routes of
\S\ref{sec:ozaki25} at their \emph{conservative} counts (L1 direct,
$c_q \approx 10.3$; A at \texttt{dp4a} grain, $c_q \approx 7.3$
with the three-plane split charged).
On Rubin the roomy $250$-\tops{} \inteight{} tensor rate keeps the
tensor-migrated routes ahead everywhere: A below $\nbar \approx 410$,
E to ${\approx}620$, and the two-limb S above, on the roof from
${\approx}670$; the \texttt{dp4a} routes are dominated there (L1
direct trails two-limb, and A-\texttt{dp4a} trails one-pass tensor
A).  On GB300 the residual \inteight{} rate (${\sim}166$~\tops{})
throttles every tensor-migrated reduction (effective knees
${\approx}290/150/210$ for S/A/E), and the small-size band
compresses: tensor A leads to its $109$-TFLOPS roof (capacity knee
${\approx}150$), with its pure-SIMT \texttt{dp4a} realisation
within ${\approx}7\%$ of it (knee ${\approx}160$; no \inteight{}
tensor use at all) as the capacity-free fallback; E carries
${\approx}180$--$210$; L1 direct takes the envelope to the full
$135$-TFLOPS roof at $\nbar \approx 220$; and the two-limb S ties
on the roof from ${\approx}290$, freeing SIMT slots there.  Modulus
codesign thus pays on \emph{both} platforms---the all-byte set leads
both small-size bands---and on GB300 the \texttt{dp4a} realisation
offers nearly the same throughput with zero \inteight{}-tensor
demand.
Should the L1 reuse schedule reach its ${\approx}7.5$ target
(\S\ref{sec:ozaki25}), L1 would take the \emph{entire} GB300
sub-roof envelope and the Rubin mid-band above ${\approx}530$;
nothing below rests on that.  Caveats as elsewhere in this paper:
greedy set selection is not proven optimal, the accuracy contract
must be re-verified per set against the source analysis, the 3-pass
schedule is assumed for balanced-digit Karatsuba, and
$\eta_{\text{red}}$, tile shape, and \inteight{}/\fp{8} concurrency
remain measurement questions; the candidate sets therefore join the
validation sweep of \S\ref{sec:validation}.  Reverting to
\inteight{} as the \emph{substrate} is not advantageous on GB300: at
$\alpha_{\inteight} = r_{\textsc{i}}{+}1 = 16$ (the all-byte set's
$r_{\textsc{i}}{=}15$) the residual rate caps emulation at
${\sim}10$~TFLOPS, so \inteight{} serves the reduction GEMM only.
Figure~\ref{fig:switch} shows the resulting
dispatch on both platforms: the switched envelope is at least as
fast as any single route everywhere, and the arithmetic roof is
conceded nowhere.

\subsection{Hardware Co-Design: Lifting the Deconstruction Floor to the Roof}
\label{sec:hw}

\begin{figure}[t]
\centering
\includegraphics[width=\linewidth]{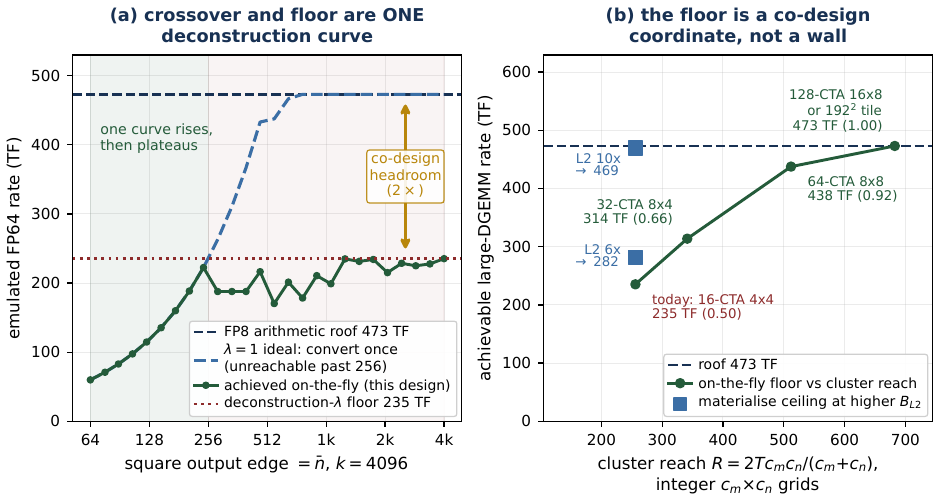}
\caption{(a) The crossover and the floor are one deconstruction curve: the
$\lambda{=}1$ ideal (convert-once, the arithmetic-roof crossover) is
unreachable past the $256$-element cluster reach, and the achieved
on-the-fly rate plateaus at the deconstruction-$\lambda$ floor
(${\approx}235$~TFLOPS, $0.50$ of the $473$-TFLOPS roof; the gap is co-design
headroom).
(b) The floor is a co-design coordinate: raising the cluster reach (bigger
cluster, or bigger TMEM tile) lifts the envelope floor to the roof
($64$-CTA $\to0.92$, $128$-CTA or $192^2$ tile $\to$ roof); a
${\approx}10\times$-HBM L2 reaches it through materialisation instead.}
\label{fig:codesign}
\end{figure}

The preceding two sections leave a single, sharp question, and it is a
co-design question. One term---deconstruction---governs the whole story,
and it does so through one quantity: the per-useful-FLOP conversion load
$\approx c_q\,r\,\lambda/\bar n$. Inside a single cluster $\lambda=1$, so as
$\bar n$ grows this load \emph{falls}---the crossover of
\S\ref{sec:puzzle}, along which the arithmetic roof is approached. But the
load cannot keep falling. Growing $\bar n$ past the cluster's
$256$-element reach forces $\lambda=\bar n/256$, at which point
$c_q\,r\,\lambda/\bar n = c_q r/256$: \emph{constant}. The crossover curve and
the ``$0.50$-of-roof floor'' ($235$ of Rubin's $473$~TFLOPS) are therefore
not two claims but one
curve---the switched envelope rises to ${\approx}235$~TFLOPS at
$\bar n{=}256$ and then plateaus (Fig.~\ref{fig:codesign}a). Its
$\lambda{=}1$ extrapolation would reach the roof near $\bar n{\approx}700$
(and the shipping $c_q{=}16$ path's near $\bar n{\approx}1200$,
\S\ref{sec:puzzle}), but that extrapolation is unreachable: any
$\bar n{>}256$ is multi-cluster, so $\lambda{>}1$ and the curve plateaus
instead. \textbf{The practical effect} is that a real large dense-DGEMM
benchmark on Rubin sees ${\approx}235$~TFLOPS, half the $473$-TFLOP
arithmetic roof---while the traced \emph{applications}, governed by their
tall/skinny, small-$k$, or sub-cluster shapes, are largely spared
(\S\ref{sec:apps}), and the whole question vanishes on GB300, whose
$135$-TFLOP roof already sits at or below its own floor.

Before acting on that floor, Table~\ref{tab:limits} lays out \emph{every}
limitation identified for Ozaki~2.5---what each binds, its magnitude, and its
fix---so the design space is explicit, and so the reader can see that the
deconstruction-$\lambda$ floor is the \emph{one} limitation that binds large
dense DGEMM, the rest being small-size, platform-, or design-space-specific.
Read with the route menu (Table~\ref{tab:algs}), it is the whole picture on
two pages.

\begin{table}[t]
\centering\footnotesize
\setlength{\tabcolsep}{4pt}
\caption{Every limitation identified for Ozaki~2.5: what it binds, its
magnitude, and what lifts it (software / hardware). The
deconstruction-$\lambda$ floor (row 2) is the one that binds large dense
DGEMM; the rest are small-size, platform-, or design-space-specific.}
\label{tab:limits}
\begin{tabular}{@{}>{\raggedright\arraybackslash}p{2.7cm}>{\raggedright\arraybackslash}p{2.4cm}>{\raggedright\arraybackslash}p{3.2cm}>{\raggedright\arraybackslash}p{4.4cm}@{}}
\toprule
limitation & binds & magnitude & lifted by (software / hardware) \\
\midrule
deconstruction $c_q$ (4th TME term) & everything & $16{\to}5$--$6{\to}0$ & modulus codesign (SW); in-flight convert (HW) \\
\textbf{deconstruction-$\lambda$ floor} & \textbf{large dense DGEMM (HPL)} & $0.50$ of $473$ (Et); $0.21$--$0.61$ of own roof & reach$\uparrow$; L2$+$blocking; $c_q{\to}0$ \\
reconstruction (Garner) tax & small-$k$ kernels & SIMT $111\%$@$256$; INT8 $2.2\%$@$1024$ & INT8 mixed-radix hosting (SW); large NB \\
TMEM capacity ($256$ KiB) & route set $+$ reach & at $n_{\text{blk}}{=}2$: $240/224/256$ S/E/A (A exact); D $288$ needs $n_{\text{blk}}{=}3$ & larger TMEM (HW) \\
plane-feed / materialise & schedule choice & needs $222$ TB/s; L2 $88$ & on-the-fly (SW); L2$+$blocking (HW) \\
cluster reach ($256$) & the $\lambda$ itself & $R{=}2Tc_mc_n/(c_m{+}c_n)$ & larger cluster / TMEM (HW) \\
SIMT--\fp{8} overlap $\rho$ & \texttt{dp4a} routes & serialised ${\approx}0.5\times$ & measure (Part 3); scheduling (SW) \\
INT8 residual cap & GB300 tensor routes & compresses small-size band & all-byte \texttt{dp4a} (SW) \\
moduli supply ($111.8$ bit) & modulus-set design & squares $94$, sub-$64$ $90$ (short) & hybrid/all-byte-tail dispatch (SW) \\
small-$\bar n$ knee (crossover) & small outputs & $n^{*}{\approx}480$--$1211$ & lower $c_q$ \\
\bottomrule
\end{tabular}
\end{table}

The floor is $R\,P_{\text{int}}/(c_q r)$---Eq.~\eqref{eq:pdense} evaluated
at $\nbar\!\to\!R$, which is Eq.~\eqref{eq:decfloor}---so what pins is a
\emph{per-element deconstruction load} $L_{\text{dec}} = c_q r/R$, and the
rate is $P_{\text{int}}/L_{\text{dec}}$ under the cap and the host-pipe
$\min(\cdot)$ of Eq.~\eqref{eq:decfloor}. We write it with $r$ explicit
because the load is per \emph{modulus} as well as per element: the shorthand
``$c_q/\text{reach}$'' used in earlier drafts drops a factor $r$ and is
dimensionally incomplete, though it never entered the engine, whose
Eq.~\eqref{eq:pdense} second term has always been
$\nbar P_{\text{int}}/(c_q r)$. Here the cluster's \emph{reach} $R$ is the
harmonic mean of the two output edges a $c_m\!\times\!c_n$ cluster of
$T\!\times\!T$ accumulator tiles spans (Eq.~\eqref{eq:reach} below; $T$ the
accumulator-tile side, $c_mc_n = C$ the CTAs per cluster), so the floor is a
\emph{co-design
coordinate, not a wall}, along three hardware axes and one algorithmic
(Fig.~\ref{fig:codesign}b; the concrete targets are collected in
Table~\ref{tab:codesigntargets}). These are \emph{potential} co-design
points---none exists on shipping or preliminary-Rubin silicon---and the
value of the model is that it localises them precisely: the closed form
$R\,P_{\text{int}}/(c_q r)$ names exactly the four quantities a change can
move---$R$ and $P_{\text{int}}$ hardware, $c_q$ the deconstruction datapath,
$r$ the modulus set---and no other.

\emph{Reach} is the largest-headroom axis, with two hardware routes to the
same on-the-fly reach---raising $C$ or raising $T$. Reach is an
\emph{integer-geometry} quantity, and we state it as one rather than through
the continuous idealisation $T\sqrt{C}$ used in earlier drafts: a
$c_m\!\times\!c_n$ cluster of square $T\!\times\!T$ tiles reaches
\begin{equation}
R \;=\; \frac{2\,T\,c_m c_n}{c_m+c_n},
\label{eq:reach}
\end{equation}
the harmonic mean of its two output edges, which coincides with $T\sqrt{C}$
exactly when $c_m{=}c_n$ and falls short otherwise (a $2{:}1$ grid loses
$5.7\%$). ``Realizable'' here carries two constraints, not one. Besides
$c_mc_n \le C$, the grid must be \emph{placeable}: CUDA requires each grid
dimension to be divisible by the corresponding cluster
dimension~\cite{nvidia_cuda_guide}, so the admissible $(c_m,c_n)$ are those
dividing the CTA-grid extents $\lceil m/T\rceil\!\times\!\lceil n/T\rceil$ of
the launch. A $4096^2$ output at $T{=}64$ is a $64\!\times\!64$ CTA grid, and
$64$ is divisible by $4$, $8$ and $16$ but by none of $5$, $6$ or $11$. The
distinction is load-bearing rather than pedantic, because the product
constraint \emph{alone} admits better reaches than the ones we tabulate:
$5\!\times\!6$ gives $R = 349.09$ against the $8\!\times\!4$ rung's $341.33$
at $C{=}32$, and $11\!\times\!11$ gives $R = 704$ against $16\!\times\!8$'s
$682.67$ at $C{=}128$. Neither divides a $64\!\times\!64$ grid. A shape whose
CTA-grid extents happened to be divisible by $5$, $6$ or $11$ could use them;
power-of-two cluster edges are the ones that stay admissible across the whole
shape sweep, which is why those are what we tabulate, and on the shapes this
paper reports the rungs of Table~\ref{tab:rungs} are therefore the admissible
optima rather than convenient round numbers. Every entry there is the
engine's own floor at that reach, not an interpolation. The two routes are
\emph{not} physically
interchangeable: more CTAs per cluster ($C{=}16$ today---Blackwell allows $8$
portable, $16$ opt-in---to $64$--$128$) requires a cross-GPC
distributed-shared-memory fabric the current within-GPC crossbar does not
provide, while a larger accumulator tile $T$ is bounded by the fixed
$256$-KiB per-CTA TMEM SRAM.

What is being asked for on the $C$ axis is fabric \emph{reach}, not a new
instruction. At today's $C{=}16$ the sharing is expressible in shipping
PTX: the producing CTA publishes its converted planes into its own
\texttt{shared::cta} window and each consumer pulls with
\texttt{cp.async.bulk.shared::}\allowbreak\texttt{cluster.shared::cta.}\allowbreak\texttt{mbarrier::complete\_tx::bytes},
addressing the peer window through \texttt{mapa} and completing on a
\texttt{.cluster}-scope \texttt{mbarrier}; all three primitives were
introduced in PTX ISA 8.0 and require target \texttt{sm\_90} or
later~\cite{nvidia_ptx_isa}. Because each consumer's MMA then issues
against its \emph{own} shared memory, the plain \texttt{.cta\_group=1}
descriptor suffices and no multi-CTA operand descriptor is invoked. The
$C{=}64$/$128$ rungs of Table~\ref{tab:rungs} therefore ask the fabric to
carry the same instruction sequence further, across GPC boundaries---a
physical-reach change, which is why we book it as hardware co-design rather
than as software the reader could write today. Part 1~\cite{matsuoka2026fp8part1}
gives the per-CTA byte accounting for this exchange.

That TMEM bound deserves a precise statement, because there is a tempting
argument that it is not a bound at all. Under the modulus blocking that
Part~1 freezes, the $r$ moduli are cut into $n_{\text{blk}}$ blocks and only
one block is live at a time, so the resident tile count falls from
$N_{\text{acc}} = \sum_i w(i)$ to
\begin{equation}
N_{\text{acc}}^{\text{live}}(n_{\text{blk}})
 \;=\; \min_{\mathcal{B}} \; \max_{B \in \mathcal{B}} \sum_{i \in B} w(i)
 \;\;\ge\;\; \big\lceil N_{\text{acc}}/n_{\text{blk}} \big\rceil ,
\label{eq:nblk}
\end{equation}
the minimum over partitions $\mathcal{B}$ of the modulus set into
$n_{\text{blk}}$ blocks. The inequality is generally \emph{strict}: blocks
partition moduli, not accumulators, so a modulus contributes its whole width
$w(i) \in \{2,3\}$ to whichever block holds it. At $n_{\text{blk}}{=}2$ the
true counts are $15/14/16$ tiles for S/E/A, not the $15/13/15$ the ceiling
suggests, because $r$ is odd for E and~A
(\texttt{verify/\allowbreak blocking.py}); an earlier version of this argument quoted
the bound as though it were the count. One might then raise
$n_{\text{blk}}$ until any tile fits. That does not work, and the
reason is worth recording: each extra block re-traverses the $k$-slab and
re-reads the \fp{64} operand at $8/T$ B per useful FLOP, so the
\textsc{l2} bill is $(n_{\text{blk}}{-}1)\,(8/T)\!\times\!\text{rate}$---and
the $T$ in the denominator is cancelled by the rate the larger tile unlocks,
while the $n_{\text{blk}}$ needed to fit a fixed TMEM grows like $T^2$. At
$T{=}96$ the fit needs $n_{\text{blk}}{=}5$ and \textsc{l2} affords $4$; at
$T{=}128$ it needs $7$--$9$ and affords $4$; at $T{=}192$ no
$n_{\text{blk}}\le r$ fits at all, and beyond $r$ blocking stops being
operand-free. So the area ask is real, and at the $n_{\text{blk}}{=}2$
operating point it is larger than earlier drafts claimed: a $128^2$ tile is a
$3.5$--$4.0\times$ TMEM ask (not ${\approx}3\times$) and a $192^2$ tile
$7.9$--$9.0\times$ (not ${\approx}6.75\times$), the range running over
S/E/A. The tile route does still cut the materialised plane feed to $p/T$
($222\to111$~TB/s at $128^2$), so of the two it remains the more leveraged
axis---but it is bought with SRAM area, not with scheduling.

\begin{table}[htbp]
\caption{Reach rungs on both escape axes, from the realizable integer
geometries of Eq.~\eqref{eq:reach}.  Rates are the engine's Rubin
reconstruction-aware floor $P_{\text{svc}}$ at $m{=}n{=}k{=}4096$ evaluated
\emph{at that reach}, against the $473$-TFLOPS roof; they are modeled
projections, not measurements.  The cluster axis holds $T{=}64$; the tile
axis holds the shipping $C{=}16{=}4{\times}4$.  TMEM is the
$n_{\text{blk}}{=}2$ live-accumulator footprint over S/E/A against the
$256$-KiB per-CTA budget.  Reproduced by \texttt{verify/cluster\_map.py}
and \texttt{verify/blocking.py}.}
\label{tab:rungs}
\centering
\footnotesize
\begin{tabular}{llrrrl}
\toprule
axis & configuration & reach $R$ & TFLOPS & of roof & TMEM ask \\
\midrule
\multirow{4}{*}{cluster $C$}
 & $4{\times}4$ \ (shipping, opt-in) & $256.00$ & $235$ & $0.50$ & $0.9$--$1.0\times$ \\
 & $8{\times}4$                       & $341.33$ & $314$ & $0.66$ & $0.9$--$1.0\times$ \\
 & $8{\times}8$                       & $512.00$ & $438$ & $0.92$ & $0.9$--$1.0\times$ \\
 & $16{\times}8$                      & $682.67$ & $473$ & $1.00$ & $0.9$--$1.0\times$ \\
\midrule
\multirow{4}{*}{tile $T$}
 & $64^2$ \ (shipping)                & $256.00$ & $235$ & $0.50$ & $0.9$--$1.0\times$ \\
 & $96^2$                             & $384.00$ & $342$ & $0.72$ & $2.0$--$2.2\times$ \\
 & $128^2$                            & $512.00$ & $438$ & $0.92$ & $3.5$--$4.0\times$ \\
 & $192^2$                            & $768.00$ & $473$ & $1.00$ & $7.9$--$9.0\times$ \\
\bottomrule
\end{tabular}
\end{table}

\emph{L2 service bandwidth} is the second axis and the
weakest: fed from L2 at the \fp{8} rate a materialised ($\lambda{=}1$)
schedule reaches the roof, but that needs ${\approx}10\times$HBM, and each set
attains \emph{its own} roof at a different multiple, because the requirement is
$B_{\mathrm{L2}} = \text{roof}\cdot p_{\text{store}}/\mathrm{TILE}$ and
$p_{\text{store}}$ differs: set~S reaches its $473$~TFLOPS at $10.08\times$
($221.7$~TB/s), set~E its $438$ at $10.25\times$ ($225.6$~TB/s), and set~A its
$380$ at $11.89\times$ ($261.5$~TB/s).  That is against the ${\approx}4\times$
assumed, and it is subject to the L2 \emph{capacity} predicate
$p\,k(m{+}n)\le C_{\mathrm{L2}}$ (\S\ref{sec:storagemodes}): the large
multi-cluster outputs the floor governs need ${\sim}1$~GB of resident planes
at $k{=}4096$, $8\times$ over the $126$-MB L2, so the planes spill to HBM and
the ceiling collapses to ${\approx}47$~TF. The L2 escape thus reaches the
roof only in the moderate-$k$ window ($N\lesssim1449$ square, set S) where
the floor scarcely bites---the least useful of the three hardware asks.
\emph{$c_q$}, the algorithmic-and-silicon axis, is in fact the
highest-leverage single lever: it scales the whole curve at \emph{every}
reach, and an Option-C-class datapath (\S6.5.2 of~\cite{matsuoka2026fp8part1})
that drives $c_q\to0$ removes the deconstruction term outright---lifting the
dense floor to the roof \emph{and} un-binding the conversion-bound sparse
kernels, which meet $c_q$ directly with no re-split floor to escape. Because
it is the sparse-critical lever, the companion Part~1 develops it in full
(the three options A/B/C: a rebalanced narrow-integer issue, a
\texttt{cvt.fp64.residues} instruction, and a residue-decompose copy-engine
mode); this paper cross-references rather than repeats it.
Table~\ref{tab:codesigntargets} collects the four axes as concrete,
measurable asks, and Fig.~\ref{fig:codesign}b plots the two reach
routes and the L2 escape against the floor they have to clear.

\begin{table}[t]
\centering\footnotesize
\setlength{\tabcolsep}{4pt}
\caption{\emph{Potential} hardware co-design targets for dense emulated
DGEMM (Rubin); none exists on shipping silicon. The floor is
$R\,P_{\text{int}}/(c_q r)$ (Eq.~\eqref{eq:decfloor}), so each axis is a
concrete, measurable ask, and each row here moves exactly one of its four
quantities with the other three held at the memo-derived reference constants
of Table~\ref{tab:codesign}: the two reach rows move $R$; the L2 row moves the
supply the deconstruction load is served from, by making a materialised
($\lambda{=}1$) schedule feasible; the Part~1 datapath row moves $c_q$; and
$r$ is not a hardware knob at all but a modulus-set choice (\S\ref{sec:codesign}).
Reach has two
routes (cluster size or TMEM tile) to the same on-the-fly value but they hit
different physical limits (cross-GPC DSM fabric vs.\ fixed TMEM SRAM); L2 is
gated by an additional capacity predicate; $c_q$---developed in Part~1 as the
sparse-critical lever---is the highest-leverage as it removes the term at
every reach. Reach targets are integer-geometry quantities (see text): the
tabulated $C{=}64$ ($8{\times}8$) and both tile rungs are exact square grids,
while $C{=}128$ must be $16{\times}8$ and so reaches $683$, not $724$---still
above the reach $666$ at which the floor meets the roof, so the roof entry holds.}
\label{tab:codesigntargets}
\begin{tabular}{@{}llll>{\raggedright\arraybackslash}p{4.7cm}@{}}
\toprule
axis & knob & today & target & dense-DGEMM effect \\
\midrule
reach & CTAs/cluster $C$ & 16 & 64 / 128 & floor $0.50\to0.92$ / roof; needs cross-GPC DSM \\
reach & TMEM tile $T$ & $64^2$ & $128^2$ ($3.5$--$4\times$) / $192^2$ ($7.9$--$9\times$) & floor $0.50\to0.92$ / roof; also feed $222\to111$~TB/s \\
bandwidth & L2 service & $4\times$HBM & $10$--$12\times$HBM & roof \emph{iff} planes L2-resident ($N\!\lesssim\!1449$); else $\to47$~TF \\
algorithm & $c_q$ & memo-count & options A/B/C & highest leverage: removes term at every reach; \emph{sparse} lever \\
\bottomrule
\end{tabular}
\end{table}

\emph{Every carried dispatch route is retained} (route D is not carried),
because the knob that rescues each differs
(Table~\ref{tab:algs}). The all-byte and hybrid sets, with their
lower $c_q$, already stand at $0.54$--$0.61$ of their own (lower) roofs
today---$438$~TFLOPS for E, $380$ for A---
and reach them at a $64$-CTA cluster; the high-roof published two-limb and
pure-SIMT routes sit lower ($0.32$--$0.38$ of their own, higher, $473$) and
reach that roof only
at a $128$-CTA cluster (via the switched envelope; a pure-SIMT \texttt{dp4a}
route on its own needs the next cluster step)---so the ``best'' route is
reach-dependent, and a design that can
switch modulus sets by shape (\S\ref{sec:codesign}) can also switch by the
available reach. The sparse and streaming kernels face the same $c_q$ term
from the other side---below the threshold intensity they are
conversion-bound rather than re-split-bound---and their lever is
\emph{deferred/persistent} deconstruction (convert the stationary operand
once, reuse the planes across the iteration), analysed in
Part~1~\cite{matsuoka2026fp8part1}. On shipping and preliminary-Rubin
silicon none of the hardware knobs exists yet, so the projections of
\S\ref{sec:proj} compare software routes at the $16$-CTA reach; the point
here is that the $2\times$ dense-DGEMM gap is \emph{named and closable},
and which improvement closes it is a Part-3 measurable (the per-route floors
and the cluster each needs are Table~\ref{tab:algs}).

\paragraph{A worked example: HPL on a Rubin cluster.}  To make the ceiling
and its escape concrete, take High-Performance Linpack---the \fp{64}
benchmark whose whole reputation is a high roof fraction. HPL is dominated
(${\gtrsim}90\%$ of its FLOPs) by the rank-\textrm{NB} trailing update
$C \gets C - A[M,\textrm{NB}]\,B[\textrm{NB},N]$, whose output $M\times N$
is the \emph{local} trailing submatrix under the two-dimensional
block-cyclic distribution: it fills the GPU and is therefore large and
multi-cluster---exactly the regime the deconstruction-$\lambda$ floor
governs.  So HPL does \emph{not} see the $473$-TFLOP roof.

It is worth being precise about \emph{why} the panel width \textrm{NB}
matters, because it does two unrelated jobs and an earlier version of this
passage credited it with the wrong one.  As the depth $k$ of the trailing
update, \textrm{NB} does \emph{not} move the deconstruction floor at all:
the deconstruction load is $c_q r (\lambda_A m k + \lambda_B k n)$ against
$2mnk$ useful FLOPs, and $k$ cancels.  What it does is amortise
\emph{reconstruction}, which is charged per output element and therefore
falls as $1/\textrm{NB}$.  Separately, and in its second role as the
block size of the block-cyclic layout, \textrm{NB} sets the granularity of
the local edges $M, N$, so an \textrm{NB} that is a multiple of the reach
$R{=}256$ keeps those edges cluster-aligned.  That second effect is real but
small at HPL's scale: at a local edge of $8192$ the misalignment penalty for
a ragged $M{=}N{=}8000$ is ${\approx}2\%$, not the sharp ragged worst case
that afflicts small matrices.  The first effect is the large one, and
Table~\ref{tab:hplnb} quantifies it.

\begin{table}[!ht]
\caption{Sensitivity of the modeled per-GPU HPL trailing-update rate to the
panel width \textrm{NB}, at a local output edge $M{=}N{=}8192$ on Rubin.
Values are $P_{\text{svc}}$ (reconstruction-aware, \S\ref{sec:twofunc}) in
TFLOPS of useful \fp{64}; \textbf{bold} marks the better route at that
\textrm{NB}.  Reported GPU HPL practice is
$\textrm{NB}\approx892$--$1024$~\cite{chpc_hpl_gpu,kempner_hpl}, i.e.\ near
the $\textrm{NB}{=}1024$ column of this table.  Claim status: modeled projection from
\texttt{params.py}, not measurement.}
\label{tab:hplnb}
\centering
\footnotesize
\begin{tabular}{lrrrrrr}
\toprule
route & $\textrm{NB}{=}128$ & $256$ & $512$ & $1024$ & $2048$ & $4096$ \\
\midrule
S (published set, RN contract) & \textbf{186} & \textbf{186} & 182 & 182 & 182 & 182 \\
E (hybrid, codesigned)         & 169 & 159 & \textbf{202} & \textbf{234} & \textbf{235} & \textbf{235} \\
\bottomrule
\end{tabular}
\end{table}

\noindent Two things in that table deserve to be said plainly rather than
left for a reader to notice.  First, the $0.50$-of-roof headline
($235$ of Rubin's $473$~TFLOPS) is an
\textrm{NB}-conditional claim: E reaches $235$ only for
$\textrm{NB}\gtrsim1024$, and at $\textrm{NB}{=}256$ it delivers $159$,
\emph{below} the published set.  The claim survives because reported GPU
HPL practice sits at $\textrm{NB}\approx892$--$1024$~\cite{chpc_hpl_gpu,%
kempner_hpl}, but it survives on an empirical convention, not on a
theorem.  Second, the two routes carry different accuracy contracts, and
the faster one carries the weaker: S is covered by the round-to-nearest
Ozaki~II theorem, while E's guarantees are the per-set obligations of
Appendix~\ref{app:obligations}.  The box below therefore quotes both, and
a reader who wants the theorem should read the S row---$182$~TFLOPS, $0.38$
of the $473$-TFLOPS roof---as the contract-backed number.

The box is an explicit
order-of-magnitude \emph{reference}, not a submission: Top500 does not
currently accept Ozaki-emulated DGEMM for \fp{64} Linpack, so the numbers
below indicate only what the per-GPU floor implies at scale.

\begin{center}\small
\fbox{\parbox{0.93\linewidth}{%
\textbf{HPL per Rubin GPU} (emulated \fp{64}, today's $16$-CTA silicon,
$\textrm{NB}{=}1024$, $P_{\text{svc}}$ at a local edge of $8192$), quoted
for both accuracy contracts:
\\[2pt]
\begin{tabular}{@{}llll@{}}
route & per GPU & of $473$-TF roof & accuracy contract \\
S (published set) & ${\approx}182$~TF & $0.38$
  & round-to-nearest Ozaki~II theorem \\
E (hybrid, codesigned) & ${\approx}235$~TF & $0.50$
  & per-set obligations, App.~\ref{app:obligations} \\
\end{tabular}
\\[3pt]
Both fractions are taken against the \emph{common} $473$-TFLOPS arithmetic roof
$P_{\fp{8}}/37$, so that the two routes are directly comparable on one scale.
Table~\ref{tab:algs} instead reports each route against \emph{its own} roof;
there the E floor reads $0.54$ of set~E's $438$~TFLOPS, which is the same
$235$~TF.\\[3pt]
That is ${\approx}6\times$ and ${\approx}8\times$ the ${\approx}30$-TFLOP
native \fp{64}, and both bracket NVIDIA's announced ${\sim}200$-TFLOP
``Emulated DGEMM.''  The S contract is componentwise \fp{64} with exact
residue-domain accumulation, for which HPL's residual check is routinely
passed on GPUs.\\[3pt]
\textbf{$\mathbf{10{,}000}$-GPU cluster (ballpark):} DGEMM peak
$1.82$~EFLOPS (S) or $2.35$~EFLOPS (E); at the HPL efficiency empirically
established across a decade of large GPU clusters ($75$--$85\%$ over panel
factorisation, pivoting, look-ahead, and communication)
$R_{\max}\approx1.4$--$1.5$~EFLOPS (S) or $1.8$--$2.0$~EFLOPS (E)
\fp{64} sustained---reference figures, no bespoke system model
claimed.\\[3pt]
\textbf{With the preferred co-design target below} (conditional on the
Option~C requirement table, Table~\ref{tab:optionc}, being met)\textbf{:}
each route reaches \emph{its own} roof---not a common one.  Because
Option~C removes the deconstruction term $c_q$ outright, the ranking is
then set by $\alpha$ alone, and the \emph{theorem-backed} published set~S
($\alpha{=}37$) becomes the fastest at ${\approx}473$~TFLOPS, ahead of the
hybrid~E ($\alpha{=}40$, ${\approx}438$) and the all-byte~A ($\alpha{=}46$,
${\approx}380$).  On the S row the cluster DGEMM peak becomes
$4.73$~EFLOPS and $R_{\max}\approx3.6$--$4.0$~EFLOPS: HPL roughly
\emph{doubles} against today's E row on the same $10{,}000$ GPUs and
reaches ${\approx}2.6\times$ today's S row---while \emph{gaining} the
round-to-nearest contract rather than trading it away.}}
\end{center}

\begin{figure}[t]
\centering
\includegraphics[width=\linewidth]{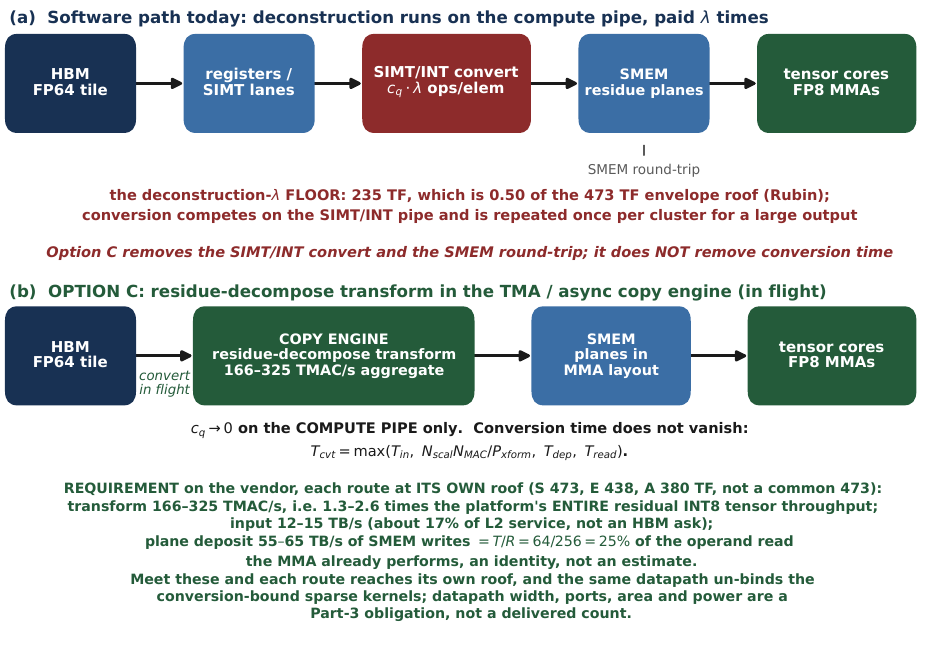}
\caption{The central hardware co-design target, illustrated. \emph{(a)}~Today
the residue conversion runs on the SIMT/integer pipe, competes with the \fp{8}
MMA stream, round-trips through SMEM, and---for a large output---is repeated
once per cluster: this is the deconstruction-$\lambda$ floor ($0.50$ of the
$473$-TFLOPS roof on Rubin). \emph{(b)}~\textbf{Option~C} performs the residue-decompose transform
\emph{in the async copy engine}, converting each \fp{64} tile in flight and
depositing planes in MMA operand layout; deconstruction leaves the
\emph{compute} budget ($c_q{\to}0$ on the SIMT/INT pipe), re-splitting becomes
free ($\lambda$-free), and---\emph{once the requirements of
Table~\ref{tab:optionc} are met}---the emulated rate reaches each route's own
roof at today's cluster, while the same datapath un-binds the
conversion-bound sparse kernels.  Option~C does not zero conversion
\emph{time}, only the compute-pipe term: the transform must sustain
$166$--$325$~TMAC/s ($1.3$--$2.6\times$ the residual \inteight{} pipe) and the
deposit $55$--$65$~TB/s of SMEM writes.  The codesigned modulus sets (byte
extraction plus a few mod-by-invariant multiply--truncates) keep it a
fixed-function residue block rather than a general converter; everything else
stays in software.}
\label{fig:optionc}
\end{figure}

\paragraph{The preferred co-design target to reach the roof.}  Constrain the
problem the way a vendor would---minimise silicon and new datapaths, and let
the proposed \emph{software} carry the complexity---and one option dominates
\emph{under this model}; area, power, routing, and concurrency remain a
Part-3 obligation (Table~\ref{tab:optionc}), so we prioritise rather than
establish minimality:
\textbf{Option~C}, a residue-decompose transform mode in the
TMA/asynchronous copy engine (\S6.5.2 of Part~1~\cite{matsuoka2026fp8part1};
illustrated in Fig.~\ref{fig:optionc}).
It converts each \fp{64} tile to residue planes \emph{in flight},
HBM${\to}$SMEM, so deconstruction leaves the compute budget
entirely---$c_q{\to}0$ on the SIMT/integer pipes---and the deconstruction load
$c_q r/R$ that sets the floor vanishes at \emph{every} reach: the emulated rate reaches
the roof at \emph{today's} $16$-CTA cluster, with no larger cluster, TMEM
tile, or L2 bandwidth, because re-splitting ($\lambda$) is now free (the copy
engine re-converts at stream rate).  This is conditional on the plane-deposit
path: Option~C zeroes the compute-pipe $c_q$ but not conversion \emph{time},
which becomes $T_{\text{cvt}}=\max(T_{\text{input}},\,
N_{\text{scalar}}N_{\text{MAC}}/P_{\text{transform}},\,T_{\text{deposit}},\,
T_{\text{tensor-read}})$; ``reaches the roof'' holds once the transform
sustains the stream rate \emph{and} the deposit sustains its
$(p_{\text{store}}/8){\times}$ share.  Those are provisioning
\emph{requirements} on a vendor, not properties we derive, so we state them
as a bill of materials: Table~\ref{tab:optionc}.  At HPL's panel widths the
model puts the result at the roof to within the ${\approx}1$--$4\%$ Garner
reconstruction epilogue, which overlaps on the residual \inteight{} pipe
(\S\ref{sec:recon}).

\paragraph{What Option~C actually costs, stated as a requirement.}
Table~\ref{tab:optionc} is the ask.  Three of its rows deserve comment
because two of them are smaller than a reader would guess and one is
larger.  \emph{The deposit is small, and exactly so}: the SMEM plane-write
rate is $p_{\text{store}}/R$ bytes per useful \fp{64} FLOP while the
operand read the MMA already performs is $p_{\text{store}}/T$, so the new
write traffic is exactly $T/R = 64/256 = \tfrac14$ of a read that already
happens---an identity, not an estimate, and the strongest honest argument
for Option~C on the bandwidth axis.  \emph{The input side is an L2 ask, not
an HBM ask}: $8/R$ bytes per FLOP is $12$--$15$~TB/s, ${\approx}17\%$ of L2
service, because the slab panels are L2-resident under the fused schedule.
\emph{The transform datapath is the large ask}: at each route's own roof it
must sustain $166$--$325$~TMAC/s, which is $1.3$--$2.6\times$ the
platform's \emph{entire} residual \inteight{} tensor throughput.  Earlier
drafts described this block as ``${\approx}300$ narrow MACs per copy
engine''; that is a per-engine \emph{width}, and it is not derivable
without a clock and a copy-engine count, neither of which is a published
quantity.  We therefore state the aggregate rate, which is derivable, and
leave the width, ports, area and power to the vendor---a Part-3 obligation,
not a delivered count.  What the software does buy is that the datapath is
\emph{narrow}: because our codesigned modulus sets reduce conversion to
byte extraction plus a handful of mod-by-invariant multiply--truncates, it
is a fixed-function residue block, not a general converter, and the moduli,
the two-limb reduction, and the exact reconstruction all stay in software.
Two further properties make it the preferred co-design choice under
the present performance model.  \emph{It is
precedented}: the copy engine already performs layout transforms (TMA
swizzle, Blackwell decompress-on-copy).  \emph{It also fixes sparse}: the
conversion-bound SpMV/GEMV kernels of Part~1 meet $c_q$ directly, so the
same datapath un-binds them---one change closes both regimes, which no
other row of Table~\ref{tab:optionc} does.  This is \textbf{Plan~A}: the option prioritised by the present model,
the only lever that
reaches the roof at \emph{unchanged} reach, and the only one that
closes the sparse side as well.

\begin{table}[!ht]
\centering\footnotesize
\setlength{\tabcolsep}{4pt}
\caption{\textbf{Option~C as a vendor requirement, not as a result.}  Every
row is what a vendor must provision, evaluated per Rubin GPU at
\emph{each route's own} roof under the $4{\times}4$ cluster of $64^2$ tiles
(reach $R{=}256$); none of it exists on shipping silicon.  Script:
\texttt{verify/optionc\_req.py}.  The three sets do not share a roof: with
$c_q$ removed the ranking is set by $\alpha$ alone, so S ($\alpha{=}37$)
leads at $473$, E ($\alpha{=}40$) at $438$, A ($\alpha{=}46$) at $380$~TF.
\emph{Correction to earlier drafts:} the deposit band was printed as
$55$--$83$~TB/s.  That $83$ folded two errors: it priced set~A at set~S's
$473$-TF roof, which set~A cannot reach, and it used
$p_{\text{store}}(\text{A}) = 45$, which counted no squares in a set whose
first modulus is $256 = 16^{2}$; the correct value is $44$
(\S\ref{sec:planecounts}, Table~\ref{tab:tuples}).  Evaluated
route-consistently at the corrected plane count the band is
$55$--$65$~TB/s.  The two \textsc{tmem}-tile alternatives are likewise
corrected: the multipliers here are what a vendor must \emph{provision},
i.e.\ the worst case over the three sets ($4\times$ and $9\times$, both
set~A's), against the per-route ranges $3.5$--$4.0\times$ and
$7.9$--$9.0\times$ of \S\ref{sec:hw}; earlier drafts printed
$3\times$ and $6.75\times$, which are neither.  Claim status: modeled
requirement derived from
\texttt{params.py}; no measurement, and no vendor commitment, is implied.}
\label{tab:optionc}
\begin{tabular}{@{}l>{\raggedright\arraybackslash}p{4.3cm}rrr@{}}
\toprule
requirement & what it is & S & E & A \\
\midrule
route roof (TF) & each route's own $P_{\text{FP8}}/\alpha$ & $473$ & $438$ & $380$ \\
input scalar rate (Gscalar/s) & unique \fp{64} scalars the transform consumes, $P/R$ & $1848$ & $1709$ & $1486$ \\
\quad as bandwidth (TB/s) & $8/R$ B per FLOP, served from L2 ($88$~TB/s) & $14.8$ & $13.7$ & $11.9$ \\
transforms per scalar (MAC) & frozen-kernel narrow multiply--adds & $176$ & $136$ & $112$ \\
\textbf{transform rate (TMAC/s)} & the datapath ask & $\mathbf{325}$ & $\mathbf{232}$ & $\mathbf{166}$ \\
\quad vs residual \inteight{} & $\times$ the platform's $125$-TMAC/s pipe & $2.60\times$ & $1.86\times$ & $1.33\times$ \\
plane deposit (TB/s) & SMEM writes, $p_{\text{store}}/R$ B per FLOP & $55.4$ & $56.4$ & $65.4$ \\
\quad as a share of the read & $=T/R$ exactly, on every route & \multicolumn{3}{c}{$25\%$} \\
tensor-side operand read (TB/s) & $p_{\text{store}}/T$ B per FLOP (unchanged by C) & $222$ & $226$ & $262$ \\
\quad of which remote (DSM) & $(1{-}1/c)\times$ the above & $166$ & $169$ & $196$ \\
staging buffer (KiB/tile) & planes of one $64^2$ \fp{64} tile ($32$~KiB in) & $120$ & $132$ & $176$ \\
latency / ports & drain the staging buffer at the deposit rate & \multicolumn{3}{c}{Part-3 obligation} \\
area / power & fixed-function residue block in the copy engine & \multicolumn{3}{c}{Part-3 obligation} \\
\midrule
\multicolumn{5}{@{}l}{\emph{the alternatives, for the same dense effect}} \\
$C{=}64$ ($8{\times}8$) cluster & cross-GPC DSM at $4\times$ the fabric reach & \multicolumn{3}{c}{floor $0.50\to0.92$} \\
$C{=}128$ ($16{\times}8$) cluster & cross-GPC DSM, reach $683$ & \multicolumn{3}{c}{floor $\to$ roof} \\
TMEM tile $128^2$ & TMEM SRAM $256\to1024$~KiB ($4\times$) & \multicolumn{3}{c}{floor $0.50\to0.92$} \\
TMEM tile $192^2$ & TMEM SRAM $256\to2304$~KiB ($9\times$) & \multicolumn{3}{c}{floor $\to$ roof} \\
faster L2 (Plan~B) & L2 service $4\times$HBM $\to10$--$12\times$HBM & \multicolumn{3}{c}{roof, \emph{dense only}} \\
\bottomrule
\end{tabular}
\end{table}

\paragraph{Plan B, independent of the conversion datapath: a faster L2 with
software-blocked materialisation.}  If a programmable copy-engine transform
is judged too invasive, the datapath-independent fallback scales an
\emph{existing} structure---L2 service bandwidth---and lets software carry
the complexity. Materialise the residue planes, but \emph{blocked}: tile the
output so each block's planes fit the $126$-MB L2 (at $\textrm{NB}{=}2048$ a
$1024^2$ block's planes are almost exactly $126$~MB), keep the row-panel's
planes L2-resident, and stream both operands to the tensor cores at the
\fp{8} rate ($\lambda{=}1$ within the block). The materialised ceiling is
then $B_{\mathrm{L2}}/(p/\mathrm{TILE})$: $0.60$ of the $473$-TFLOPS roof at
$6\times$HBM,
$0.79$ at $8\times$, and the roof at ${\approx}10$--$12\times$. The lone
hardware change is L2 bandwidth---no new datapath, instruction, or
fabric---and the blocking that satisfies the L2 \emph{capacity} predicate is
pure software, so this is the hardware-lightest path \emph{second} to
Option~C. Two honest caveats: ${\approx}10\times$-HBM L2 is a real bandwidth
investment (today ${\sim}3$--$5\times$), and, unlike Option~C, it is
\emph{dense-only}---sparse kernels have no operand reuse to amortise the
materialised feed, so it does nothing for SpMV/GEMV. Two further ``scale an
existing structure'' options round out the menu: a larger \emph{TMEM tile}
(a $3.5$--$4.0\times$ SRAM capacity bump $\to0.92$ on-the-fly, no
materialisation, and it would let D run at $n_{\text{blk}}{=}2$ rather
than $3$---though that halves D's re-read charge without touching its
roof, which is why D stays out of the menu either way), and a pure
\emph{panel} ($\textrm{NB}$)
increase---free, but it reaches only the $0.50$ floor, never the roof. The
whole ladder is Table~\ref{tab:limits}.

\section{Projected Performance}
\label{sec:proj}

Table~\ref{tab:proj} evaluates Eq.~\eqref{eq:pdense} across the
Ozaki~2.5 ladder---S0, the shipping path; L1, the \texttt{dp4a}
route of \S\ref{sec:ozaki25}; the two-limb tensor route; and the
codesigned sets of \S\ref{sec:codesign}---on Rubin and GB300.  Three
readings summarise it.

\begin{table}[!ht]
\caption{\emph{Reduced-model upper envelopes}, software routes only
(TFLOPS of useful \fp{64} $2mnk$ work; $\eta_{\fp{8}}{=}1$;
\inteight{}-capacity-aware effective rates; conditional projections,
\emph{not} delivered performance; generated from \texttt{params.py}).
``Switched'' dispatches the route by $\nbar$ (Rubin: A below
${\approx}410$, E to ${\approx}620$, two-limb S above; GB300:
tensor A below ${\approx}180$---its \texttt{dp4a} realisation
within ${\approx}7\%$---E to ${\approx}210$, L1 to the
$135$-TFLOPS roof at ${\approx}220$, two-limb S from ${\approx}290$;
Figure~\ref{fig:switch}).  The aspirational L1 count
(${\approx}7.5$) is excluded from the envelope pending its reuse
schedule.  Hardware is deliberately absent: Option-C-class conversion
hardware (Part~1) would remove the conversion term at every size, but
on current GPUs these software routes are what is deployable.  All
route values are ideal-overlap envelopes; schedule eligibility
(storage mode, shape, capacity) per \S\ref{sec:storagemodes};
serialisation endpoints in the text.  Claim status: modeled
projection.}
\label{tab:proj}
\centering
\footnotesize
\setlength{\tabcolsep}{4.5pt}
\begin{tabular}{lcccc|cccc}
\toprule
 & \multicolumn{4}{c}{$\nbar$: \textbf{achieved today} (single cluster)}
 & \multicolumn{3}{c}{$\nbar$: convert-once envelope$^{\dagger}$} & knee \\
\cmidrule(lr){2-5}\cmidrule(lr){6-8}
software route & 32 & 64 & 128 & 256 & 512 & 1024 & 2048 & $n^{*}$ \\
\midrule
\multicolumn{9}{@{}l}{\emph{Rubin ($17.5$-\pflops{} \fp{8}; \inteight{} cap $250$~\tops{})}} \\
\addlinespace[1pt]
S0: shipping path, $c_q{=}16$ & 12 & 25 & 50 & 100 & 200 & 400 & \textbf{473} & 1211 \\
L1: \texttt{dp4a} direct, $c_q{\approx}10.3$ & 19 & 39 & 77 & 155 & 310 & \textbf{473} & \textbf{473} & 782 \\
Ozaki 2.5 on S (two-limb) & 23 & 45 & 91 & 182 & 364 & \textbf{473} & \textbf{473} & 666 \\
Ozaki 2.5 on A (all-byte) & 30 & 61 & 121 & 243 & \textbf{380} & \textbf{380} & \textbf{380} & 401 \\
Ozaki 2.5 on E (hybrid) & 29 & 59 & 118 & 235 & \textbf{438} & \textbf{438} & \textbf{438} & 476 \\
A at \texttt{dp4a} grain, $c_q{\approx}7.3$ & 22 & 44 & 88 & 176 & 352 & \textbf{380} & \textbf{380} & 553 \\
\emph{Switched (A/E/S by $\nbar$)} & \emph{30} & \emph{61} & \emph{121} & \emph{243} & \emph{438} & \emph{473} & \emph{473} & --- \\
\midrule
\multicolumn{9}{@{}l}{\emph{GB300 ($5$-\pflops{} \fp{8}; \inteight{} cap $166$~\tops{})}} \\
\addlinespace[1pt]
S0: shipping path, $c_q{=}16$ & 12 & 25 & 50 & 100 & \textbf{135} & \textbf{135} & \textbf{135} & 346 \\
L1: \texttt{dp4a} direct, $c_q{\approx}10.3$ & 19 & 39 & 77 & \textbf{135} & \textbf{135} & \textbf{135} & \textbf{135} & 223 \\
Ozaki 2.5 on S (two-limb) & 15 & 30 & 60 & 121 & \textbf{135} & \textbf{135} & \textbf{135} & 287 \\
Ozaki 2.5 on A (all-byte) & 24 & 47 & 95 & \textbf{109} & \textbf{109} & \textbf{109} & \textbf{109} & 147 \\
Ozaki 2.5 on E (hybrid) & 20 & 39 & 78 & \textbf{125} & \textbf{125} & \textbf{125} & \textbf{125} & 205 \\
A at \texttt{dp4a} grain, $c_q{\approx}7.3$ & 22 & 44 & 88 & \textbf{109} & \textbf{109} & \textbf{109} & \textbf{109} & 158 \\
\emph{Switched (A/E/L1/S by $\nbar$)} & \emph{24} & \emph{47} & \emph{95} & \emph{135} & \emph{135} & \emph{135} & \emph{135} & --- \\
 
\bottomrule
\end{tabular}
\\[2pt]
{\footnotesize $^{\dagger}$\emph{Clipping status.} Rates are tabulated for
\emph{square} outputs, for which $\nbar{=}$edge and single-cluster
$\Leftrightarrow$ edge${\le}256$. Columns $\nbar{\le}256$ thus fit one cluster
and are \textbf{achieved today} (throughout, in the \S\ref{sec:intro} sense:
modeled on shipping silicon with no hardware change, not measured), irrespective
of clipping (a non-square shape
at the same $\nbar$ is instead governed by the size-weighted
$\lambda_{\mathrm{eff}}$ of \S\ref{sec:codesign}). Columns $\nbar{>}256$ are
\emph{convert-once envelopes}: achieved today for tall/skinny or multi-panel
shapes, whose $\lambda_{\mathrm{eff}}$ stays $O(1)$ (${\approx}1$--$1.5$---the
real-application regime of Table~\ref{tab:apps}, where they are realised), but
for a large \emph{square} output ($\lambda_{\mathrm{eff}}{=}$edge$/256$) they
require the co-design of \S\ref{sec:hw} and are otherwise \emph{clipped} to the
deconstruction-$\lambda$ floor (${\approx}235$~TFLOPS on Rubin,
Table~\ref{tab:algs}). GB300 is the exception: its floor equals its
$135$-TFLOP roof, so its $\nbar{>}256$ cells are roof-bound (achieved today),
not clipping-limited. Every Rubin rate in this paper is thus either
\emph{fundamental} (achieved today) or \emph{clipping-limited} (a convert-once
envelope that needs the hardware of \S\ref{sec:hw} for large square DGEMM); the
two are distinguished wherever a number appears.}
\end{table}

\emph{(i) The announced operating region---and where the gains need no
hardware change.}  The $\nbar{\le}256$ column is achieved today,
single-cluster, with no hardware change: the switched dispatch reaches
$243$~TFLOPS at $\nbar{=}256$ (${\approx}2.4\times$ the shipping path). Above
$256$, the reduced model projects $\nbar{=}512$ rising from ${\approx}200$ to
$364$ (two-limb on the published set), $380$ (all-byte), or $438$~TFLOPS
(hybrid/switched)---a conditional $1.8$--$2.2\times$. These larger-$\nbar$
figures are \emph{convert-once envelopes}: they are realised \emph{today} for
the tall/skinny and small-batch shapes of real applications (Table~\ref{tab:apps}),
where the switched dispatch is worth ${\approx}1.6$--$1.9\times$
(${\approx}2\times$ on the block-Krylov rows) over simple deconstruction with no
hardware---but for a large \emph{square} output they require the co-design of
\S\ref{sec:hw} and are otherwise clipped to the ${\approx}235$-TF floor. The
practical reading is therefore \emph{positive}: on Rubin, Ozaki~2.5 is
already useful---near the crossover, only lightly clipped ($0.89$--$0.94\times$,
not the $0.50$ floor)---for the block-Krylov, batched, and panel work that
dominates real solvers; only the large-square-DGEMM headline waits on
hardware. These are projections at
$\eta_{\fp{8}}{=}1$; every loss the roof ignores subtracts from both
numerator and baseline in ways only measurement can apportion.  On
GB300 the switched dispatch runs tensor A in the block-width band
(its \texttt{dp4a} realisation within ${\approx}7\%$), E and L1
through the middle, and the two-limb route on the roof (from
$\nbar \approx 290$): the $135$-TFLOPS roof is reached from
$\nbar \approx 220$, with $95$ modeled at $\nbar{=}128$
($1.9\times$ the shipping path there)---against GB300's
$1.4$-TFLOPS native reference, a ${\approx}17$--$96\times$ band
throughout the evaluated range $\nbar \ge 32$, which is the Part~1
argument in its starkest form.
\emph{(ii) One-shot large DGEMM.}  This is precisely the regime the
deconstruction-$\lambda$ floor governs (\S\ref{sec:hw}): a single large
square output is multi-cluster, so the convert-once crossover of
Eq.~\eqref{eq:pdense}---roof approached above $\nbar{\approx}1200$ for the
shipping path---is realised only as an \emph{upper envelope}, and the
achieved rate plateaus at ${\approx}0.50$ of the $473$-TFLOPS roof (Rubin), not at the
roof. That convert-once reading holds only inside a single cluster or under
materialisation; for the deployable on-the-fly schedule the binding term is
the re-split floor, and lifting it to the roof (larger cluster, TMEM tile,
or L2 bandwidth) is the co-design question of \S\ref{sec:hw}.
\emph{(iii) Structurally sub-crossover workloads} are where
the routes change the picture qualitatively; they are treated in
\S\ref{sec:apps}.

\paragraph{Tensor-service concurrency: the two-endpoint bracket.}
The tensor-migrated routes charge their reduction GEMMs to the
\inteight{} tensor rate as if it were a concurrent capacity beside
near-peak \fp{8} work; as \S\ref{sec:ozaki25} notes, that
independence is not a proven property, and the honest statement is a
bracket between two analysable endpoints.  Under full independence
the tensor-migrated routes deliver the Table~\ref{tab:codesign}
values.  Under complete serialisation on a shared tensor service the
delivered rate is the harmonic composition
$P_{\text{ser}} = (1/P_{F} + 1/P_{I})^{-1}$ of the \fp{8}-side and
\inteight{}-side rates: at $\nbar{=}512$ on Rubin this gives
${\approx}206$ (S), ${\approx}228$ (A), and ${\approx}227$~(E)
TFLOPS.  The dispatch's floor, however, is
\emph{\inteight{}-tensor-contention-free}:
the pure-\texttt{dp4a} routes use no tensor \inteight{} capacity at
all, so even a fully serialised tensor service leaves
A-at-\texttt{dp4a} at ${\approx}352$ and L1 direct at
${\approx}310$~TFLOPS at $\nbar{=}512$ on Rubin.  The honest
software-route bracket at $\nbar{=}512$ is therefore $[352, 438]$
for the switched dispatch and $[310, 364]$ for the published set
(\texttt{dp4a} versus two-limb), degrading the conditional uplift
over the published $200$ from ${\approx}2.2\times$ to a floor of
${\approx}1.76\times$; at $\nbar{=}256$ the bracket is $[176, 243]$.
On GB300 the \texttt{dp4a} route reaches the full $135$-TFLOPS
ceiling from $\nbar \approx 223$, so the GB300 conclusions are
\inteight{}-tensor-contention-free.  Between the endpoints we
interpolate as
$T_{\text{tensor}} = \max(T_F, T_I) + \theta\, \min(T_F, T_I)$,
$\theta \in [0,1]$ ($\theta{=}0$ independent, $\theta{=}1$
serialised); measuring $\theta(\text{shape}, \text{occupancy})$
joins the validation plan of \S\ref{sec:validation}.

\paragraph{SIMT--\fp{8} overlap: the second serialisation axis.}
Avoiding \inteight{}-tensor contention is not the whole story: the
\texttt{dp4a} routes still assume their SIMT conversion work
overlaps the \fp{8} MMA stream.  We parameterise that overlap by
$\rho_{\text{simt,fp8}} \in [0,1]$ ($0$ ideal overlap, $1$ complete
serialisation on the shared issue path).  At complete serialisation
the Rubin $\nbar{=}512$ endpoints are the harmonic compositions
$(1/352 + 1/380)^{-1} \approx 183$~TFLOPS for A-\texttt{dp4a} and
$(1/310 + 1/473)^{-1} \approx 187$ for L1.  Read those numbers as an \emph{equal-overlap-degradation
sensitivity}, not an unconditional floor---it conditions on both
paths sharing one overlap coefficient---and keep its three
ingredients separate: \emph{(i)} the route endpoints (ideal
$352/310$, serialised $183/187$ at $\nbar{=}512$); \emph{(ii)} the
announced $200$, whose size and algorithm provenance are unknown; and
\emph{(iii)} the matched-degradation ratio.  For (iii): complete
SIMT--tensor serialisation collapses the shipping baseline too,
because S0 embeds the same overlap assumption---the memo's
$P_{\text{int}}$ normalisation and the shipping path's own SIMT
conversion presume it---so the consistently serialised S0 is
$(1/200 + 1/473)^{-1} \approx 141$ at $\nbar{=}512$ and ${\approx}83$
at $256$, and the like-for-like ratio is ${\approx}1.30\times$
($512$) and $1.46\times$ ($256$).  Comparing serialised routes
against the announced $200$ mixes worlds and is not claimed.  Structurally, a tensor MMA occupies a shared issue
slot only once per many tensor-busy cycles, so the shared-issue
serialisation mechanism is weak---an argument, not a measurement,
and $\rho_{\text{simt,fp8}}$ joins $\theta$ in the paired-rate
measurements of \S\ref{sec:validation}.

\subsection{Application-Geometry Scenarios: Measured Call-Shape Traces}
\label{sec:apps}

The value of the switched dispatch is best seen against the workloads
that actually occupy the sub-crossover region, and against the
alternative they would otherwise fall back to: Rubin's \emph{native}
\fp{64} path, illustratively ${\approx}30$~TFLOPS of blocked DGEMM
(Figure~\ref{fig:knee}, flat line; to be measured).  The geometry
matters: rank-$n_b$ trailing updates have large $\nbar$ and were never
the problem; the structurally bounded class is where one \emph{output}
dimension is a block width---supernodal/frontal update panels in
SuperLU\_DIST- and MUMPS-class sparse direct
solvers~\cite{li2003superlu,mumps}, panel-internal factorisation
kernels, LOBPCG and block-Krylov eigensolvers~\cite{knyazev2001lobpcg}
(block widths $16$--$128$), and batched small GEMMs in tensor
contraction.  Under the switched dispatch the reduced model projects,
at $\nbar = 64/128/256/384$: $61/122/243/364$~TFLOPS on Rubin, versus
$25/50/100/150$ on the shipping path---a uniform
${\approx}2.4\times$ across the band, with the
emulation-versus-native crossover pushed from $\nbar \approx 80$
down to $\nbar \approx 32$ against the illustrative $30$-TF
baseline (${\approx}35$ against the official $33$-TFLOPS
specification~\cite{nvidia_hgx})---at or below typical block widths.
On GB300 the same band runs $47/95/135/135$ versus $25/50/100/135$
(${\approx}1.9\times$ at block widths $64$--$128$, converging at
the $135$-TFLOPS roof beyond the shipping knee of ${\approx}350$)---and with
native \fp{64} at ${\approx}1.4$~TFLOPS there is no crossover to
push within the evaluated range: every plotted route exceeds the
native reference for $\nbar \ge 32$, and the dispatch question is
only \emph{which} route.

\begin{figure}[t]
\centering
\includegraphics[width=\linewidth]{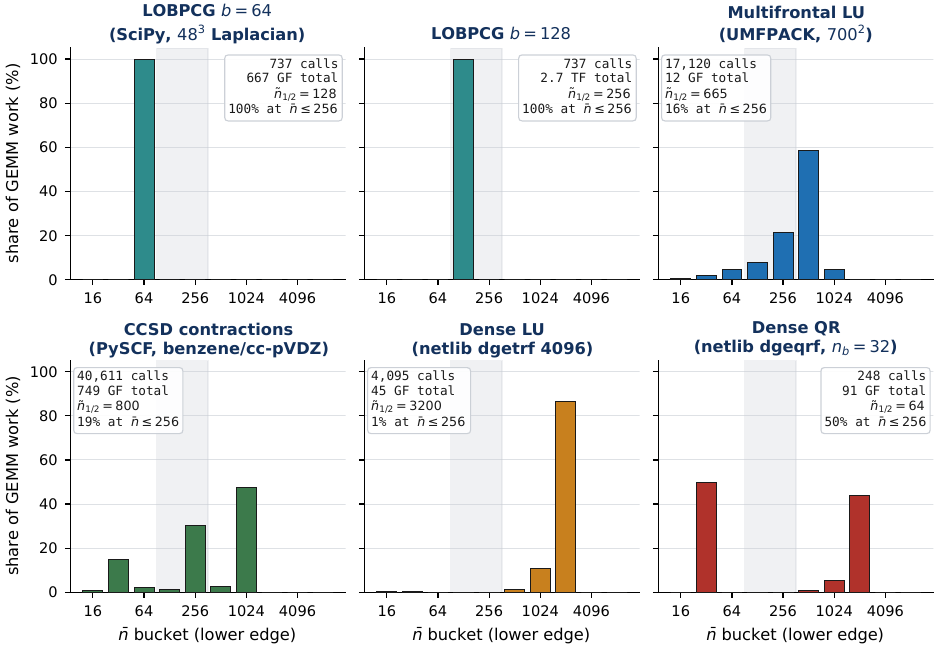}
\caption{Measured call-shape distributions (share of traced
GEMM/SYRK work per $\log_2 \nbar$ bucket) for the six instrumented
configurations, from the interposer logs (\texttt{traces/}, in the artifact).  The shaded band marks
$\nbar \in [128, 512]$.  LOBPCG and QR-panel work is
block-width-bounded; multifrontal fronts and CCSD contractions
straddle the crossover band; dense LU trailing updates sit far right.
Geometry is hardware-independent; only the projections built on it
are model-conditional.}
\label{fig:traces}
\end{figure}

\paragraph{Measured call-shape traces.}  Whole-application gains
depend on each code's mix of GEMM geometries.  Rather than assume
mixes, we \emph{measured} them: using an \texttt{LD\_PRELOAD}
interposer (\texttt{traces/blas\_shim.c}, supplied in the artifact with every runner script), we
captured the $(m, n, k)$ of each \texttt{dgemm}/\texttt{dsyrk} call
from \emph{six instrumented configurations across four
library/application classes}, all running unmodified library
implementations on author-constructed inputs---SciPy LOBPCG on a
$48^3$ Laplacian at block widths $64$ and $128$; multifrontal sparse
LU (UMFPACK) on a $700^2$ convection--diffusion operator;
coupled-cluster tensor contractions (PySCF CCSD on benzene in
cc-pVDZ, capped at six iterations); and netlib LAPACK blocked LU and QR at
$n{=}4096$.  The fixed captured traces measure geometry for the
stated software stack; replaying them through the route model
produces GEMM/SYRK-portion \emph{scenario projections}, not
application benchmarks or executions of an Ozaki-2.5 kernel
(Table~\ref{tab:apps}, Figure~\ref{fig:traces}; work-weighted
harmonic composition over Eq.~\eqref{eq:pdense}, the same machinery
as Table~\ref{tab:proj}).
Three measured facts stand out.  \emph{First}, LOBPCG at $b{=}64$
places $100\%$ of its GEMM work at $\nbar \le 256$ (median $128$:
the $2mn/(m{+}n) \to 2b$ geometry, measured), and netlib blocked QR
places half its BLAS-3 work at $\nbar \approx 64$---the
block-width-bounded class is real and load-bearing.  \emph{Second},
multifrontal fronts are fatter than commonly assumed: UMFPACK's
median GEMM sits at $\nbar \approx 670$ (quartiles $380$--$850$),
squarely in the crossover band of \S\ref{sec:puzzle}---real
sparse-direct work lives exactly where the deconstruction term
bites, which independently motivates the size sweep.  \emph{Third},
sequential supernodal SuperLU (SciPy's \texttt{splu}) issues
essentially no BLAS-3 at all (its panel updates are
\texttt{dgemv}-based), a reminder that the BLAS-3 supernodal
geometry this paper targets is the multifrontal/distributed class.
Amdahl fractions for non-GEMM work further dilute whole-run numbers,
and the mixes below are GEMM-portion composites only; the native
columns compare against flat illustrative references and should be
read as order-of-magnitude bands, not predictions.

\paragraph{Composite model conventions.}  The geometry of
Table~\ref{tab:apps} is measured; the composites are model
projections over it, computed as follows.  Each record is assigned
its eligible storage schedule (split-$k$ / fused-$\lambda$ /
L2-materialised / HBM-materialised planes at $2p$ bytes,
\S\ref{sec:storagemodes}) and its route rate is capped by the
record's memory roof at its \emph{additive} traffic
$Q_0 + Q_{\text{plane}}$ with $Q_0 = 8\{k(m{+}n)+mn\}$
(Eq.~\eqref{eq:modeMadd}); the reconstruction tax of
Eq.~\eqref{eq:recongamma} is applied on the dispatched host; and the
shipping baseline S0 is capped by the \emph{same} memory roof and
tax, so the comparison is like-for-like.  \textsc{dsyrk} records are
charged as one unique operand (the Gram factor $A$ of
$C = A A^{\mathsf T}$): triangular work $m\,n\,k$, a single converted
operand $n_{\text{uniq}} = mk$ in the conversion and plane ledgers,
and $\nbar = n_C$; the transposed operand view is a marshalling, not
a second conversion; the output $C$ is charged a single write under the
overwrite convention $\beta_{\text{BLAS}}{=}0$ that $Q_0$ uses
throughout (a nonzero $\beta_{\text{BLAS}}$ would add a read of the
triangular $C$, which the interposer does not currently record).  (These traces contain no \textsc{dsyrk}-tagged
records---the LOBPCG Gram products were interposed as GEMM---so this
convention is stated for completeness and leaves the present
composites unchanged.)  Quantiles and composite speedups are
FLOP-weighted, in the harmonic form
$P_{\text{comp}} = \big(\sum_i W_i\big)/\big(\sum_i W_i/P(\nbar_i)\big)$
with speedup $S = \big(\sum_i W_i/P^{\text{base}}_i\big)/\big(\sum_i
W_i/P^{\text{route}}_i\big)$; per-record predictions ship as
\texttt{traces/sched\_*.csv}.

\begin{table}[!ht]
\caption{Measured call-shape traces and their model-projected
GEMM-portion composites under the switched dispatch (both platforms).
Geometry is \emph{measured} (BLAS-3 interposer; hardware-independent);
the composites are \emph{model projections} over that geometry under
the per-record schedule-, reconstruction-, and memory-aware model
detailed in the preceding paragraph.  ``work $\nbar{\le}256$'' =
fraction of traced GEMM flops at $\nbar \le 256$.  Claim status:
modeled projection over measured geometry.  Composites are reported at
the conservative SIMT-Garner reconstruction cost floor; the ${\approx}r$
codesign target of \S\ref{sec:recon} would raise the small-$k$ Rubin
native multipliers as noted.  The final three columns are the
\emph{counterfactual} of \S\ref{sec:hw}: the same per-record
schedule, and the same memory roof and plane-feed caps, but with the
deconstruction floor lifted to the compute roof
($473$~TF on Rubin, $135$~TF on GB300).  It is a ceiling bought by
hardware that does not exist, not a delivered rate.  Its two multipliers
are that ceiling over the shipping baseline and over the native
\fp{64} unit---the same pair as the delivered columns to their left, so
the two cases can be read against one another rather than only the
co-design case against today's software; \emph{delivered} is the
composite rate so obtained and the fraction of the roof it represents.
Because both pairs are ratios of the same two rates to the same two
references, each pair carries the identical headroom factor, which is
the quantity the next sentence reports.
\emph{No row reaches $100\%$}---once deconstruction is free the traced
geometry is bandwidth-bound, so the \emph{additional} headroom the
co-design buys over Ozaki~2.5 as modeled today is
${\approx}2.5$--$3.3\times$ on Rubin but only ${\approx}1.1$--$1.5\times$
on GB300, whose memory roof is much nearer its compute roof.}
\label{tab:apps}
\centering
\footnotesize
\setlength{\tabcolsep}{2pt}
%
\begin{tabular}{>{\raggedright\arraybackslash\hspace{0pt}}p{2.95cm}ccccccc>{\raggedright\arraybackslash\hspace{0pt}}p{1.55cm}}
\toprule
& & & \multicolumn{2}{c}{as modeled today} & \multicolumn{3}{c}{\emph{if} co-design reaches roof} & \\
\cmidrule(lr){4-5}\cmidrule(lr){6-8}
workload (traced config.) &
\begin{tabular}[b]{@{}c@{}}med.\ $\nbar$\\(q25--q75)\end{tabular} &
\begin{tabular}[b]{@{}c@{}}work\\$\nbar{\le}256$\end{tabular} &
vs ship & vs native & vs ship & vs native &
\begin{tabular}[b]{@{}c@{}}delivered\\(\% of roof)\end{tabular} &
dominated by \\
\midrule
\multicolumn{9}{@{}l}{\emph{Rubin (native ${\approx}30$~TF illustrative; roof $473$~TF)}} \\
\addlinespace[1pt]
LOBPCG $b{=}64$ ($48^3$ Laplacian) & 128 (64--128) & 100\% & ${\approx}1.94\times$ & ${\approx}2\times$ & ${\approx}6.00\times$ & ${\approx}6\times$ & {\scriptsize 176~TF (37\%)} & block width \\
LOBPCG $b{=}128$ (same operator) & 256 (128--256) & 100\% & ${\approx}1.83\times$ & ${\approx}4\times$ & ${\approx}6.00\times$ & ${\approx}12\times$ & {\scriptsize 352~TF (74\%)} & block width \\
Multifrontal LU (UMFPACK, $700^2$) & 665 (383--850) & 16\% & ${\approx}1.19\times$ & ${\approx}1\times$ & ${\approx}3.80\times$ & ${\approx}4\times$ & {\scriptsize 132~TF (28\%)} & frontal panels \\
CCSD contractions (benzene/cc-pVDZ) & 800 (353--1302) & 19\% & ${\approx}1.69\times$ & ${\approx}4\times$ & ${\approx}4.30\times$ & ${\approx}10\times$ & {\scriptsize 288~TF (61\%)} & contractions \\
Dense LU (netlib, $n{=}4096$) & 3200 (2560--3712) & 1\% & ${\approx}1.14\times$ & ${\approx}3\times$ & ${\approx}3.68\times$ & ${\approx}10\times$ & {\scriptsize 299~TF (63\%)} & trailing updates \\
Dense QR (netlib, $n_b{=}32$) & 64 (63--3232) & 50\% & ${\approx}1.60\times$ & ${\approx}2\times$ & ${\approx}5.19\times$ & ${\approx}6\times$ & {\scriptsize 172~TF (36\%)} & panel factors \\
\midrule
\multicolumn{9}{@{}l}{\emph{GB300 (native ${\approx}1.4$~TF; roof $135$~TF)}} \\
\addlinespace[1pt]
LOBPCG $b{=}64$ ($48^3$ Laplacian) & 128 (64--128) & 100\% & ${\approx}1.67\times$ & ${\approx}35\times$ & ${\approx}2.18\times$ & ${\approx}46\times$ & {\scriptsize 64~TF (47\%)} & block width \\
LOBPCG $b{=}128$ (same operator) & 256 (128--256) & 100\% & ${\approx}1.59\times$ & ${\approx}67\times$ & ${\approx}2.18\times$ & ${\approx}91\times$ & {\scriptsize 128~TF (95\%)} & block width \\
Multifrontal LU (UMFPACK, $700^2$) & 665 (383--850) & 16\% & ${\approx}1.19\times$ & ${\approx}29\times$ & ${\approx}1.38\times$ & ${\approx}34\times$ & {\scriptsize 48~TF (36\%)} & frontal panels \\
CCSD contractions (benzene/cc-pVDZ) & 800 (353--1302) & 19\% & ${\approx}1.41\times$ & ${\approx}58\times$ & ${\approx}1.62\times$ & ${\approx}66\times$ & {\scriptsize 93~TF (68\%)} & contractions \\
Dense LU (netlib, $n{=}4096$) & 3200 (2560--3712) & 1\% & ${\approx}0.87\times$ & ${\approx}50\times$ & ${\approx}1.34\times$ & ${\approx}78\times$ & {\scriptsize 109~TF (80\%)} & trailing updates \\
Dense QR (netlib, $n_b{=}32$) & 64 (63--3232) & 50\% & ${\approx}1.41\times$ & ${\approx}33\times$ & ${\approx}1.89\times$ & ${\approx}45\times$ & {\scriptsize 63~TF (46\%)} & panel factors \\
 
\bottomrule
\end{tabular}
\end{table}

The reading is not that every solver triples, but that the measured
geometry confirms the structural claim: the workloads for which
emulated \fp{64} was least usable---LOBPCG and QR panels, whose
traced work is block-width-bounded---are exactly where the
software-only switched dispatch is modeled to be worth multiples, on
current GPUs, with no hardware change; and the traced multifrontal
fronts populate the crossover band itself.  Relative to the earlier
draft, the schedule/memory-roof filter and the on-the-fly
deconstruction-$\lambda$ correction together move several composites
materially---on Rubin, multifrontal $2.07 \to 1.19\times$ and QR
$2.27 \to 1.60\times$; on GB300, LOBPCG $1.90/1.64 \to
1.67/1.59\times$, multifrontal $\to 1.19\times$, and dense LU $\to
0.87\times$ (below unity: at $\bar n{=}3200$ the rank-$64$ trailing
update is bandwidth-bound and the residue reconstruction makes emulation
marginally slower than the shipping baseline there)---because the
multifrontal small-front tail and the GB300 tall-skinny block shapes
(operational intensity ${\approx}8$) are bandwidth-bound, where the
emulated and shipping paths nearly coincide at the same memory roof; the
GB300 LOBPCG composites are capped by exactly that roof.  The GB300
block sharpens
the point from the other side: per-route gains are more modest (the
roof is nearer), but with a ${\approx}1.4$-TFLOPS native reference
every traced workload is an order-of-magnitude argument for emulating
at all---led, below the roof, by pure-SIMT routes that any CUDA
kernel can implement today.

The final three columns answer, per workload, the question the co-design of
\S\ref{sec:hw} raises: \emph{if} the deconstruction floor were lifted
all the way to the compute roof ($473$~TFLOPS on Rubin, $135$ on
GB300), what would each of these workloads
then get---against the shipping baseline, and against the native
\fp{64} unit a user would otherwise run on?  The answer is bounded well
short of the roof---$28$--$74\%$
of it on Rubin, $36$--$95\%$ on GB300---because once deconstruction is
free the binding constraint on this measured geometry is the memory
roof, not the arithmetic one.  The \emph{additional} headroom the
co-design buys over Ozaki~2.5 as modeled here is therefore
${\approx}2.5$--$3.3\times$ on Rubin, where the two roofs are far
apart, and only ${\approx}1.1$--$1.5\times$ on GB300, where they are
not.  Two things follow.  The co-design case is a Rubin-class
argument, not a general one; and even granting the co-design in full,
these traced applications would remain bandwidth-limited, which is the
honest reason we book the deconstruction floor as the \emph{first}
target rather than the only one.  These are scenario projections over
measured geometry, not application benchmarks; replaying the to-be-released
traces through measured kernels is step (vi) of
\S\ref{sec:validation}.

\subsection{Storage Modes: Where the Planes Live, and What Each Mode
Charges}
\label{sec:storagemodes}

The projections above lean on two disciplines at once---O1's
convert-once accounting and a fused traffic model in
which the residue planes generate no HBM traffic of their
own---and these are not simultaneously free: a plane that is never
written to memory may have to be converted more than once, and a
plane converted exactly once must live somewhere.  We make the
coupling explicit.  For an $m{\times}k$ by $k{\times}n$ product with
$n_{\text{uniq}} = k(m{+}n)$ unique operand elements and $p$ stored
plane bytes per scalar ($30/44/33$ for S/A/E), the storage modes
are: \emph{mode F (fused)}---the planes live in
the SMEM of the owning CTA or thread-block cluster, conversion
multiplicity $\lambda \ge 1$ (a tile shared by several clusters may
be converted by each), no plane HBM traffic; \emph{mode M
(transient materialisation)}---the planes are staged through a
workspace, L2-resident or in HBM, and converted exactly once; and
\emph{mode P (persistent)}---the planes are retained across calls
(operand-stationary O1), each call still \emph{reading} $p$ bytes
per touched scalar.

The rest of this section derives, mode by mode, the bound each of
these charges.  Because those bounds are not evaluated in isolation
but as a per-call \emph{selection}---the engine tries every eligible
schedule on every route and keeps the best---we state that selection
once, as Algorithm~\ref{alg:schedule}, before deriving its
ingredients.  It is the exact content of \texttt{rate\_rec} in
\texttt{params.py}: every projection in this paper, every composite in
Table~\ref{tab:apps}, and every per-record tag in
\texttt{traces/sched\_*.csv} is an evaluation of it.  That last
sentence is a claim about the correspondence between a printed
algorithm and a program, which is exactly the kind of claim this
paper has already had to retract once, so it is a test and not a
promise: \texttt{verify/alg4\_fidelity.py} transcribes the algorithm
\emph{as printed}, independently of the engine, and requires the two
to agree on both the winning rate and the winning tag at every point
of a grid that includes the tall/skinny corner where the ledger
omissions below used to hide.  Reading the algorithm alongside the
derivations that follow is how those omissions became visible in the
first place.

\begin{algorithm}[tp]
\caption{\textsc{Select-Schedule}$(\pi, \nu, m, n, k)$ --- the
per-call storage-mode and route selection. $\pi$ is the platform,
$\nu$ a route of the menu (\S\ref{sec:codesign}). Every schedule is
capped by its own compute rate \emph{and} by its additive memory
traffic $Q_0 + Q_{\text{plane}}$ \emph{and} by any bandwidth it must
be fed at; the best surviving rate wins. The envelope quoted
throughout the paper is $\max_\nu$ of this function.}
\label{alg:schedule}
\begin{algorithmic}[1]
\State \textbf{input:} platform $\pi$, route $\nu$; shape $m,n,k$;
   set $\mathcal{S}(\nu)$ with $p$ stored plane bytes/scalar and
   $r$ moduli; $B = B_{\text{mem}}(\pi)$,
   $B_{\mathrm{L2}} = 4B$, $C_{\mathrm{L2}} = 126$~MB,
   $\text{TILE} = 64$, $C = 16$ CTAs/cluster, $R_\ast = 256$,
   $K_{\text{slab}} = 4096$
\State $W \gets 2mnk$;\quad
   $Q_0 \gets 8\{k(m{+}n) + mn\}$;\quad
   $n_{\text{uniq}} \gets k(m{+}n)$
   \Comment{\textsc{syrk}: $W{\gets}mnk$, one operand, $mn$ terms halved}
\State $S_k \gets \lceil k/K_{\text{slab}} \rceil$;\quad
   $Q_{\text{red}} \gets 8\,r\,mn\,(S_k{-}1)$
   \Comment{split-$k$ residue partials, written and read}
\State $T_m, T_n \gets \lceil m/\text{TILE} \rceil,
   \lceil n/\text{TILE} \rceil$;\quad
   $\mathcal{P} \gets \emptyset$
   \Comment{$\mathcal{P}$: (rate, tag) candidates}
\Statex
\If{$T_m T_n \le C$}
   \Comment{\textbf{mode F, one cluster}: DSM-shared, $\lambda = 1$}
   \State $\rho \gets \textsc{ComputeRate}(\nu, \lambda{=}1)$
   \If{$S_k = 1$}
      \Comment{$Q_{\text{red}}{=}0$: one slab, no partials}
      \State add $(\min\{\rho,\ BW/Q_0\},\
         \textsf{grid-fused-nosplit})$ to $\mathcal{P}$
   \ElsIf{$4\,r\,mn > C_{\mathrm{L2}}$}
      \Comment{partials spill: charge them to HBM}
      \State add $(\min\{\rho,\ BW/(Q_0{+}Q_{\text{red}})\},\
         \textsf{splitk})$ to $\mathcal{P}$
   \Else
      \Comment{partials stay resident: charge them to L2}
      \State add $(\min\{\rho,\ BW/Q_0,\
         B_{\mathrm{L2}}W/Q_{\text{red}}\},\ \textsf{splitk})$
         to $\mathcal{P}$
   \EndIf
\Else
   \Comment{\textbf{multi-cluster}: mode F pays $\lambda$, mode M pays traffic}
   \State $\lambda_A, \lambda_B \gets
      \lceil n/R_\ast \rceil, \lceil m/R_\ast \rceil$;\quad
      $\lambda_{\text{eff}} \gets
      (m\lambda_A + n\lambda_B)/(m{+}n)$
      \Comment{size-weighted multiplicity; $R_\ast$ from Eq.~\eqref{eq:reach}}
   \State $\rho_\lambda \gets
      \textsc{ComputeRate}(\nu, \lambda_{\text{eff}})$;\quad
      $\rho_1 \gets \textsc{ComputeRate}(\nu, 1)$
   \State $Q_{\text{re}} \gets 8\{mk(\lambda_A{-}1) + kn(\lambda_B{-}1)\}$
      \Comment{mode F: \fp{64} operand re-reads, \emph{not} planes}
   \State add $(\min\{\rho_\lambda,\
      BW/(Q_0{+}Q_{\text{red}}{+}Q_{\text{re}})\},\
      \textsf{on-the-fly})$ to $\mathcal{P}$
   \If{$p\,k\min(m,n) \le C_{\mathrm{L2}}$}
      \Comment{hybrid: small operand in L2, large on the fly}
      \State add $(\min\{\rho_\lambda,\
         BW/(Q_0{+}Q_{\text{red}}),\
         B_{\mathrm{L2}}/\Lambda(p\,k\min(m,n),\,
         \tfrac{p}{2\,\text{TILE}})\},\
         \textsf{hybrid-L2})$ to $\mathcal{P}$
   \EndIf
   \If{$p\, n_{\text{uniq}} \le C_{\mathrm{L2}}$}
      \Comment{\textbf{mode M in L2}, Eq.~\eqref{eq:l2cap}: no plane HBM traffic}
      \State add $(\min\{\rho_1,\
         BW/(Q_0{+}Q_{\text{red}}),\
         B_{\mathrm{L2}}/\Lambda(p\,n_{\text{uniq}},\,
         \tfrac{p}{\text{TILE}})\},\
         \textsf{mat-L2})$ to $\mathcal{P}$
   \Else
      \Comment{\textbf{mode M in HBM}: planes cross the \emph{same} bus}
      \State $Q_{\text{plane}} \gets Q_{\text{red}} +
         2p\, n_{\text{uniq}}$
         \Comment{$c = 2p$, Eq.~\eqref{eq:modeMadd}: write \emph{and} read}
      \State add $(\min\{\rho_1,\
         BW/(Q_0{+}Q_{\text{plane}}),\
         B\,\text{TILE}/p\},\
         \textsf{mat-HBM})$ to $\mathcal{P}$
         \Comment{both terms: the feed cap alone is
         looser below $\nbar{=}128$}
   \EndIf
\EndIf
\State \Return $\max \mathcal{P}$, with the maximising tag
   \Comment{the tag emitted to \texttt{traces/sched\_*.csv}}
\Statex
\Function{$\Lambda$}{plane bytes $Z$, feed $\phi$}
   \Comment{L2 bytes/useful FLOP, L2-resident workspace}
   \State \Return $Z/W + \max\{\phi,\ Z/W\}$
      \Comment{write once, read ${\ge}$ once: $\phi$ is under
      one pass below $\nbar{=}64$}
\EndFunction
\Statex
\Function{ComputeRate}{route $\nu$, multiplicity $\lambda$}
   \State $P_{\text{roof}} \gets P_{\fp{8}}/\alpha$;\quad
      $\nbar \gets 2mn/(m{+}n)$
      \Comment{roofs tabulated in Table~\ref{tab:codesign}}
   \State \textbf{if} co-design roof assumed \textbf{then return}
      $P_{\text{roof}}$
      \Comment{overlay: co-design column of Table~\ref{tab:apps}}
   \State $\sigma \gets \lambda\, c_q^{\text{res}}\, r/\nbar +
      N_\gamma(r)/2k$
      \Comment{SIMT ops/FLOP: deconstruction residual $+$ Garner}
   \State $\delta \gets \lambda\, a_{\inteight}/\nbar$
      \Comment{\inteight{} MACs/FLOP; $a_{\inteight}, c_q^{\text{res}}$
      from Table~\ref{tab:codesign}}
   \State \Return $\min\{P_{\text{roof}},\ I_{\text{SIMT}}/\sigma,\
      (P_{\text{I8TC}}/2)/\delta\}$
      \Comment{Eq.~\eqref{eq:tensorknee}, Eq.~\eqref{eq:decfloor};
      an unloaded pipe contributes $\infty$}
\EndFunction
\end{algorithmic}
\end{algorithm}

\paragraph{Mode M through HBM: the plane-only obstruction bound.}
Transient materialisation through an HBM workspace mandates the plane
\emph{write} and at least one \emph{read}: the plane traffic is
$Q_{\text{plane}} = 2p\, n_{\text{uniq}}$ bytes, so
\begin{equation}
P_{\text{useful}} \;\le\; \frac{B_{\text{mem}}\, \nbar}{2p},
\qquad \nbar = \frac{2mn}{m+n},
\label{eq:modeM}
\end{equation}
against $2mnk$ useful flops.  (An earlier draft charged only the
read; the write is not optional, and the crossovers below are the
corrected values, doubled accordingly.)  The bound stops binding
above the plane-only crossover
$\nbar_{\text{mat}} = 2p \cdot (\text{set roof})/B_{\text{mem}}$: on
Rubin ($22$~TB/s) $1290/1522/1312$ for
S/A/E, on GB300 ($8$~TB/s) $1014/1196/1031$.  This is a
\emph{necessary} obstruction bound only: below $\nbar_{\text{mat}}$
HBM materialisation cannot support the set roof whatever the compute
engines do, but above it the plane-only obstruction merely
\emph{ceases}---roof attainment is a stronger, additive condition
stated next.

\paragraph{The additive full-traffic bound and the roof-attainment
crossover.}  The obstruction bound above isolates the plane term; the
schedule must in fact move the plane traffic \emph{and} the call's own
operand/output traffic across the \emph{same} bus.  Charging both
additively---plane bytes $Q_{\text{plane}}$ plus
$Q_0 = 8\{k(m{+}n) + mn\}$---the memory roof is
\begin{equation}
P_{\text{useful}} \;\le\; \frac{B_{\text{mem}}}
{\dfrac{8 + c}{\nbar} + \dfrac{4}{k}},
\qquad
c = \begin{cases} 2p & \text{both operands materialised},\\
p & \text{one operand (fused-hybrid)},\end{cases}
\label{eq:modeMadd}
\end{equation}
and roof attainment requires $\nbar$ above the \emph{additive}
crossover $\nbar_{\text{add}} = (8 + c + 4\,\mathbb{1}_{\text{sq}})\,
P_{\text{roof}}/B_{\text{mem}}$ (with $P_{\text{roof}}$ the set roof
and $\mathbb{1}_{\text{sq}}{=}1$ for square shapes).  For full
materialisation ($c=2p$)
this is $1462/1695/1472$ (large-$k$) and $1548/1764/1551$ (square)
on Rubin, and $1149/1332/1156$ and $1216/1386/1219$ on GB300, for
S/A/E.  The fused-hybrid one-operand crossovers ($c=p$) are lower, at
$817/917/815$ (Rubin) and $642/720/641$ (GB300), which is why the
switched dispatch attains the roof earlier than the full-materialised
threshold would suggest.  These one-operand values use $c=p$, hence
the additive term $(8{+}p)/\nbar$, which is the \emph{square} case;
for a general shape the additive memory cost per useful FLOP is
\[
\frac{Q_0 + 2p\,k\,s}{2mnk}
\;=\; \frac{8}{\nbar} + \frac{4}{k} + \frac{p}{\ell},
\qquad s = \min(m,n),\quad \ell = \max(m,n)
\]
(the per-record engine uses the exact $2p\,k\min(m,n)$ plane term),
reducing to $(8{+}p)/\nbar$ only when $m = n$.
We report both: $\nbar_{\text{mat}}$
(plane-only, necessary) and $\nbar_{\text{add}}$ (additive,
roof-attainment).

\paragraph{A model defect this exposition exposed, and its repair.}
Writing the additive bound down carefully made it possible to audit
the engine against it, and the audit failed.  Earlier revisions of
this paper asserted that the per-record engine had \emph{always}
capped each materialised mode by its additive traffic
$B_{\text{mem}}W/(Q_0{+}Q_{\text{plane}})$, and that an audit of the
schedule tags showed \emph{no} record taking an HBM-materialised
branch.  Both statements were false, and we correct them here rather
than quietly restate them.  The HBM-materialisation branch charged
only the $k$-slab reduction traffic into its additive roof, together
with a plane-\emph{feed} cap of $p/\text{TILE}$ bytes per useful FLOP;
it never charged the $c = 2p$ write-plus-read of
Eq.~\eqref{eq:modeMadd} at all.  Those two terms cross at
$\nbar = 2\,\text{TILE} = 128$, so on square shapes the feed cap is
the binding one and the omission is invisible---which is why it
survived---but on tall-and-skinny shapes it is not.  In the released
trace set, $276$ CCSD records (all Rubin, all route~$E_t$, all at
$\nbar \approx 42$, $6.4\%$ of that application's FLOPs) took the
branch, and the worst of them,
$(m,n,k) = (1953, 21, 4325)$, was credited $38.2$~TF where
Eq.~\eqref{eq:modeMadd} permits $12.3$~TF---a $3.1\times$ escape from
this paper's own obstruction bound.  A second instance of the same
omission class sat in the L2-resident branch: \S\ref{sec:storagemodes}
says the planes there are ``written and read at cache rates'', but the
engine charged the streaming feed alone, which for $\nbar < 64$ is
\emph{less than a single pass} over the workspace---$6167$ of $8121$
L2-resident records.  The engine now charges $c = 2p$ additively on
the HBM branch (in addition to, not instead of, the feed cap) and
write-plus-at-least-one-pass on the L2 branch.

The repair is contained and moves every affected number the
conservative way.  All $276$ HBM-materialised records fall back to the
on-the-fly schedule, so the tag histogram of
\texttt{traces/\allowbreak sched\_*.csv} now genuinely contains no
HBM-materialised record---a property that is asserted by test rather
than by prose (\texttt{verify/paper\_numbers.py}), precisely because
asserting it by prose is what failed here.  A further $68$ records
move from L2-resident materialisation to on-the-fly.  Exactly one
composite in Table~\ref{tab:apps} changes: CCSD on Rubin falls from
${\approx}1.73\times$ to ${\approx}1.69\times$ over shipping
($116 \to 114$~TF delivered, co-design headroom
${\times}2.49 \to {\times}2.54$).  Every other application, both
platforms, the projection and co-design tables, and the whole
co-design column of Table~\ref{tab:apps} are unchanged to the printed
precision.

\paragraph{Mode M in L2: the capacity predicate.}  The workspace
need not touch HBM at all.  Whenever the whole plane workspace fits
in L2,
\begin{equation}
p\, k\,(m{+}n) \;\le\; C_{\mathrm{L2}},
\label{eq:l2cap}
\end{equation}
the planes are written and read at cache rates and generate \emph{no}
plane HBM traffic whatever.  We assume $C_{\mathrm{L2}} = 126$~MB---a
stated, checkable parameter, not a measured one---under which the
square-problem reach is $k \le 1449/1196/1381$ for S/A/E; L2-resident
materialisation thus covers precisely the moderate-$k$ region in
which the HBM bound of Eq.~\eqref{eq:modeM} would otherwise bind.
This ``no HBM plane traffic'' claim is an \emph{L2-capacity-eligible}
hypothesis, testable but unmeasured: it presumes usable (not merely
nominal) capacity net of competing footprints, a producer--consumer
lifetime that keeps the planes resident between write and read, and a
write-back policy that does not spill them.  As a safety-factor
sensitivity, halving the effective capacity to
$C_{\text{eff}} = \tfrac12 C_{\mathrm{L2}}$ lowers the S-set square
reach from $1449$ to $1024$, keeping the L2-resident materialisation
predicate honest under a $2\times$ residency haircut (the single-cluster
grid-fused reach is the separate, smaller $T_mT_n\le C$ bound of
\S\ref{sec:storagemodes}, ${\approx}256$ square).

\paragraph{Grid-tiled split-$k$ (single-cluster, unmeasured).}  When
the output fits \emph{one} thread-block cluster---$T_m T_n \le C$
output tiles, with $T_m = \lceil m/64\rceil$, $T_n = \lceil n/64\rceil$
and $C$ the CTAs per cluster ($8$ portable, $16$ opt-in on
Blackwell), i.e.\ roughly $\max(m,n) \lesssim 256$ for square
outputs---the cluster's CTAs share each $k$-slab's converted $A$/$B$
residue planes through distributed shared memory (DSM): every operand
element is converted exactly once ($\lambda = 1$), the sharing stays
on-chip, and there is \emph{no} HBM or L2 plane storage at all.  This
covers Gram-type products ($C = A^{\mathsf T}\!A$) and the small-front
shapes of the traces.  Its ledger is a \emph{proposed,
resource-eligible, unmeasured} schedule: with $S_k$ $k$-slabs each held
by a CTA, the per-slab partials are kept in the residue domain (exact,
no rounding on reduction) and combined through an L2-resident reduction
tree.  \emph{Larger} outputs span multiple clusters, which cannot share
converted planes across clusters without either repeated conversion or
an L2-materialised plane workspace.  The repetition is
\emph{cluster-granular}, not tile-granular: a $c_m\!\times\!c_n$ cluster
spans $64c_m\!\times\!64c_n$ output elements, which is $256\!\times\!256$
at the $4\!\times\!4$ grid of $C{=}16$ and $256\!\times\!128$ at the
portable $C{=}8$ grid $4\!\times\!2$ (Eq.~\eqref{eq:reach}), so an
operand block is re-split once per cluster that consumes it---
$\lambda_A = \lceil n/64c_n\rceil$ (the $A$ panel reused down the $n$ axis),
$\lambda_B = \lceil m/64c_m\rceil$, i.e.\ $\lceil n/256\rceil$ and
$\lceil m/256\rceil$ at $C{=}16$---for a size-weighted per-element
multiplicity $\lambda_{\mathrm{eff}} = (m\lambda_A + n\lambda_B)/(m{+}n)$
(Appendix figures; \texttt{params.py}: \texttt{\_lam\_eff}).  For a square
output this is $E/256$; those shapes take the on-the-fly (or, for the
smaller operand, L2-materialised) schedule above, not this branch.  The $k$-slab reduction ($S_k > 1$) is orthogonal to this
spatial-cluster question and is charged under \emph{either} tiling as
L2 service traffic ${\approx}\,2\cdot 4r\,mn\,(S_k{-}1)$ bytes,
i.e.\ ${\approx}\,4r/K_{\text{slab}}$ bytes per useful FLOP (useful
work $2mnk$ with $k = S_k K_{\text{slab}}$; the earlier
${\approx}\,8r/K_{\text{slab}}$ dropped the factor of two in the
work).  This traffic is L2-resident and generates \emph{no} HBM plane
traffic, but it is \emph{not} below $1\%$ of $Q_0$: as a fraction of
$Q_0$ it is ${\approx}\,rn/(2K_{\text{slab}})$, e.g.\
${\approx}\,150\%$ for a width-$n{=}1024$ square output at
$K_{\text{slab}}{=}4096$.  What must instead be checked is the L2
\emph{service} rate: at the achieved emulation rates the worst traced
record demands ${\approx}\,2.44$~TB/s of L2 service (the CCSD
$(441,441,4324)$ record on the E route; ${\approx}\,6.9$~TB/s were any
record run at the full $473$-TFLOP S roof).  That $B_{\mathrm{L2}} = 4B_{\text{mem}}$ ($4\times$ HBM) is a
model \emph{assumption}: NVIDIA documents the $126$-MB L2 capacity, not
a $4\times$ sustained reduction bandwidth.  Under
$1\times/2\times/4\times$ L2:HBM ratios ($22/44/88$~TB/s on Rubin,
$8/16/32$ on GB300) the worst-case demand stays under the cap in every
case, so no traced record is L2-service-bound \emph{within the assumed
$4\times$ service model}---the conclusion does not depend on the boost;
the companion engine charges this as an explicit L2 service cap
(\texttt{params.py}: \texttt{B\_L2}, \texttt{K\_SLAB}).  The engine
gates the $\lambda = 1$ branch on the single-cluster predicate
$T_m T_n \le C$ (\texttt{params.py}: \texttt{CLUSTER\_CTAS}${=}16$) and
tags $S_k = 1$ records \texttt{grid-fused-nosplit} to distinguish pure
spatial tiling from actual $k$-splitting; the launched-CTA count is
$T_m T_n S_k$, not merely the $T_m T_n$ spatial tiles.  The residue
accumulators live in per-CTA TMEM (PTX: a $512$-column $\times$
$128$-lane logical array, $256$~KiB), and they cost $4N_{\text{acc}}$
bytes per output, not $4r$: $N_{\text{acc}}$ is counted per modulus
(\S\ref{sec:planecounts}) and is $30/26/30/34$ for S/E/A/D, so an
unblocked $64^{2}$ tile asks $480/416/480/544$~KiB and \emph{nothing}
fits.  What fits is the blocked schedule of \eqref{eq:nblk}: at
$n_{\text{blk}}{=}2$ the live counts are $15/14/16/18$ tiles
$= 240/224/256/288$~KiB, so S, E and A fit (A with no headroom at all)
and D does not.  D fits only at $n_{\text{blk}}{=}3$ ($12$ tiles,
$192$~KiB), which doubles the \fp{64} slab re-read charge from
${\approx}1/3$ to ${\approx}2/3$ of \textsc{l2}; that, together with
its $337$~TF roof---the lowest in the menu---is why D is not carried as
a design point (\texttt{verify/blocking.py}).
This distinguishes the MMA $K$-chunk $K_{\text{mma}}{=}64$ from the
split slab $K_{\text{slab}} \ge 4096$.

\paragraph{Fused-hybrid with integer $\lambda$, and the surcharge
predicate.}  For $k$-large shapes outside the L2 predicate with one
small output dimension, the stream-side operand is converted in the
owning cluster with integer multiplicity
$\lambda = \lceil \min(m,n)/256 \rceil$ (the $256$-element square reach of
a $16$-CTA cluster), delivering
$\text{roof} \cdot \min\!\big(1,\, \nbar/(\lambda \cdot
\text{knee})\big)$ under ideal overlap.  The small-side operand's
planes are materialised once---in L2 if
$p\,k \min(m,n) \le C_{\mathrm{L2}}$, else in HBM at
$2p\,k \min(m,n)$ bytes---and the HBM surcharge relative to the
call's full traffic $Q_0 = 8\{k(m{+}n)+mn\}$ is
\begin{equation}
\delta \;=\; \frac{2p\,k\,\min(m,n)}{Q_0}.
\label{eq:hybsurcharge}
\end{equation}
For square shapes $\delta = p/12 = 2.50/3.75/2.75$ for S/A/E---the
surcharge \emph{dwarfs} the call traffic---and requiring
$\delta \le 0.2$ in the $k \gg n$ limit forces an aspect ratio
$m/n \ge 36.5/55.25/40.25$: the hybrid is a genuinely tall/skinny
schedule, eligible only where $\delta \le \varepsilon$ or its small
planes are L2-resident.  An earlier draft claimed a blanket
``$\le$10--20\% of call traffic'' surcharge; that claim is
\emph{withdrawn} for general shapes---it was a tall-skinny value.

\paragraph{L2 traffic, plane-feeding, and why the design generates on
the fly.}  The residue planes must reach the tensor cores somehow, and
there are two ways.  \emph{Materialise} them---write the $r$ planes to
memory once and stream them back---or \emph{generate} them on the fly,
splitting each operand block into planes in registers/TMEM immediately
before the MMA that consumes it (the fused deconstruction schedule of the
companion Part~1~\cite{matsuoka2026fp8part1}).  Materialisation looks cheaper (convert
once) but is \emph{feed-starved}: at the $64\times64$ TMEM-limited output
tile the operand reuse is $64$-fold, so streaming both materialised
operands to the cores costs $p/\mathrm{TILE}$ bytes per useful FLOP, and
to hold set~S's roof that feed must sustain $p/\mathrm{TILE}\cdot 473 =
222$~TB/s.  The companion figures for E and A, $226$ and $262$~TB/s, are
each taken at \emph{that set's own} roof ($438$ and $380$~TF), not at
S's $473$: the three sets do not share a roof, and pricing one set's
plane count at another set's roof is exactly the error corrected in
Table~\ref{tab:optionc}.  L2 at the assumed $4\times$HBM supplies
$88$~TB/s and HBM $22$; a materialised schedule is therefore capped at
$B_{\mathrm{L2}}/(p/\mathrm{TILE}) = 188/171/128$~TFLOPS (S/E/A, L2-resident)
or $B_{\mathrm{HBM}}/(p/\mathrm{TILE}) = 47/43/32$ (HBM-resident)---both
\emph{below} the on-the-fly rate derived next.  These three are the
\emph{feed} term alone; the per-record engine additionally charges the
one-off write of the planes into L2, so its realised L2-resident cap
sits a little under them (hence $186/169$ rather than $188/171$ at the
small-\textrm{NB} end of Table~\ref{tab:hplnb}), which tightens the
inequality in the same direction.  So the design generates
on the fly, and the residue planes never occupy L2 or HBM.  One
consequence is worth stating plainly, because it reverses a claim of our
own earlier draft: the emulated throughput is now \emph{insensitive} to
the L2 service bandwidth (the planes are not L2-resident), so the
$B_{\mathrm{L2}}$ assumption that a previous version made load-bearing no
longer conditions the result (the trace composites move by $\le 1$
native-$\times$ across $B_{\mathrm{L2}}=4/2/1\times$HBM).  What
on-the-fly generation \emph{does} cost is re-splitting.  A single
$16$-CTA cluster of $64\times64$ tiles covers a $256\times256$ output
($T_mT_n\le C$), inside which every operand block is split once and
shared through distributed shared memory ($\lambda=1$).  A larger output
spans several clusters, and---planes being on-chip and unshared across
clusters---each operand block is re-split once per cluster that consumes
it: $A[m,k]$ (reused down the $n$ axis) $\lceil n/256\rceil$ times,
$B[k,n]$ $\lceil m/256\rceil$ times, for a size-weighted multiplicity
$\lambda_{\mathrm{eff}}=(m\lceil n/256\rceil+n\lceil m/256\rceil)/(m{+}n)$.
For a square output of edge $E$ this is $E/256$; for a tall/skinny one it
stays $O(1)$---bounded, not unity: the large operand is split once
($\lambda_A{=}1$) and the small one many times but contributes little, giving
$1.25/1.50$ at LOBPCG's $110{,}592\times64/128$.  Those values are for the
square $4{\times}4$ dispatch the engine models, which is a conservative upper
bound rather than a property of the shape: a cluster \emph{shaped to the
output} ($c_mc_n\le C$; here $16{\times}1$ and $8{\times}2$) keeps every CTA
productive and lowers them to $1.06/1.25$ at no hardware cost, so the
tall/skinny entries below are quoted against the pessimistic dispatch.  The
split-$k$ \emph{reduction}
service---the term a previous round flagged---is by contrast negligible:
its worst demand across all trace records is $2.44$~TB/s, far under
$88$.  The engine charges all of this (\texttt{params.py}:
\texttt{rate\_rec}, on-the-fly vs.\ materialise, best wins).

\begin{figure}[t]
\centering
\includegraphics[width=\linewidth]{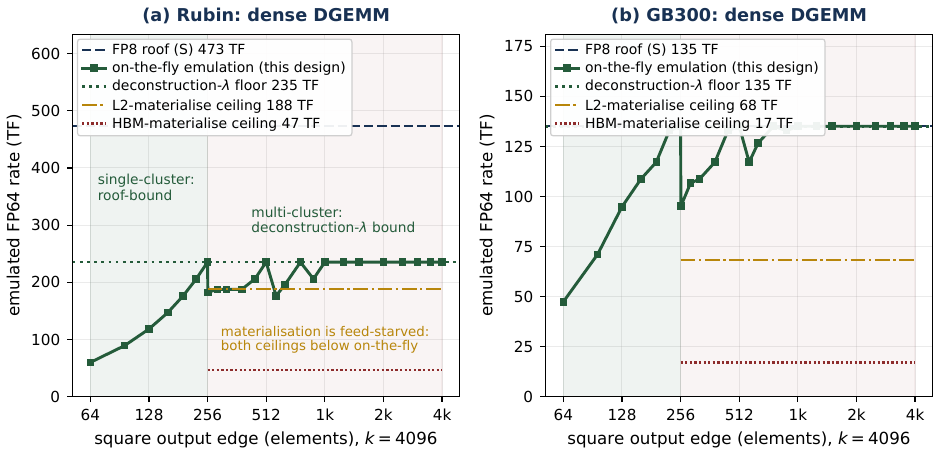}
\caption{Emulated dense DGEMM under on-the-fly generation.  Outputs up to
one cluster's reach ($256^2$) keep $\lambda=1$; larger square outputs are
pinned at the deconstruction-$\lambda$ floor (${\approx}235$~TFLOPS Rubin,
$0.50$ of set~S's $473$-TFLOP roof; the GB300 floor coincides with its lower
$135$-TFLOP roof, so GB300 is roof-bound throughout).  Every materialised
alternative (L2- or HBM-resident planes, dash-dot/dotted) is feed-starved
below the on-the-fly curve, which is why the design generates on the fly.}
\label{fig:dgemm}
\end{figure}

\paragraph{How the limit manifests, kernel by kernel.}  The
re-splitting cost has a scale-invariant consequence for dense
multiplication that
Figure~\ref{fig:dgemm} makes concrete.  For a square output the continuous idealisation $\lambda =
E/256$ grows with the edge exactly as the work does ($\bar n = E$), so
the per-FLOP deconstruction load is constant and the emulated rate settles at
${\approx}235$~TFLOPS on Rubin, $0.50$ of set~S's $473$-TFLOPS roof, for
cluster-aligned large
squares.  Under the realizable integer schedule $\lambda=\lceil E/256\rceil$
this is a \emph{sawtooth}: its aligned upper edge is the $0.50$ plateau, and
its ragged worst case (a $+1$ one-cluster sliver) is the $0.36$ rigid-schedule
minimum reported below.
(The floor is the deconstruction re-split load; it binds on the INT8/TMAC
deconstruction-MAC pipe for the winning E- and S-routes, not on the SIMT
residual.)  An output that fits one cluster ($\le 256^2$) keeps
$\lambda=1$ and follows the $\bar n$-crossover of \S\ref{sec:tme};
between the two, a narrow band is served by materialising the smaller
operand in L2.  Strikingly, the ${\approx}235$-TFLOPS large-DGEMM floor
lands on NVIDIA's preliminary ${\sim}200$-TFLOPS ``Emulated DGEMM''
figure from an \emph{independent} direction to the $\bar n{=}500$--$900$
crossover of contribution~(1): a DGEMM benchmark runs large square
shapes, exactly the regime the re-split floor governs.
Table~\ref{tab:kernman} carries this through the traced workloads and the
streaming kernels.  The picture is benign wherever real shapes live:
block-Krylov (LOBPCG) GEMMs are tall/skinny, so $\lambda_{\mathrm{eff}}$
stays $O(1)$ ($1.25/1.50$ at $b{=}64/128$ under square dispatch, $1.06/1.25$
shape-matched) and the hit is mild ($0.89$--$0.94\times$); multifrontal LU
and the rank-$64$ trailing updates of blocked dense LU are
memory-bound (small $k$), sitting below the floor and untouched
($1.00\times$); dense QR has no large-square/large-$k$ product
($0.95\times$); only CCSD, whose contractions are genuinely large-$k$
($k{\sim}1953$) and compute-bound, takes a real $0.82\times$.  Sparse and
streaming kernels (SpMV, SpMM, stencils) are exempt outright: each value
is converted once and consumed immediately on-chip, with no reuse to
re-split.  The limit, in short, is a large-\emph{dense}-DGEMM
phenomenon, not an application-throughput one.

\begin{table}[t]
\centering\small
\caption{How the deconstruction-$\lambda$ limit manifests per kernel
(Rubin).  Only large \emph{dense} square GEMMs with large $k$ sit at the
floor; the traced workloads are governed by their real shapes and are
largely immune.  ``B/A'' is the composite ratio of the honest on-the-fly
schedule (B) to the materialised one (A).
The tall/skinny $\lambda_{\mathrm{eff}}$ entries assume the square
$4{\times}4$ cluster dispatch the engine models; a shape-matched rectangular
cluster lowers them to $1.06/1.25$, so that row is conservative.}
\label{tab:kernman}
\setlength{\tabcolsep}{4pt}
\footnotesize
\begin{tabular}{@{}llll@{}}
\toprule
kernel / regime & shape & mechanism & outcome \\
\midrule
Dense DGEMM $\le 256^2$ & 1 cluster & $\lambda{=}1$; SMEM-fed on-chip & crossover model \\
Dense DGEMM $\gtrsim 512^2$, lg.\ $k$ & multi-cluster & re-split $\lambda{=}E/256$; feed-starved & $\approx$235 TF ($0.50$ of $473$) \\
LOBPCG ($b{=}64/128$) & tall-skinny & $\lambda_A{=}1$; $\lambda_{\mathrm{eff}}{\approx}1.25/1.5$ (bounded) & mild ($0.89$--$0.94\times$) \\
Multifrontal LU & $\bar n$ med.\ 665 & memory-bound fronts, below floor & immune ($1.00\times$) \\
CCSD contractions & $\bar n$ 441, $k{\sim}1953$ & large-$k$, compute-bound ($53\%$ work) & \textbf{hit} ($0.82\times$) \\
Dense LU trailing & \emph{rank-}$64$ & small $k{=}64\Rightarrow$ memory-bound & immune ($1.00\times$) \\
Dense QR & $k\in\{32,3232\}$ & no large-square/large-$k$ product & near-immune ($0.95\times$) \\
Sparse SpMV/SpMM, stencil & streaming & converted once, consumed on-chip & exempt \\
\bottomrule
\end{tabular}
\end{table}

\paragraph{Coverage, restated: a deconstruction-$\lambda$ bound, not an
L2 one.}  The earlier draft's ``no matrix-size regime is left uncovered''
and its blanket $\ge 0.83$ figure are withdrawn; so, one round later, is
the intermediate ``$0.79$, L2-bandwidth-conditional'' figure of the
draft that first charged plane-feeding---it charged the feed to a
\emph{materialised} schedule the design does not use, and (through a
$1024$-element cluster reach in place of the correct $256$) undercounted
the re-split multiplicity that actually binds.  Both are replaced by the
on-the-fly statement.  The enumeration domain is the same
boundary-oriented suite (clamped to $\le 4096$, $\pm1$ probes around every
$\lambda$ boundary $\min(m,n){=}256j$ and the L2-capacity widths), the
coverage ratio uses one reconstruction world in numerator and
denominator, and the numerator is now the best of on-the-fly,
hybrid, and (correctly feed-charged) materialised routes.  The operative
bound on the multi-cluster shapes is the on-the-fly re-split load: the
switched coverage minimum is the deconstruction-$\lambda$ floor,
\textbf{$0.50$ of the top ($473$-TFLOPS) roof, cluster-aligned}, on Rubin
(at large squares such as $m{=}n{=}666$), and $1.00$ on GB300---whose
$135$-TFLOP \fp{8} roof
sits at or below its own re-split floor, so the roof binds first and no dip
appears.\footnote{The \emph{strict} enumerated minimum under rigid
integer-$\lceil\cdot\rceil$ re-splitting is $0.36$ (Rubin) / $0.80$
(GB300), at $(513,513)$---a $+1$ ragged boundary where a one-element
sliver is charged a full extra re-split (area-weighted
$\lambda{=}2.004$ but $\lceil 513/256\rceil{=}3$).  A ragged-edge-aware
schedule (predicated partial clusters, standard in production GEMM)
is \emph{modeled} to hold $0.54$ there---a Part-3 validation target, since
we enumerate it but do not construct it as a minimum.  We therefore headline
the cluster-aligned $0.50$ plateau and report $0.36$ as the only
rigid-schedule worst case the enumeration establishes.}  Two conditions on the floor should be
stated, both Part-3 measurables: it is a \emph{large-$k$} value (the
Garner residual drags it to ${\approx}0.35$ at $k{=}256$, recovering by
$k{\gtrsim}1024$), and it assumes the $16$-CTA cluster opt-in (an
$8$-CTA-portable cluster is a $4{\times}2$ grid, which by
Eq.~\eqref{eq:reach} lowers the reach to $170.67$ and the floor to
${\approx}157$~TFLOPS; a $32$-CTA cluster is $8{\times}4$, reach $341.33$,
floor ${\approx}314$---see Table~\ref{tab:rungs}).
Crucially, because the planes are never L2-resident, the minimum is
\emph{no longer conditioned on} $B_{\mathrm{L2}}$: it is a compute
(deconstruction) bound, not a bandwidth one, and is therefore robust to
the $B_{\mathrm{L2}}$ assumption the previous round leaned on.  The
per-record engine and enumerator (\texttt{params.py}:
\texttt{rate\_rec}/\texttt{env\_rec}/\texttt{s0\_rec},
\texttt{coverage\_min}) apply these predicates to every trace record and
emit per-record predictions (\texttt{traces/sched\_*.csv}), from which
Table~\ref{tab:apps} is computed.

\paragraph{Persistent sparse operators.}  Mode P applied to a sparse
value array (operand-stationary O1) changes the dominant
value-plus-index stream from ${\approx}12$~B/nnz ($8{+}4$) to
${\approx}34$~B/nnz ($30{+}4$) for set S---$\times 2.83$---capping
delivered throughput near $0.35$ of the raw-value memory roof.
Operand-stationary conversion is therefore a
\emph{bandwidth-for-instructions} trade, not a free amortisation:
strictly worse than fused conversion on GB300 (where the fused SIMT
residual fits the keep-up budget), comparable on Rubin ($0.35$
against the fused $0.36$--$0.39$), and preferable mainly when the
plane array is L2-resident or SIMT issue is contended.

\section{Validation Plan and Falsifiable Predictions}
\label{sec:validation}

Everything above is a model projection; its constants come from the
NVIDIA memo's counted kernel~\cite{bayraktar2026tme} (stage provenance in Appendix~\ref{app:memo}) and from official
specifications~\cite{nvidia_dgx_rubin_nvl8,nvidia_rubin_blog}.  In the
two-way spirit of the TME model~\cite{matsuoka2026fp8part1}, we state
what measurement would confirm or refute, and note that each outcome
identifies its cause:

\begin{prediction}[Linearity]
\label{pred:linear}
Below the knee, the delivered throughput of emulated DGEMM on the
CRT/\fp{8} path is linear in $\nbar$---not flat---with slope
$P_{\text{int}}/(c_q r) \approx 0.39$~TFLOPS per unit $\nbar$ at the
reference constants.
\end{prediction}

\begin{prediction}[Knee position]
\label{pred:knee}
The signature is storage-mode-dependent, and the sweep must fix and report
the mode. On a \emph{convert-once} sweep---single-cluster tiling, or a
materialised kernel with L2-resident planes---the transition to the ceiling
occurs at $\nbar \approx n^{*}(c_q)$ of Eq.~\eqref{eq:crossover} and the
fitted knee measures the delivered $c_q$. On the \emph{deployable
on-the-fly} large-square path the curve instead knees at the cluster reach
($\nbar \approx 256$) into the deconstruction-$\lambda$ floor
(${\approx}0.50$ of the $473$-TFLOPS roof, \S\ref{sec:hw}) and never reaches
$n^{*}$; there
$c_q$ is read from the rising-branch slope $P_{\text{int}}/(c_q r)$ of
Prediction~\ref{pred:linear}, not the knee.
\end{prediction}

\begin{prediction}[Substrate structure]
\label{pred:control}
Under the fourth-term reading, a \emph{convert-once} size sweep shows
\emph{two} knees at the positions each path's constants predict---$\nbar
\approx 730$ on the B200 CRT/\inteight{} path, $\nbar \approx 1{,}211$ on the
Rubin CRT/\fp{8} path---whereas the deployable on-the-fly large-square sweep
shows, on each path, the same rising slope up to a plateau at the
deconstruction floor near the cluster reach; either signature is
constant-predicted. An error-free-slicing \inteight{} kernel on B200
(deconstruction by byte extraction; crossover below $\nbar \approx 25$)
shows neither at practical sizes and serves as the control arm.  Flat
sub-ceiling margins on \emph{all three} paths, in \emph{either} storage
mode, refute the branch hypothesis and indict generic efficiency
artifacts instead.
\end{prediction}

\begin{prediction}[Ladder shift, hierarchical]
\label{pred:ladder}
An Ozaki~2.5 prototype is validated in stages, each falsifiable
before the next is attempted: (i)~\emph{exact residue equivalence}
of the two-limb reduction against a trusted integer modulo,
exhaustively sampled over signed 64-bit inputs per modulus;
(ii)~\emph{isolated} conversion throughput at the logical
paired $K{=}8, N{=}r$ (or single $K{=}8, N{=}2r$, separate limb
outputs) shape; (iii)~\emph{concurrent} conversion/\fp{8}
throughput (the Pareto frontier, not two isolated peaks);
(iv)~end-to-end knee movement toward $\nbar \approx 480$--$730$ on
Rubin-class parts; (v)~application-level benefit on traced call
geometries.  A negative control---injecting the same volume of dummy
SIMT/\inteight{} work---separates deconstruction service demand from
generic small-GEMM inefficiency.
\end{prediction}

\noindent
The decisive experiment is a single DGEMM size sweep per path
(square and tall-skinny shapes to separate $\nbar$ from matrix
volume), over the three arms of
Prediction~\ref{pred:control}---Rubin CRT/\fp{8}, B200
CRT/\inteight{}, and a slicing \inteight{} control---each sweep fixing
and reporting its storage mode (single-cluster/convert-once vs.\
multi-cluster on-the-fly), since the two expose different but
constant-predicted signatures (the $n^{*}$ knee vs.\ the reach-$256$
deconstruction floor of \S\ref{sec:hw}); it decides the
branch hypothesis, fits each path's $c_q$, and prices the recovery
in one run.  The conversion-kernel
derby of Part~1---including the \texttt{DP4A} and tensor-migrated
variants against the memo's ${\sim}7$-instruction
criterion~\cite{bayraktar2026tme}---and the integer-pipe census that
fixes the true $P^{\text{eff}}_{\text{int}}$ complete the set.  These
are the opening entries of the follow-up implementation and
measurement work (Part~3), and are being automated on RIKEN's Rikyu GB200~NVL4
system so that the sweep re-runs continuously as kernels evolve (the
GB200 testbed calibrates the model; it does not by itself validate
GB300's reduced integer-tensor balance or Rubin's concurrency, which
require those parts).

\paragraph{Pass/fail criteria.}  Each stage carries an explicit
criterion: \emph{(correctness)} bit-exact agreement with an
arbitrary-precision reference over exhaustive 16-bit and stratified
random signed 64-bit inputs per modulus, including even-modulus
boundaries ($-m/2$), $\pm 0$, subnormals, and the API's
infinity/NaN contract; \emph{(accounting)} SASS instruction counts
per output residue, achieved \texttt{dp4a} issue rate, register
count, and occupancy---this directly adjudicates the L1 direct count
of \S\ref{sec:ozaki25} and its ${\approx}7.5$ target;
\emph{(reduction shape)} useful \emph{and} issued operations at the
exact paired-limb shapes, isolated and concurrent, fixing
$\eta_{\text{red}}$; \emph{(concurrency)} the paired-rate
measurements: isolated \fp{8}, \inteight{}-MMA, and \texttt{dp4a}
rates; then \fp{8}$+$\inteight{} and \fp{8}$+$\texttt{dp4a} at
matched occupancy, each pair reported against its serialised,
ideal-overlap, and measured predictions---fixing $\theta$ and
$\rho_{\text{simt,fp8}}$; \emph{(knee)} a piecewise fit with confidence
intervals against the size-independent-efficiency alternative on
held-out shapes, not visual inspection; \emph{(numerics)} the source
accuracy suite re-run per candidate set with adversarial
scaling/conditioning; \emph{(applications)} the to-be-released traces
replayed through measured kernels, GEMM-portion and whole-run Amdahl
numbers reported separately.

\paragraph{Claim status.}  For the record, the paper's claims
separate as follows.  \emph{Algebraically proved:} the
harmonic-mean crossover under the reduced model; the two-limb byte
identity and its $2^{19}$ accumulation bound; coprimality and exact
CRT products of the released candidate sets.  \emph{Script-checked
(scripts in the artifact:
\texttt{lemma\_search.py}, \texttt{verify\_candidates.py},
\texttt{stail\_layout.py}):} the
exhaustive supply bounds; per-set coprimality and exact CRT products;
canonical centred digit maps and E4M3 representability of all A/E/D
planes and the six square S moduli; the carry-corrected layout of the
six nonsquare S-tail moduli, exhaustive over all $m^{2}$ ordered
centred pairs per modulus, together with the canonical ($321$) and
redundant ($577$) envelopes that bracket
it; the signed two-limb identity (exhaustive 16-bit plus
stratified 64-bit).  \emph{Ledger-projected (uncompiled instruction
counts):} the SIMT residuals (the uniform ${\approx}6.3$ and the
per-set $5.8$/$5.3$/$5.2$ for S/E/D, $5.27$ for A) and the \texttt{dp4a} route counts
($10.3$/$7.3$)---instruction-ledger projections, neither compiled
nor measured.  \emph{Memo-counted:} the $c_q{=}16$ shipping-path
SASS estimate (Appendix~\ref{app:memo}).  \emph{Script-generated
(\texttt{params.py}):} every number in
Tables~\ref{tab:codesign}--\ref{tab:apps}, as deterministic
evaluation of the stated constants---generation is not independent
verification of those constants.  \emph{Measured (fixed software
stack):} the \texttt{LD\_PRELOAD} call-shape traces, and nothing
else.  \emph{Modeled
(conditional):} all route knees, envelopes, and composite speedups.
\emph{Hypothesized:} the deconstruction-limited reading of the
Rubin specification.  \emph{To be measured:} the L1 reuse schedule,
$\eta_{\text{red}}$, the concurrency parameter $\theta$, storage-mode
boundaries, end-to-end knees, per-set
numerical behaviour, and application replay.  The artifact
(\texttt{params.py}, \texttt{lemma\_search.py},
\texttt{verify\_candidates.py}, the traces, the per-record
predictions \texttt{traces/sched\_*.csv}, and the table manifests),
with a manifest mapping every table and figure cell to its generating
command, \emph{is supplied with this submission} as a reproduction
artifact---an archival Zenodo DOI and commit hash will be added in the first arXiv revision
and cited in the camera-ready.  In the revised artifact the per-record CSVs
(\texttt{traces/sched\_*.csv}) emit the selected route, storage mode,
reconstruction host, $Q_0$, $Q_{\text{plane}}$, $Q_{\text{red}}$, and
the selected rate as audit fields, and a single-command driver
regenerates every table, figure, and report from the raw traces with
recorded tool versions.  Trace quantiles and composites are FLOP-weighted.
\texttt{params.py} is the single parameter source.

\section{Conclusion}
\label{sec:conclusion}

Ozaki~II made \fp{64} dense matrix multiplication a schedule of \fp{8}
tensor-core operations; Rubin made the result a product specification.
This paper has argued that part of the distance between that
preliminary specification ($\sim\!200$~TFLOPS) and the raw arithmetic
roof ($\approx\!473$~TFLOPS) may have a specific, measurable, and
largely removable cause: the deconstruction of streamed \fp{64}
operands into CRT residues, a cost the \inteight{} substrate could
largely avoid---by error-free slicing, or at least Karatsuba-free---and
that the \fp{8} substrate, mandatory once \inteight{} tensor
throughput is deprecated, pays in full per modulus on every element.
The size crossover $n^{*} = c_q r P_{\fp{8}}/(\alpha P_{\text{int}})$
makes the hypothesis quantitative: at the memo-derived constants the
published figure is consistent with the deconstruction-limited branch
at model-inferred sizes $\nbar \approx 500$--$900$, and a single size
sweep---with a slicing \inteight{} kernel as control---decides the
question while measuring $c_q$.

Honesty about the ceiling must close the paper as it opened it. On today's
silicon the deliverable for \emph{large dense DGEMM}---the regime HPL runs
in---is not the roof but the deconstruction-$\lambda$ floor,
${\approx}235$~TFLOPS on Rubin ($0.50$ of the common $473$-TFLOPS roof, or
equivalently $0.54$ of set~E's own $438$; ${\approx}8\times$
native, ${\approx}1.9$~EFLOPS \fp{64} sustained on a $10{,}000$-GPU
cluster) on the codesigned set~E at $\textrm{NB}\gtrsim1024$, or
${\approx}182$~TFLOPS ($0.38$, ${\approx}1.4$--$1.5$~EFLOPS) on the published
set~S, which is the number backed by the round-to-nearest theorem
(Table~\ref{tab:hplnb}).
The rising envelopes of Fig.~\ref{fig:knee} are achievable only up to one
cluster's reach; larger outputs plateau there. That floor is
$R\,P_{\text{int}}/(c_q r)$, liftable to the roof only by the co-design of
\S\ref{sec:hw}---minimally an in-flight-convert copy-engine datapath
(Plan~A), with a higher-bandwidth L2 under software-blocked materialisation
as the datapath-independent Plan~B. The equally important positive is that
this floor is a \emph{large-square} phenomenon: the tall/skinny and
small-batch matrices that dominate real solvers---block-Krylov, batched
GEMMs, panel factorisations---re-split their big operand only once
($\lambda_A{=}1$; size-weighted $\lambda_{\mathrm{eff}}{\approx}1.25$--$1.5$,
bounded, not the $\propto$edge of a large square), so they stay near the
crossover (a mild $0.89$--$0.94\times$, not the $0.50$ floor) and Ozaki~2.5 is
\emph{already useful on Rubin today}, worth ${\approx}1.6$--$1.9\times$
(${\approx}2\times$ on the block-Krylov rows) over simple deconstruction with no
hardware change (Table~\ref{tab:apps}), and GB300 is roof-bound outright. Only the headline
large-square-DGEMM number is half the roof until the hardware changes, and we
state it that way---clipping-limited, not fundamental---rather than leaving
the roof as the last impression.

The constructive contributions are the Ozaki~2.5 method and the
modulus/encoding codesign it led to.  The method adds no new
mathematics: convert once, reduce exactly on the integer tensor side
(two limbs, because the actual moduli exceed a byte), keep only the
irreducible nonlinearities on SIMT, and hide the rest behind the MMA
stream.  Under stated assumptions the reduced model projects the
Rubin knee moving from $\approx\!1{,}211$ to $\approx\!480$--$730$
and a conditional $1.8$--$2.2\times$ at the announced operating
region ($364$--$438$~TFLOPS at $\nbar{=}512$, route-dependent---a convert-once
envelope, delivered to within a mild clip
($\lambda_{\mathrm{eff}}{\approx}1.25$--$1.5$) for tall/skinny but clipped to the
${\approx}235$-TF floor for a large square); the
measured result, whatever it is, will replace these projections.  The codesign study of \S\ref{sec:codesign} then shows
the modulus set itself is a first-class, runtime-switchable
performance parameter: an all-byte system legalises the one-pass
reduction and is projected faster below $\nbar \approx 540$ despite a
$20\%$ lower roof, and the hybrid concedes only $7.5\%$ asymptotically
while leading the ${\approx}410$--$620$ band---so a library can
dispatch by shape and lose almost nothing anywhere.  None of this is
Rubin-specific: instantiated at Blackwell~Ultra (GB300) rates the same
machinery gives $135/109/125$-TFLOPS roofs but a different
winner---the ${\sim}166$-\tops{} residual integer-tensor rate
throttles the tensor-migrated reductions, compressing the small-size
band, which the all-byte set leads (its \texttt{dp4a} realisation
within ${\approx}7\%$, needing no integer tensor at all), with the
two-limb route taking the roof from $\nbar \approx 290$---so even
the dispatch itself is
platform-dependent, which is precisely what a parameterised model is
for.  And because GB300's native \fp{64} pipe delivers only
${\approx}1.4$~TFLOPS, every modeled route exceeds the native
reference throughout the evaluated range $\nbar \ge 32$: Ozaki~2.5
is a proposition for the Blackwell generation already shipping, not
one that waits for Rubin.  The measured call-shape traces of
\S\ref{sec:apps} ground this where it matters: block-Krylov and
QR-panel work is block-width-bounded in real runs, and multifrontal
fronts sit in the crossover band itself.  For workloads
whose geometry keeps one output dimension at a block width
(tall-skinny and block-vector GEMMs), these gains, if confirmed,
apply throughout that band; for what software cannot
reach, Part~1's hardware options remain the outlook---Option-C-class
conversion would retire the term at every size---but on current GPUs
the switched software dispatch already does well across realistic
matrix sizes.  Whether the branch
hypothesis survives measurement or falls to it, the outcome
identifies the responsible term---which is exactly what a performance
model is for.

\section*{Acknowledgments}

The author is indebted to Harun Bayraktar, John Gunnels, and Peter
Caday of NVIDIA, whose technical note~\cite{bayraktar2026tme}
(cited with permission) identified the deconstruction term on which
this paper rests, and to Dan Ernst and Matthew Martineau for
discussions that helped shape the experimental programme; the
follow-up implementation and
measurement work that will test the predictions made here is in
preparation (Part~3).  The author also thanks the
RIKEN R-CCS teams standing up the Rikyu GB200~NVL4 measurement
harness.  This work was undertaken as part of the FugakuNEXT project
and related R-CCS initiatives on AI for Science.

\paragraph{Disclosure of AI-assisted writing.}  This manuscript was
prepared with assistance from large language models: Anthropic's
Claude (Opus~4.8 and Fable~5, with the final pre-submission pass of
Draft~27 by Fable~5.1) served as authoring assistants for
drafting, the derivation checks, figure generation, and \LaTeX{}
mechanics, and OpenAI's GPT-5.6 (Codex) provided critical technical
review across drafts, under the author's direction.  All scientific arguments,
performance projections, and conclusions were directed, reviewed, and
validated by the author, who takes full responsibility for the
content, including any errors of fact or judgment.

\appendix
\section{Proof Obligations per Candidate Modulus Set}
\label{app:obligations}

Changing the modulus set leaves the Ozaki-II reconstruction
\emph{framework} intact but does not inherit the source paper's
numerical argument automatically.  Table~\ref{tab:variants} fixes
the numerical-contract variants by name; every accuracy statement in
this paper is scoped to one of them.

\begin{table}[!ht]
\caption{Named numerical-contract variants and their status.}
\label{tab:variants}
\centering
\footnotesize
\begin{tabular}{lp{0.33\linewidth}p{0.40\linewidth}}
\toprule
variant & conversion rule & status \\
\midrule
Ozaki-II-RN & round-to-nearest & \emph{theoretical}: the cited error
theorem~\cite{ozaki_error_analysis_2026} holds under its own
hypotheses \\
Ozaki-II-TZ & implemented truncation, plus the stated endpoint rule
& \emph{proposed/unvalidated}: no automatic inheritance from RN; a
proof or a measured error distribution is required \\
Codesigned A/E/D & per-set digit and endpoint rules & \emph{obligation}:
performance case made here; accuracy is an explicit per-set
obligation, items (i)--(vi) below \\
\bottomrule
\end{tabular}
\end{table}

For each released candidate
(A, E, D) the following obligations are stated here and
script-checked (\texttt{verify\_candidates.py}, in the artifact with its
output manifest);
SASS-level and
accuracy-suite confirmation belongs to \S\ref{sec:validation}.  The
checker reports \textsc{pass}/\textsc{skip}/\textsc{fail} per
obligation and never returns an all-pass verdict while any obligation
is skipped.
(Candidate D, the one set not detailed in \S\ref{sec:codesign}, is
the 7-bit system $\{128$, $127$, $125$, $123$, $121$, $119$, $113$,
$109$, $107$, $103$, $101$, $97$, $89$, $83$, $79$, $73$, $71\}$
($r{=}17$, $\alpha'{=}52$, one-pass reduction with signed-byte
constants); the list ships in \texttt{verify\_candidates.py} and the
artifact manifest.)
\emph{(i)~Range:} pairwise coprimality and the exact CRT product
($\log_2 P = 117.8$ for A, $113.3$ for E, $113.5$ for D; the
published set's $111.8$; $\log_2 P$---total CRT product bits---is
the single range metric used throughout this paper, the usable
symmetric range being $P/2$, exactly one bit less)---each meets or
exceeds the source
requirement.  \emph{(ii)~Digit map:} residues carried centred in
$[-m/2, m/2)$, even-modulus endpoint included at $-m/2$ (for
$m{=}256$, the value $-128$); under the canonical tie rule
$d_0 \in [-8, 7]$, $d_1 = (x{-}d_0)/16$ the decomposition is unique
on that interval (without a tie rule balanced digits are redundant at
ties, \eg $8 = 0 \cdot 16 + 8 = 1 \cdot 16 - 8$).  \emph{(iii)~Representability:} every
stored digit and every Karatsuba sum actually used is
\textsc{e4m3}-exact (checked exhaustively per modulus).
\emph{(iv)~Accumulation:} all reduction dot products bounded by
$8 \cdot 255 \cdot 255 < 2^{19}$, exact in \intthirtytwo{}; the
\fp{32} MMA accumulation bound and $k$-blocking rule of the source
apply unchanged per set.  \emph{(v)~Signed equivalence:} the limb
formation with the $(-2^{64}) \bmod m_i$ correction equals a trusted
signed integer modulo on exhaustive 16-bit and stratified 64-bit
samples.  \emph{(vi)~Contract:} scaling, truncation, ESC/ADP
escalation, and reconstruction conditions of the source analysis are
re-verified per set as part of the validation suite---until then the
accuracy contract for A/E/D is a stated obligation, not a result.

\section{Provenance of the Memo-Counted Constants}
\label{app:memo}

The model's anchor constants---$c_q = 16$ and the
$P_{\text{int}} \approx 75$~\tops{} normalisation---derive from a
private NVIDIA technical note~\cite{bayraktar2026tme}, cited with
permission.  To make the anchor auditable we reproduce the
non-confidential stage structure; the note's authors have been asked
to confirm this summary for publication.  The confirmation request
also asks which emulation realisation the counted SASS belongs
to---the released Scheme-I slicing path or a CRT (Scheme-II-class)
path---since the stage structure is CRT-flavoured while the released
cuBLAS path is publicly described as
Scheme-I~\cite{emugemm2026,nvidia_cuest}; \S\ref{sec:validation}
carries the reading as conditional pending that answer.  Stages, per streamed
element and per modulus on the shipping conversion path
(SASS-counted \emph{estimates}, not throughput measurements;
architecture and toolchain preliminary): scale and
truncate-to-integer, ${\sim}4$ per element (amortised $4/r$ per
modulus); per-modulus byte-plane reduction, ${\sim}8$--$10$
(constant division compiling to multiply-high sequences); final
reduction to the residue interval, ${\sim}2$; Karatsuba split into
three \fp{8} pieces, ${\sim}3$; totalling $c_q \approx 16$ under
the stated amortisation.  One \texttt{dp4a} counts as one issued
SIMT instruction throughout ($I_{\text{SIMT}} = 75 \cdot 10^{12}$
instructions/s in the memo's normalisation); \inteight{} tensor
capacity is counted separately in operations/s
($P_{\text{I8TC}}$, two operations per MAC).  Nothing else from the
note is used.

\paragraph{Reconciliation with the spectral companion.}  The FFT
companion~\cite{matsuoka2026fft} normalises the same pipe at
$41.7$~T~inst/s, the published B300 \fp{32} TFLOP/s divided by two on
the assumption that \fp{32} and integer issue share lanes on
Blackwell.  This paper keeps the memo's $75$ as its reference constant
because it is the normalisation the memo's counts were made against;
the $1.8\times$ gap between the two is exactly what the Part-3
integer-pipe census measures.  Every $I_{\text{SIMT}}$-bounded quantity
here---the SIMT keep-up budget ($6.25 \to {\approx}3.5$ instructions
per modulus on GB300, $2.3 \to {\approx}1.3$ on Rubin at $41.7$), the
SIMT-residual knees, and the Garner keep-up thresholds---tightens
monotonically under the spec-derived value, so the $41.7$ case is a
uniform downside to the software-only picture and a uniform
strengthening of the co-design case.  The tensor-side quantities
($473$~TFLOPS; the $0.497$ ratio of the deconstruction-$\lambda$ term
to the roof in Eq.~\eqref{eq:whyhalf}; the \inteight{}-tensor knees)
do not depend on it, but the \emph{floor itself} does, because it is
a minimum over terms: re-running the engine at $41.7$~T~inst/s, the
SIMT-residual term becomes binding on Rubin and the large-square floor
falls from $235$ to ${\approx}148$~TFLOPS ($0.31$ of $473$; routes~E and
S at $148$ and $147$, A at $128$), and on GB300 the \texttt{dp4a}
route~L1d that supplies the $135$~TFLOPS floor drops to $84$, leaving
route~E at its own $125$~TFLOPS roof as the floor.  Route~E leads on
both parts under either normalisation; what the census settles is
whether the floor is one half of the roof or one third of it, which
makes $P_{\text{int}}$ the single largest sensitivity this paper
carries.  The $75$-normalised figures are the ones this paper quotes;
the $41.7$ case is the stated sensitivity, and both floor values are
engine-checked (\texttt{paper\_numbers.py}, \texttt{check\_pint\_sensitivity}).

\bibliographystyle{plain}
\bibliography{references}

\end{document}